\documentclass[12pt,article,nofootinbib]{revtex4}%
\usepackage{amsfonts}
\usepackage{amsmath}
\usepackage{amssymb}
\usepackage{graphicx}%
\providecommand{\U}[1]{\protect \rule{.1in}{.1in}}

\renewcommand{\thesection}{\arabic{section}}

\makeatletter \@addtoreset{equation}{section}
\renewcommand{\theequation}{\thesection.\arabic{equation}}
\begin{document}
\preprint{ }
\title{\rightline{\mbox{\small {LPHE-MS-16-01}} \vspace
{1cm}} \textbf{Type II seesaw supersymmetric neutrino model for }$\theta
_{13}\neq0$\textbf{\ }}
\author{R. Ahl Laamara, M.A Loualidi, and E.H Saidi}
\affiliation{{\small 1. LPHE-Modeling and Simulations, Faculty of \ Sciences,}}
\affiliation{{\small Mohammed V University, Rabat, Morocco}}
\affiliation{{\small 2. Center of Physics and Mathematics, CPM- Morocco, Rabat-10090,
Morocco}}

\begin{abstract}
Using the type II seesaw approach and properties of discrete flavor symmetry
group representations, we build a supersymmetric $A_{4}\times A_{3}$\ neutrino
model with $\theta_{13}\neq0$. After describing the basis of this model--which
is beyond the minimal supersymmetric Standard Model--with a superfield
spectrum containing flavons in $A_{4}\times A_{3}$\ representations, we first
generate the tribimaximal neutrino mixing which is known to be in agreement
with the mixing angles $\theta_{12}$ and $\theta_{23}$. Then, we give the
scalar potential of the theory where the $A_{3}$ discrete subsymmetry is used
to avoid the so-called sequestering problem. We \textrm{next} study the
deviation from the tribimaximal mixing matrix which is produced by perturbing
the neutrino mass matrix with a nontrivial $A_{4}$ singlet. Normal and
inverted mass hierarchies are discussed numerically. We also study the
breaking of $A_{4}$\ down to $Z_{3}$\ in the charged lepton sector, and use
the branching ratio of the decay $\tau \rightarrow \mu \mu e$--which is allowed
by the residual symmetry $Z_{3}$--to get estimations on the mass of one of the
flavons and the cutoff scale $\Lambda$ of the model.\newline Key words:
Neutrino family symmetry, supersymmetry, deviation from TBM

\end{abstract}
\email{h-saidi@fsr.ac.ma}
\volumeyear{ }
\volumenumber{ }
\issuenumber{ }
\eid{ }
\maketitle


\section{Introduction}

In the Standard Model (SM) of electroweak interactions, neutrinos $\left(
\nu_{i}\right)  _{{\small i=1,2,3}}$ are left-handed and massless; this is
because in the SM there are no right-handed neutrino singlets $\nu_{iR}$ that
allow gauge-invariant Yukawa couplings to the Higgs doublet $y\left(
H.L_{i}\right)  \nu_{iR}$. However, recent experimental data on neutrino
oscillations have shown that they have very tiny masses $m_{i}$ and that the
different flavors $\nu_{1},\nu_{2},\nu_{3}$ are mixed with some mixing angles
$\theta_{ij}$, as shown in Table \ref{t} below. This important discovery led
to awarding the Nobel Prize in Physics for 2015 to Takaaki Kajita
(SUPER-KAMIOKANDE Collaboration) and Arthur B. McDonald (SNO Collaboration).
Although we cannot determine the exact masses $m_{i}$ of the neutrinos, many
experiments performed in\ the last few years measured the squared-mass
differences $\Delta m_{ij}^{2}=m_{i}^{2}-m_{j}^{2}$ and mixing angles
$\theta_{ij}$, as reported by several global fits of neutrino data
\textrm{\cite{A1,A2,A3}, }the most recent of which can be found in
Ref.\textrm{\  \cite{A4}}.

\begin{table}[h]
\centering \renewcommand{\arraystretch}{1.2}
\begin{tabular}
[c]{ll||ll}%
$\  \  \  \  \  \  \  \text{Parameters}$ &  &  & $\  \  \  \  \ $B$\text{est
fit}_{\left(  -1\sigma,-2\sigma,-3\sigma \right)  }^{\left(  +1\sigma
,+2\sigma,+3\sigma \right)  }$\\ \hline
$\  \  \  \  \Delta m_{21}^{2}\left[  10^{-5}\text{eV}^{2}\right]  $ &  &  &
$\  \  \  \  \  \ 7.60_{\left(  -0.18,-0.34,-0.49\right)  }^{\left(
+0.19,+0.39,+0.58\right)  }\left.
\begin{array}
[c]{c}%
\text{ }\\
\text{ }%
\end{array}
\right.  $\\ \hline
$%
\begin{array}
[c]{c}%
\Delta m_{31}^{2}\left[  10^{-3}\text{eV}^{2}\right]  \text{(NH)}\\
\Delta m_{31}^{2}\left[  10^{-3}\text{eV}^{2}\right]  \text{(IH)}%
\end{array}
$ &  &  & $\  \  \
\begin{array}
[c]{c}%
2.48_{\left(  -0.07,-0.13,-0.18\right)  }^{\left(  +0.05,+0.11,+0.17\right)
}\\
-2.38_{\left(  -0.06,-0.12,-0.18\right)  }^{\left(  +0.05,+0.10,+0.16\right)
}%
\end{array}
\left.
\begin{array}
[c]{c}%
\text{ }\\
\text{ }\\
\text{ }%
\end{array}
\right.  $\\ \hline
$\  \  \  \  \  \  \  \  \  \  \sin^{2}\theta_{12}$ &  &  & $\  \  \ 0.323_{\left(
-0.016,-0.031,-0.045\right)  }^{\left(  +0.016,+0.034,+0.052\right)  }\left.
\begin{array}
[c]{c}%
\text{ }\\
\text{ }%
\end{array}
\right.  $\\ \hline
$\  \  \  \
\begin{array}
[c]{c}%
\sin^{2}\theta_{23}\text{(NH)}\\
\sin^{2}\theta_{23}\text{(IH)}%
\end{array}
$ &  &  & $\  \
\begin{array}
[c]{c}%
0.567_{\left(  -0.128,-0.154,-0.175\right)  }^{\left(
+0.032,+0.056,+0.076\right)  }\\
0.573_{\left(  -0.043,-0.141,-0.170\right)  }^{\left(
+0.025,+0.048,+0.067\right)  }%
\end{array}
\left.
\begin{array}
[c]{c}%
\text{ }\\
\text{ }\\
\text{ }%
\end{array}
\right.  $\\ \hline
$\  \  \  \
\begin{array}
[c]{c}%
\sin^{2}\theta_{13}\text{(NH)}\\
\sin^{2}\theta_{13}\text{(IH)}%
\end{array}
$ &  &  & $%
\begin{array}
[c]{c}%
0.0234_{\left(  -0.0020,-0.0039,-0.0057\right)  }^{\left(
+0.0020,+0.004,+0.006\right)  }\\
0.0240_{\left(  -0.0019,-0.0038,-0.0057\right)  }^{\left(
+0.0019,+0.0038,+0.0057\right)  }%
\end{array}
\left.
\begin{array}
[c]{c}%
\text{ }\\
\text{ }\\
\text{ }%
\end{array}
\right.  $\\ \hline
\end{tabular}
\caption{The global fit values for the mass squared differences $\Delta
m_{ij}^{2}$ and mixing angles $\theta_{ij}$ as reported by Ref. \cite{A2}. NH
and IH stand for normal and inverted hierarchies respectively. }%
\label{t}%
\end{table}To deal with the small masses and mixing of neutrinos we need to go
beyond the SM framework; for this purpose many neutrino models have been
proposed in recent years, and it is common that the observed mixing angles
$\theta_{12}$ and $\theta_{23}$ are close to the tribimaximal mixing matrix
(TBM), which predicts them to be in the $2\sigma$\ and $3\sigma$ ranges, as in
Table \ref{t} \textrm{\cite{A5}}. The remaining $\theta_{13}$ is however not
compatible with TBM, as announced by recent experiments
\textrm{\cite{A6,A7,A8,A9}}, although TBM still remains a good approach to the
present data. We recall that one way to reproduce TBM at leading order (LO) is
to go beyond the usual spectrum of the Standard Model via discrete non-Abelian
groups like the alternating $A_{4}$ symmetry, which is admitted as the most
natural discrete group that captures the family symmetry, as motivated in the
literature. Following Altarelli and Feruglio\textrm{\  \cite{AF}, }$A_{4}%
$\textrm{\ }models have a particularly economical and attractive structure,
e.g., in terms of group representations and field content \cite{ER,ZE,MB,BE}.
For neutrino models based on other discrete groups see, for instance, Ref.
\textrm{\cite{A12}},\textrm{\ }and for\textrm{\ }an introduction to
non-Abelian discrete symmetries and representations see Ref.
\textrm{\cite{A10}} and references therein. Recall also that there are several
ways to generate masses for neutrinos beyond the standard model, such as the
implementation of dimension-five nonrenormalizable operators
\textrm{\cite{A13}} or by using the three types of the seesaw mechanism: type
I with extra SU(2) singlet fermions, type II with an extra SU(2) triplet
scalar, and type III with an extra SU(2) triplet fermion
\cite{14,15A,15,16,17}.

\  \  \  \  \newline In this paper, we propose a supersymmetric neutrino model
with discrete flavor symmetry $A_{4}\times$\ $A_{3}$ that extends the minimal
supersymmetric SM (MSSM), and whose theoretical predictions for $\Delta
m_{ij}^{2}$ and $\sin^{2}\theta_{ij}$ are compatible with
experiments\textrm{\  \cite{A6,A7,A8,A9}. }This field theory prototype is a
supersymmetric type II seesaw neutrino theory based on a particular extension
of the (MSSM). In addition to the usual MSSM superfield spectrum and the
chiral superfield triplets of the type II seesaw model, our model involves the
extra flavon chiral superfields $\{ \vec{\chi},\vec{\chi}^{\prime},\Phi
,\Phi^{\prime}\}$ carrying quantum numbers under $A_{4}\times$\ $A_{3}$
discrete symmetry. $\vec{\chi}$\ is needed by the $A_{4}$\ symmetry in charged
sector, while the three others concern the chargeless sector: $\vec{\chi
}^{\prime}$ to realize the tribimaximal texture, $\Phi$\ to reproduce the
correct mass squared difference $\Delta m_{31}^{2}\neq0$, and $\Phi^{\prime}$
to generate $\theta_{13}\neq0$. By giving vacuum expectation values (VEVs) to
these flavons, one generates Majorana mass terms and induces neutrino mixing
compatible with the observations listed in Table I. Notice that supersymmetry
plays a crucial rule in our construction; it is needed to have the right
vacuum alignment and to overcome the sequestering problem, as was first
noticed in Refs.\textrm{\ }\cite{31,32}. Without supersymmetry there is no way
to forbid terms of the form $\lambda_{\chi \chi^{\prime}}\left \vert
\chi \right \vert ^{2}\left \vert \chi^{\prime}\right \vert ^{2}$ in the scalar
potential which destroys the desired VEV structure in four-dimensional
renormalizable theories. With supersymmetry, the scalar potential is derived
from complex $F$ terms in the chiral superpotential $W=W\left(  \chi
,\chi^{\prime};...\right)  $ sector, and Hermitian $D$ terms of the Kahler
$K\left(  \chi,\chi^{\dag},\chi^{\prime},\chi^{\prime \dag};......\right)
$\ involving gauge interactions; terms like undesirable $\left \vert
\chi \right \vert ^{2}\left \vert \chi^{\prime}\right \vert ^{2}$ come from
complex $W$ and may be eliminated by an extra discrete symmetry having complex
representations. Notice also that aspects of the type II seesaw mechanism for
neutrinos with an $A_{4}$ flavor symmetry were considered before in Ref.
\cite{21} but without supersymmetry. In our supersymmetric extension, the two
$A_{4}$ flavon superfield triplets $\vec{\chi}$ and $\vec{\chi}^{\prime}$, act
respectively, in the charged lepton sector and neutrino sector; they carry
different charges under the extra $A_{3}$ discrete subsymmetry which is needed
to exclude unwanted terms in the superpotential $W$ and to avoid the
communication between charged and chargeless sectors. To engineer appropriate
squared mass differences $\Delta m_{ij}^{2}$ and mixing angles $\sin^{2}%
\theta_{ij}$ in the chargeless sector, we find that we also need to implement
two $A_{4}$ scalar flavon chiral superfields $\Phi$ and $\Phi^{\prime}$. By
giving them VEVs, we obtain TBM consistent with the experimental data on
$\Delta m_{ij}^{2}$ and $\sin^{2}\theta_{13}$. In this regard, we recall that
several models use different approaches to generate a $\theta_{13}$ deviation
from the TBM pattern; for instance, in Ref. \cite{20}, the deviation of TBM is
obtained by adding a nonleading contribution coming from charged lepton mass
diagonalization. In\textrm{\ Ref. }\cite{21}, the TBM was generated at LO with
the type I seesaw mechanism and the deviation was made by perturbing the
neutrino mass matrix with the type II seesaw mechanism. In our approach, we
borrow techniques from the method used in Ref. \cite{22} before $\theta
_{13}=0$ was ruled out. This method relies on perturbing the neutrino mass
matrix by adding nontrivial $A_{4}$ singlets and has been used recently in
Ref. \cite{23} where neutrino masses were generated by dimension-five
operators. After a numerical study, we show that normal and inverted
hierarchies are both permitted. The VEV of the triplet $\vec{\chi}$\ breaks
$A_{4}$\ down to $Z_{3}$\ in the charged lepton sector; because of this
residual symmetry, only the lepton-flavor-violating decays $\tau \rightarrow
ee\mu$ and $\tau \rightarrow \mu \mu e$ are allowed in our model. We find that
these decays are mediated by the flavon triplet $\chi_{i}$,\ and by using the
experimental upper bound of the branching ratio of the decay $\tau
\rightarrow \mu \mu e$ \ we obtain an estimation on the mass of the flavon as
well as the cutoff scale $\Lambda$ of our model.

\  \  \  \  \newline The presentation is as follows. In Sec. II, we present the
superfield content of the extended MSSM we are interested in here, and give
their $A_{4}$\ representations. Useful tools on $A_{4}$ tensor calculus,
superpotential building, and the lepton charged sector are also given. In Sec.
III, we first introduce our supersymmetric $A_{4}\times$\ $A_{3}$ model and
make some comments. Then, we focus on the chargeless sector; we first study
the neutrino mass matrix and its diagonalization with TBM matrix, then we
analyze the scalar potential of flavons and describe the motivation beyond the
need for the extra $A_{3}$ discrete symmetry. In Sec. IV, we study the
deviation\ of the TBM matrix with the help of the $A_{4}$ flavon singlets and
give numerical results for both normal hierarchy (NH) and inverted hierarchy
(IH). In Sec. V, we study the lepton flavor violation (LFV) in the charged
lepton sector to constrain the mass of the flavons $\chi_{i}$\ and the cutoff
scale $\Lambda$. In Sec. VI we give our conclusion and comments. In the three
appendices, we report some relevant details and extra tools. In Appendix A, we
recall useful properties of the $A_{4}$\ group and irreducible
representations. In Appendix B, we derive the vacuum alignments of $\vec{\chi
}$ and $\vec{\chi}^{\prime}$ used in this paper, and show that they are
obtained without having to add extra superfields. In this regard, recall that
in many models in the literature, the problem of vacuum alignment is resolved
by adding the so-called driving fields \textrm{\cite{A18,A19}}. In Appendix C,
we give explicit details on the tensor product of $A_{4}$ invariant terms used
in the derivation of the flavon scalar potential (\ref{fp}) obtained in Sec.
III. We also give details on solving the minimum condition of the scalar
potential of the theory with respect to the two $A_{4}$ triplets $\vec{\chi}$
and $\vec{\chi}^{\prime}$.

\section{Flavor symmetry in supersymmetric models}

We begin by noticing that it is quite commonly admitted that the family
symmetry relating flavors belonging to different generations of the SM might
be behind the neutrino mass hierarchy and their mixing. This hypothetical
flavor symmetry $\Gamma$ is a discrete invariance that has been the subject of
several studies, and particular interest has been focused on those $\Gamma$'s
given by non-Abelian discrete symmetries \cite{A10,S}. In this study, we
consider the interesting case where flavor symmetry is given by $A_{4}\times
A_{3}$; and describe how this discrete symmetry can be implemented in models
around the supersymmetric scale $M_{SUSY}^{2}$ where the discrete $\Gamma$'s
are expected to follow from more basic symmetries such as the breaking of
$E_{8}$ gauge invariance of heterotic string or F-theory GUTs on Calabi-Yau
manifolds \cite{HE,FT,TF}.

\subsection{Extending the MSSM}

We start with the usual chiral superfield spectrum of the MSSM; then, we
describe a particular extension of this minimal supersymmetric model by
implementing flavon superfields carrying quantum numbers under a flavor
symmetry $A_{4}\times$\ $A_{3}$. This extension is one of the results of this
paper; it will be further developed in forthcoming sections.\  \ 

\subsubsection{MSSM contents}

In addition to the usual gauge superfield sector that we will omit for
simplicity, the chiral superfield spectrum of the MSSM and their quantum
numbers under {\small SU(3)}$_{C}\times${\small SU(2)}$_{L}\times
${\small U(1)}$_{Y}$ invariance are as shown in Table \ref{1} \begin{table}[h]
\centering \renewcommand{\arraystretch}{1.2}
\begin{tabular}
[c]{|l|l|l|l|l|}\hline
{\small sector} & {\small chiral superfields} & {\small SU(3)}$_{C}$ &
{\small SU(2)}$_{L}$ & {\small U(1)}$_{Y}$\\ \hline
{\small leptons} & $L_{i}=(\nu_{i},e^{-})_{L}$ & $\  \  \  \ 1$ & $\  \  \  \ 2$ &
$-1$\\ \cline{2-5}
& $R_{i}^{c}=e_{i}^{c}$ & $\  \  \  \ 1$ & $\  \  \  \ 1$ & $+2$\\ \hline
{\small quarks} & $Q_{i}=(u_{i},d_{i})_{L}$ & $\  \  \  \ 3$ & $\  \  \  \ 2$ &
$+\frac{1}{3}$\\ \cline{2-5}
& $U_{i}^{c}=u_{i}^{c}$ & $\  \  \  \  \bar{3}$ & $\  \  \  \ 1$ & $-\frac{4}{3}%
$\\ \cline{2-5}
& $D_{i}^{c}=d_{i}^{c}$ & $\  \  \  \  \bar{3}$ & $\  \  \  \ 1$ & $+\frac{2}{3}%
$\\ \hline
{\small Higgs} & $H_{u}=({\small H}_{u}^{+}{\small ,H}_{u}^{0})$ & $\  \  \  \ 1$
& $\  \  \  \ 2$ & $+1$\\ \cline{2-5}
& $H_{d}=({\small H}_{d}^{0}{\small ,H}_{d}^{-})$ & $\  \  \  \ 1$ & $\  \  \  \ 2$
& $-1$\\ \hline
\end{tabular}
\caption{MSSM chiral superfield content}%
\label{1}%
\end{table}with $\mathrm{i}$\textrm{=1,2,3} referring to the number of matter
generations. In superspace, these chiral superfields (and similar ones to be
introduced later; see Tables \ref{2} and \ref{y}) may be generically denoted
by $\Phi_{m}$ with the usual $\theta$ expansion \cite{Wess}%
\begin{equation}
\Phi_{m}=\phi_{m}+\sqrt{2}\theta.\psi_{m}+\theta^{2}F_{m}. \label{22}%
\end{equation}
Recall that properties and theoretical predictions of the MSSM are well
established; the interacting dynamics of the MSSM spectrum is very well known,
including both spontaneous and soft supersymmetry breaking. Recall also that
this particular field theory dynamics is nicely described in superspace; we
refer to the rich literature for details \cite{SUSY,SA}. Moreover, notice that
in this study we will focus on those relevant contributions to neutrino
physics coming from couplings involving some $\phi_{m}$'s, auxiliary $F_{m}%
$'s, and the usual auxiliary $D$'s; that is, those contributions to the scalar
potential of the model that lead to the computation of neutrino masses and
mixing angles (for details, see Sec. III).

\subsubsection{Extending the MSSM}

There are several extensions of the MSSM that have been considered in
literature. The extension of the MSSM\ we are interested in here\emph{\ }%
concerns the enlargement of the \emph{Higgs sector}; it is obtained by adding
extra chiral superfields which carry quantum numbers under gauge symmetry and
also under the discrete symmetry $A_{4}\times A_{3}$. So the Higgs sector in
our proposal may be thought of as consisting of three subsectors.

(i)The $H$ subsector, involving the usual $H_{u},$ $H_{d}$of the MSSM.

(ii)The $\Delta$ subsector of the extended MSSM (type II seesaw); see Table
\ref{2}.

(iii)The $\mathcal{\chi}$ subsector. This is our subsector; see Table \ref{y}
for its content.

Before giving the full superfield spectrum of our model, let us first focus on
the $\Delta$ subsector; this is a particular extension of the Higgs sector of
the MSSM given by adding two chiral superfield triplets $\vec{\Delta}_{u}$ and
$\vec{\Delta}_{d}$ with gauge quantum numbers as in Table \ref{2}.
\begin{table}[h]
\centering \renewcommand{\arraystretch}{1.2}
\begin{tabular}
[c]{|l|l|l|l|}\hline
{\small chiral superfields} & {\small SU(3)}$_{C}$ & {\small SU(2)}$_{L}$ &
{\small U(1)}$_{Y}$\\ \hline
$\left.
\begin{array}
[c]{c}%
{\small \Delta}_{u}=({\small \Delta}_{u}^{0}{\small ,\Delta}_{u}%
^{-}{\small ,\Delta}_{u}^{--})\\
{\small \Delta}_{d}=({\small \Delta}_{d}^{++}{\small ,\Delta}_{d}%
^{+}{\small ,\Delta}_{d}^{0})
\end{array}
\right.  $ & $\left.
\begin{array}
[c]{c}%
1\\
1
\end{array}
\right.  $ & $\left.
\begin{array}
[c]{c}%
3\\
3
\end{array}
\right.  $ & $\left.
\begin{array}
[c]{c}%
-2\\
2
\end{array}
\right.  $\\ \hline
\end{tabular}
\caption{Chiral superfields added to the MSSM.}%
\label{2}%
\end{table}The $y=\pm2$ hypercharge values are required by gauge invariance of
the superfield couplings $H_{u,d}$ and $\Delta_{u,d}$ in the chiral
superpotential $W=W\left(  H,\Delta \right)  $ of the extended supersymmetric
model; this chiral superfield coupling has the form%
\begin{equation}
W=\lambda_{u}Tr\left(  H_{u}\otimes \Delta_{u}\otimes H_{u}\right)
+\lambda_{d}Tr\left(  H_{d}\otimes \Delta_{d}\otimes H_{d}\right)  ,
\end{equation}
where $\lambda_{u,d}$ are Yukawa coupling constants.\newline To describe the
$\chi$ subsector, it is interesting to first collect some useful tools on
discrete groups, in particular, on the group $A_{4}\times A_{3}$ and its representations.

\subsection{$A_{4}\times A_{3}$ symmetry}

First, notice that $A_{3}\simeq Z_{3}$ it is an Abelian group and so its
irreducible representations $\mathbf{1}_{q^{r}}$ are one dimensional with
charge $r=0,\pm1$ and $q=e^{\frac{2ir\pi}{3}}$. This group should not be
confused with the $A_{3}^{\prime}$ subgroup contained in $A_{4}$. In what
follows, we will focus on describing pertinent properties of the discrete
symmetry, in particular those concerning the non-Abelian $A_{4}$ factor and
its representations. These realizations will be used later to refine the
quantum numbers of the chiral superfield spectrum (see Tables \ref{1} and
\ref{2}) as well as the content of the $\chi$ subsector given in Table \ref{y}.\ 

\subsubsection{ $A_{4}$ and its representations}

The finite $A_{4}$ symmetry is a non-Abelian discrete group with order 12; it
is a particular subgroup of the symmetric $S_{4}$ and is generated by two
noncommuting elements $S$ and $T$ that satisfy the following cyclic
relations:
\begin{equation}
S^{2}=T^{3}=(ST)^{3}=1.
\end{equation}
Because of their noncommutativity, $S$ and $T$ cannot be diagonalized
simultaneously; later, we use the basis where $S$ is diagonal.

\emph{Representations and tensor products}\newline By using the group
character relation $12=\sum_{i}d_{i}^{2}$ relating the order $12$ of the group
$A_{4}$ to the dimensions $d_{k}$ of the irreducible representations
$\boldsymbol{R}_{i}$ of $A_{4}$, we have
\begin{equation}
12=1^{2}+1^{2}+1^{2}+3^{2}. \label{6}%
\end{equation}
From this relation we learn a set of useful features, in particular

\begin{description}
\item[$\left(  \alpha \right)  $] the group $A_{4}$ has four $\boldsymbol{R}%
_{1},$ $\boldsymbol{R}_{2},$ $\boldsymbol{R}_{3},$ $\boldsymbol{R}_{4}$ with
respective dimensions $d_{i}$ as in Eq. (\ref{6}),

\item[$\left(  \beta \right)  $] it has four conjugacy classes $\mathcal{C}%
_{1},$ $\mathcal{C}_{2},$ $\mathcal{C}_{3},$ $\mathcal{C}_{4}$ given by Eq.
(\ref{c4}) of Appendix A, and

\item[$\left(  \gamma \right)  $] it has one irreducible triplet $3$, but three
kinds of singlets $1,$ $1^{\prime},$ $1^{\prime \prime}$.
\end{description}

Though interesting, the appearance of three singlets in the $A_{4}$
representation theory makes their use somehow subtle; this difficulty is
apparent and can be overcome by using the characters \textrm{$\chi$}$_{R_{i}%
}\left(  \mathcal{C}_{j}\right)  =$\textrm{$\chi$}$_{ij}$ of the irreducible
representations. The basic table of these characters, thought of as a matrix
\textrm{$\chi$}$_{ij}\equiv$\textrm{$\chi$}$_{R_{i}}\left(  \mathcal{C}%
_{i}\right)  ,$ is given by Eq. (\ref{c5}) in\textrm{\ }Appendix A. By
restricting to the characters of the $S$ and $T$ generators of $A_{4}$, the
above four irreducible representations $\boldsymbol{R}_{i}$ can be
characterized as follows:%
\begin{equation}%
\begin{tabular}
[c]{lllllll}%
$\mathbf{1}$ & $:$ & $\mathbf{1}_{\left(  1,1\right)  },$ & $\qquad \qquad$ &
$\mathbf{1}^{\prime}$ & $:$ & $\mathbf{1}_{\left(  1,\omega \right)  },$\\
$\mathbf{3}$ & $:$ & $\mathbf{3}_{\left(  -1,0\right)  },$ & $\qquad \qquad$ &
$\mathbf{1}^{\prime \prime}$ & $:$ & $\mathbf{1}_{\left(  1,\omega^{2}\right)
,}$%
\end{tabular}
\end{equation}
where $\omega=e^{\frac{2i\pi}{3}}$ with the usual feature $1+\omega
+\bar{\omega}=0$ and $\bar{\omega}=\omega^{2}$. These irreducible
representations obey the following tensor product algebra \cite{A10,S}:%
\begin{equation}%
\begin{tabular}
[c]{lll}%
$\mathbf{3}_{\left(  -1,0\right)  }\otimes \mathbf{3}_{\left(  -1,0\right)  }$
& $=$ & $\mathbf{1}_{\left(  1,1\right)  }\oplus \mathbf{1}_{\left(
1,\omega \right)  }\oplus \mathbf{1}_{\left(  1,\omega^{2}\right)  }%
\oplus \mathbf{3}_{\left(  -1,0\right)  }\oplus \mathbf{3}_{\left(  -1,0\right)
},$\\
$\mathbf{3}_{\left(  -1,0\right)  }\otimes \mathbf{1}_{\left(  1,\omega
^{r}\right)  }$ & $=$ & $\mathbf{3}_{\left(  -1,0\right)  },$\\
$\mathbf{1}_{\left(  1,\omega^{r}\right)  }\otimes \mathbf{1}_{\left(
1,\omega^{s}\right)  }$ & $=$ & $\mathbf{1}_{\left(  1,\omega^{r+s}\right)
},$%
\end{tabular}
\  \label{re}%
\end{equation}
where the integers $r$ and $s$ take the values \textrm{0, 1, 2}
$\operatorname{mod}$3. Observe that these relations preserve total dimension
and the total character. Observe also that the tensor product $\mathbf{3}%
_{\left(  -1,0\right)  }\otimes \mathbf{3}_{\left(  -1,0\right)  }$ has a
singlet $\mathbf{1}_{\left(  1,1\right)  }$; the same feature holds for higher
product powers, in particular, for the cubic and quartic powers to be
encountered later in our construction%
\begin{equation}%
\begin{tabular}
[c]{lll}%
$\mathbf{3}_{\left(  -1,0\right)  }\otimes \mathbf{3}_{\left(  -1,0\right)
}\otimes \mathbf{3}_{\left(  -1,0\right)  }$ & $=$ & $\mathbf{1}_{\left(
1,1\right)  }\oplus \mathbf{...,}$\\
$\mathbf{3}_{\left(  -1,0\right)  }\otimes \mathbf{3}_{\left(  -1,0\right)
}\otimes \mathbf{3}_{\left(  -1,0\right)  }\otimes \mathbf{3}_{\left(
-1,0\right)  }$ & $=$ & $\mathbf{1}_{\left(  1,1\right)  }\oplus \mathbf{....}$%
\end{tabular}
\end{equation}

\emph{Superpotential}\newline The superpotential of chiral superfields
$\Phi_{i}$ in the extended MSSM is given by a superfunction $W\left(  \Phi
_{i}\right)  $ that obeys two kinds of symmetries:

\begin{description}
\item[$i)$] invariance under the \textrm{SU(2)}$_{L}\times$\textrm{U(1)}$_{Y}$
gauge group;

\item[$ii)$] invariance under the flavor group $A_{4}\times A_{3}$.
\end{description}

Since $W\left(  \Phi_{i}\right)  $ has a polynomial form in the chiral
superfields $\Phi_{i}$, the invariance of the superpotential under
$A_{4}\times A_{3}$ is obtained by performing tensor products of irreducible
representations. Seeing that the tensor product of the $\mathbf{1}_{q^{r}}$
representation of $A_{3}$ is governed by the fusion relation $\mathbf{1}%
_{q^{r}}\otimes \mathbf{1}_{q^{s}}=\mathbf{1}_{q^{r+s}}$, the main difficulty
comes from the non-Abelian $A_{4}$ when computing higher-order monomials of
the type
\begin{equation}%
{\displaystyle \prod \nolimits_{i}}
\Phi_{i}^{n_{i}}%
\end{equation}
with the fusion algebra (\ref{re}). These computations are necessary since the
$A_{4}$-invariant trace $Tr_{A_{4}}W\left(  \Phi_{i}\right)  $ is given by the
following restriction\
\begin{equation}
Tr_{A_{4}}W\left(  \Phi_{i}\right)  =\left.  W\left(  \Phi_{i}\right)
\right \vert _{1_{\left(  1,1\right)  }}. \label{tr}%
\end{equation}
To illustrate how the method works let us focus on the $A_{4}$ subsymmetry and
later extend the construction to the full discrete symmetry.

\subsubsection{$A_{4}$-invariant superpotential}

As a first step to implementing flavor symmetry in neutrino supersymmetric
model building, we consider the superfield spectrum given in Tables \ref{1}
and \ref{2} to which we add flavon chiral superfields
\begin{equation}
\mathcal{\chi}_{k}=(\mathcal{\chi}_{1},\mathcal{\chi}_{2},\mathcal{\chi}_{3}),
\end{equation}
which transform as a triplet under the discrete group $A_{4}$. Then, we
attribute the following $A_{4}$ quantum numbers to the chiral superfield
spectrum:%
\begin{equation}%
\begin{tabular}
[c]{l||l|llll|ll|l}%
{\small chiral superfields} & $\  \ L_{i}$ & $R_{i}^{c}$ & $Q_{i}$ & $U_{i}%
^{c}$ & $D_{i}^{c}$ & $H_{u,d}$ & $\Delta_{u,d}$ & $\mathcal{\chi}_{k}%
$\\ \hline
$A_{4}$ {\small symmetry} & $\mathbf{1}_{(1,\bar{\omega}^{i-1})}$ &
\multicolumn{4}{|l|}{$\  \  \  \  \mathbf{3}_{\left(  -1,0\right)  }$} &
\multicolumn{2}{|l|}{$\  \  \  \mathbf{1}_{(1,1)}$} & $\mathbf{3}_{\left(
-1,0\right)  }$%
\end{tabular}
\  \  \label{tt}%
\end{equation}
where the $L_{i}$'s refer to the left doublets $(\nu_{i},e^{-})_{L}$, the
$R_{i}^{c}$'s to the right-handed $e_{i}^{c}$, and the others are as in Tables
\ref{1} and \ref{2}. Notice the following remarkable features:

\begin{itemize}
\item The three lepton doublets $\left(  L_{1},L_{2},L_{3}\right)  $\ sit in
different $A_{4}$ singlets, while the right leptons $\left(  R_{1}^{c}%
,R_{2}^{c},R_{3}^{c}\right)  $\ sit together in an\textrm{\ }$A_{4}$ triplet
\cite{24}.

\item The implementation of the $A_{4}$ discrete symmetry is not a soft
operation; by attributing $A_{4}$ quantum numbers to leptons $L_{i}$ and
$R_{i}^{c}$, the usual superfield couplings for building the lepton mass
matrix, such as
\[
\mathrm{y}^{ij}R_{i}^{c}L_{j}H_{d},
\]
are forbidden by invariance under discrete $A_{4}$. Indeed, by focusing on the
charged lepton sector, the chiral superpotential $W_{\text{lep}^{+}}$
describing the usual gauge-invariant Yukawa couplings,%
\begin{equation}
W_{\mathrm{lep}^{+}}=\mathrm{y}^{ij}R_{i}^{c}L_{j}H_{d}, \label{ch}%
\end{equation}
is no longer invariant under $A_{4}$ transformations, since from the view of
the $A_{4}$ representation group theory this chiral superfield coupling has
the following tensor product form%
\begin{equation}
\mathbf{3}_{\left(  -1,0\right)  }\otimes \mathbf{1}_{(1,\bar{\omega}^{i-1}%
)}\otimes \mathbf{1}_{(1,1)}\text{ \  \ }\mathbf{\sim}\text{ \  \ }%
\mathbf{3}_{\left(  -1,0\right)  },
\end{equation}
which does not contain the desired $A_{4}$ singlet $\mathbf{1}_{(1,1)}$ in the
trace (\ref{tr}). We will see later that a similar feature to Eq. (\ref{ch})
also happens for the chiral superpotential $W_{\mathrm{lep}^{0}}$ describing
couplings involving neutrinos.
\end{itemize}

\  \  \newline To make the gauge-invariant $W_{\text{lep}^{+}}$ symmetric as
well under the discrete $A_{4}$, we have to modify the chiral superfield
interaction (\ref{ch}) like $\mathcal{\tilde{W}}_{\mathrm{lep}^{+}}=Tr_{A_{4}%
}(\tilde{W}_{\mathrm{lep}^{+}}),$ with%
\begin{equation}
\tilde{W}_{\mathrm{lep}^{+}}=\frac{1}{\Lambda}y^{ijk}\left(  \mathcal{\chi
}_{i}R_{j}^{c}L_{k}H_{d}\right)  , \label{wlepton}%
\end{equation}
where $y^{ijk}$ are Yukawa couplings, $\Lambda$ denotes a cutoff scaling as
mass (to be related in Sec. IV with a flavon VEV), and $\mathcal{\chi}_{i}$ is
an $A_{4}$ flavon triplet. The fourth-order superfields coupling
$\mathcal{\chi}_{i}R_{j}^{c}L_{k}H_{d}$ transforms under discrete symmetry as
\begin{equation}
\mathbf{3}_{\left(  -1,0\right)  }\otimes \mathbf{3}_{\left(  -1,0\right)
}\otimes \mathbf{1}_{(1,\bar{\omega}^{i-1})}\otimes \mathbf{1}_{(1,1)},
\end{equation}
with the reduction containing the desired $A_{4}$ singlet type $\mathbf{1}%
_{(1,1)}$. Indeed, by using the fusion algebra (\ref{re}) in particular, the
reduction $\mathbf{3}_{\left(  -1,0\right)  }\otimes \mathbf{3}_{\left(
-1,0\right)  }=\mathbf{1}_{(1,\omega^{1-p})}\oplus...$ with $p=1,2,3$ it
follows that the above chiral superfield product usually contains a term of
the form $\mathbf{1}_{(1,\omega^{1-i})}\otimes \mathbf{1}_{(1,\omega^{i-1})},$
leading precisely to the desired singlet $\mathbf{1}_{(1,1)}$. To write down
an explicit expression in terms of the superfields, it is interesting to work
in the basis of $A_{4}$ where the generator $S$ is diagonal. In this basis,
the tensor product $R^{c}\otimes \mathcal{\chi}$ between the two $A_{4}$
triplet superfields $R^{c}=\left(  e_{1}^{c},e_{2}^{c},e_{3}^{c}\right)  $ and
$\mathcal{\chi}=(\mathcal{\chi}_{1},\mathcal{\chi}_{2},\mathcal{\chi}_{3})$
reads as%
\begin{equation}
R^{c}\otimes \mathcal{\chi}=\left(
\begin{array}
[c]{ccc}%
e_{1}^{c}\mathcal{\chi}_{1} & e_{1}^{c}\mathcal{\chi}_{2} & e_{1}%
^{c}\mathcal{\chi}_{3}\\
e_{2}^{c}\mathcal{\chi}_{1} & e_{2}^{c}\mathcal{\chi}_{2} & e_{2}%
^{c}\mathcal{\chi}_{3}\\
e_{3}^{c}\mathcal{\chi}_{1} & e_{3}^{c}\mathcal{\chi}_{2} & e_{3}%
^{c}\mathcal{\chi}_{3}%
\end{array}
\right)  .
\end{equation}
It is formally given by $\mathbf{3}_{\left(  -1,0\right)  }\otimes
\mathbf{3}_{\left(  -1,0\right)  }$ with nine components transforming in the
$\mathbf{9}_{\left(  1,0\right)  }$ representation of $A_{4},$ which is
reducible as in Eq. (\ref{re}). The restrictions of this tensor product to the
three $A_{4}$ singlet components $\mathbf{1}_{\left(  1,\omega^{r}\right)  }$
are given by%
\begin{equation}%
\begin{tabular}
[c]{lll}%
$\left.  R^{c}\otimes \mathcal{\chi}\right \vert _{1_{\left(  1,1\right)  }}$ &
$=$ & $e_{1}^{c}\mathcal{\chi}_{1}+e_{2}^{c}\mathcal{\chi}_{2}+e_{3}%
^{c}\mathcal{\chi}_{3},$\\
$\left.  R^{c}\otimes \mathcal{\chi}\right \vert _{1_{\left(  1,\omega \right)
}}$ & $=$ & $e_{1}^{c}\mathcal{\chi}_{1}+\omega e_{2}^{c}\mathcal{\chi}%
_{2}+\omega^{2}e_{3}^{c}\mathcal{\chi}_{3},$\\
$\left.  R^{c}\otimes \mathcal{\chi}\right \vert _{1_{(1,\omega^{2})}}$ & $=$ &
$e_{1}^{c}\mathcal{\chi}_{1}+\omega^{2}e_{2}^{c}\mathcal{\chi}_{2}+\omega
e_{3}^{c}\mathcal{\chi}_{3},$%
\end{tabular}
\  \  \  \label{product}%
\end{equation}
satisfying the properties%
\begin{equation}%
\begin{tabular}
[c]{lllllll}%
$e_{1}^{c}\mathcal{\chi}_{1}$ & $=$ & $\frac{1}{3}\left.  R^{c}\otimes
\mathcal{\chi}\right \vert $ & $+\frac{1}{3}$ & $\left.  R^{c}\otimes
\mathcal{\chi}\right \vert _{\omega}$ & $+\frac{1}{3}$ & $\left.  R^{c}%
\otimes \mathcal{\chi}\right \vert _{\omega^{2}},$\\
$e_{2}^{c}\mathcal{\chi}_{2}$ & $=$ & $\frac{1}{3}\left.  R^{c}\otimes
\mathcal{\chi}\right \vert $ & $+\frac{\omega^{2}}{3}$ & $\left.  R^{c}%
\otimes \mathcal{\chi}\right \vert _{\omega}$ & $+\frac{\omega}{3}$ & $\left.
R^{c}\otimes \mathcal{\chi}\right \vert _{\omega^{2}},$\\
$e_{3}^{c}\mathcal{\chi}_{3}$ & $=$ & $\frac{1}{3}\left.  R^{c}\otimes
\mathcal{\chi}\right \vert $ & $+\frac{\omega}{3}$ & $\left.  R^{c}%
\otimes \mathcal{\chi}\right \vert _{\omega}$ & $+\frac{\omega^{2}}{3}$ &
$\left.  R^{c}\otimes \mathcal{\chi}\right \vert _{\omega^{2}},$%
\end{tabular}
\end{equation}
where we have used the notations
\begin{align}
\left.  R^{c}\otimes \mathcal{\chi}\right \vert  &  \equiv \left.  R^{c}%
\otimes \mathcal{\chi}\right \vert _{1_{\left(  1,1\right)  }},\nonumber \\
\left.  R^{c}\otimes \mathcal{\chi}\right \vert _{\omega}  &  \equiv \left.
R^{c}\otimes \mathcal{\chi}\right \vert _{1_{\left(  1,\omega \right)  }%
},\label{not}\\
\left.  R^{c}\otimes \mathcal{\chi}\right \vert _{\omega^{2}}  &  \equiv \left.
R^{c}\otimes \mathcal{\chi}\right \vert _{1_{\left(  1,\omega^{2}\right)  }%
}.\nonumber
\end{align}
If we choose the VEVs of the $A_{4}$ triplet $\mathcal{\chi}_{i}$ as in the
Altarelli-Feruglio model (AF) \cite{25} and the VEV of the Higgs $H_{d}$ as
usual
\begin{equation}
\left \langle \mathcal{\chi}_{i}\right \rangle =\upsilon_{\chi}\left(
1,1,1\right)  ,\qquad \qquad \left \langle H_{d}\right \rangle =\upsilon_{d},
\label{va}%
\end{equation}
then by substituting these expressions back into the superpotential
(\ref{wlepton}) we obtain the charged lepton mass matrix $M_{\text{lep}^{+}}$
as%
\begin{equation}
M_{\mathrm{lep}^{+}}=\frac{\upsilon_{\chi}\upsilon_{d}}{\Lambda}\left(
\begin{array}
[c]{ccc}%
y_{e} & y_{e} & y_{e}\\
y_{\mu} & \omega y_{\mu} & \omega^{2}y_{\mu}\\
y_{\tau} & \omega^{2}y_{\tau} & \omega y_{\tau}%
\end{array}
\right)  ,
\end{equation}
where the Yukawa couplings $y_{e,\mu,\tau}$\ are related to the ones in Eq.
(\ref{wlepton}) as follows:
\begin{equation}
y_{e}=y^{ij1},\qquad \qquad y_{\mu}=y^{ij2},\qquad \text{\qquad}y_{\tau}%
=y^{ij3},
\end{equation}
\ where $i=j=1,2,3$. Following Ref. \cite{26}, this matrix can be diagonalized
by using asymmetric left and right transformations like\textrm{\ }%
$M_{\mathrm{lep}^{+}}^{\mathrm{diag}}=U_{R}M_{\mathrm{lep}^{+}}U_{L}^{\dag}%
$\textrm{\ }with eigenvalues $m_{i}(i=e,\mu,\tau)$ given by%
\begin{equation}
M_{\mathrm{lep}^{+}}^{\mathrm{diag}}=\frac{\sqrt{3}\upsilon_{\chi}\upsilon
_{d}}{\Lambda}\left(
\begin{array}
[c]{ccc}%
y_{e} & 0 & 0\\
0 & y_{\mu} & 0\\
0 & 0 & y_{\tau}%
\end{array}
\right)  , \label{ml}%
\end{equation}
and where
\begin{equation}
U_{L}=\frac{1}{\sqrt{3}}\left(
\begin{array}
[c]{ccc}%
1 & 1 & 1\\
1 & \omega & \omega^{2}\\
1 & \omega^{2} & \omega
\end{array}
\right)  ,\qquad \qquad U_{R}=\left(
\begin{array}
[c]{ccc}%
1 & 0 & 0\\
0 & 1 & 0\\
0 & 0 & 1
\end{array}
\right)  .
\end{equation}
$\allowbreak$ $\allowbreak$In order to obtain the hierarchy among the three
families of charged leptons, one may use the Froggatt-Nielsen (FN) mechanism
which consists of adding a new U(1)$_{\text{FN}}$\ symmetry with a new charge
to be assigned to the right-handed charged leptons \cite{27}; for more details
we refer to Refs. \cite{A10,25}. Following the AF model \cite{25}, by taking
$y_{\tau}\upsilon_{d}<250$ GeV and by using the experimental value of the tau
lepton mass, we get a constraint on the lower bound of the ratio of the
triplet VEV $\upsilon_{\chi}$\ over the $\Lambda$\ cutoff scale as follows:%
\begin{equation}
\frac{\upsilon_{\chi}}{\Lambda}>0.004 \label{cu}%
\end{equation}

\section{{Supersymmetric }$A_{4}\times A_{3}$ \textbf{neutrino }model}

In this section, we use the tools introduced in the previous section to
develop our supersymmetric $A_{4}\times A_{3}$ neutrino\textbf{\ }model
describing neutrino mixing and their masses. First, we give the superfield
spectrum of the proposal; then, we study the contributions of the
$\mathcal{\chi}$ sector to the chargeless leptons of the model, in particular
the aspects regarding neutrino masses and their mixing.

\subsection{Superfield content}

The superfield spectrum of the $A_{4}\times A_{3}$ neutrino model involves--in
addition to the usual superfields of the type II seesaw picture--extra flavon
superfields with nontrivial quantum numbers under $A_{4}\times A_{3}$.

\subsubsection{Chiral superfields in type II seesaw}

In our model, the Higgs sector has three subsectors: $\left(  a\right)  $ the
$H$ subsector involving the $H_{u},H_{d}$ superfields of the MSSM, $\left(
b\right)  $ the $\Delta$ subsector given in Table \ref{2}, and $\left(
c\right)  $ an extra $\mathcal{\chi}$ subsector involving flavons. The quantum
numbers of the chiral superfields of the $H$ and $\Delta$ sectors are shown in
Table IV (with explicit content like in Tables \ref{1} and \ref{2}%
).\begin{table}[h]
\centering \renewcommand{\arraystretch}{1.2}
\begin{tabular}
[c]{|l|l|l|l|l|l|l|}\hline
{\small sector} & {\small superfields} & {\small SU(3)}$_{C}$ & {\small SU(2)}%
$_{L}$ & {\small U(1)}$_{Y}$ & $A_{4}$ & $A_{3}$\\ \hline
{\small leptons} & $\  \  \  \ L_{i}$ & $\  \  \  \ 1$ & $\  \  \  \ 2$ & $-1$ &
$1_{\left(  1,\bar{\omega}^{i-1}\right)  }$ & $1_{0}$\\ \cline{2-7}
& $\  \  \  \ R_{i}^{c}$ & $\  \  \  \ 1$ & $\  \  \  \ 1$ & $+2$ & $3_{\left(
-1,0\right)  }$ & $1_{-}$\\ \hline \hline
{\small quarks} & $\  \  \  \ Q_{i}$ & $\  \  \  \ 3$ & $\  \  \  \ 2$ & $+\frac{1}{3}$
& $3_{\left(  -1,0\right)  }$ & $1_{0}$\\ \cline{2-7}
& $\  \  \  \ U_{i}^{c}$ & $\  \  \  \  \bar{3}$ & $\  \  \  \ 1$ & $-\frac{4}{3}$ &
$3_{\left(  -1,0\right)  }$ & $1_{0}$\\ \cline{2-7}
& $\  \  \  \ D_{i}^{c}$ & $\  \  \  \  \bar{3}$ & $\  \  \  \ 1$ & $+\frac{2}{3}$ &
$3_{\left(  -1,0\right)  }$ & $1_{0}$\\ \hline \hline
{\small Higgs} & $\  \  \  \ H_{u}$ & $\  \  \  \ 1$ & $\  \  \  \ 2$ & $+1$ &
$1_{\left(  1,1\right)  }$ & $1_{0}$\\ \cline{2-7}
& $\  \  \  \ H_{d}$ & $\  \  \  \ 1$ & $\  \  \  \ 2$ & $-1$ & $1_{\left(  1,1\right)
}$ & $1_{0}$\\ \cline{2-7}
& $\  \  \  \ {\small \Delta}_{u}$ & $\  \  \  \ 1$ & $\  \  \  \ 3$ & $-2$ &
$1_{\left(  1,1\right)  }$ & $1_{0}$\\ \cline{2-7}
& $\  \  \  \ {\small \Delta}_{d}$ & $\  \  \  \ 1$ & $\  \  \  \ 3$ & $+2$ &
$1_{\left(  1,1\right)  }$ & $1_{0}$\\ \hline
\end{tabular}
\caption{$A_{4}\times A_{3}$ quantum numbers of the matter and Higgs
superfields. }%
\label{z}%
\end{table}\newline The $A_{4}\times A_{3}$-invariant superpotentials relevant
for the neutrino physics will be studied explicitly once we introduce the
superfield content of the $\mathcal{\chi}$ subsector.

\subsubsection{Flavon sector}

Flavon superfields are chiral superfields which transform as singlets under
gauge symmetry, but in general they carry nontrivial charges under the
$A_{4}\times A_{3}$ flavor symmetry; for our concern, we show the relevant
flavons in Table \ref{y}

\begin{table}[h]
\centering \renewcommand{\arraystretch}{1.2}
\begin{tabular}
[c]{|l|l|l|l|l|l|}\hline
{\small superfields} & {\small SU(3)}$_{C}$ & {\small SU(2)}$_{L}$ &
{\small U(1)}$_{Y}$ & $A_{4}$ & $A_{3}$\\ \hline
$\  \  \  \  \mathcal{\chi}_{i}$ & $\  \  \  \ 1$ & $\  \  \  \ 1$ & $\  \  \  \ 0$ &
$3_{\left(  -1,0\right)  }$ & $1_{+}$\\ \hline
$\  \  \  \  \mathcal{\chi}_{i}^{\prime}$ & $\  \  \  \ 1$ & $\  \  \  \ 1$ &
$\  \  \  \ 0$ & $3_{\left(  -1,0\right)  }$ & $1_{0}$\\ \hline
$\  \  \  \  \Phi$ & $\  \  \  \ 1$ & $\  \  \  \ 1$ & $\  \  \  \ 0$ & $1_{\left(
1,1\right)  }$ & $1_{0}$\\ \hline
$\  \  \  \  \Phi^{\prime}$ & $\  \  \  \ 1$ & $\  \  \  \ 1$ & $\  \  \  \ 0$ &
$1_{\left(  1,\omega \right)  }$ & $1_{0}$\\ \hline
\end{tabular}
\caption{The flavon superfields.}%
\label{y}%
\end{table}These flavons couple to the lepton superfields of the model; for
instance, the chiral superfield triplet $\mathcal{\chi}_{i}$, which was
introduced previously in Eq. (\ref{wlepton}), is needed to build the mass
matrix for the charged leptons. The other chiral superfield triplet
$\mathcal{\chi}_{i}^{\prime}$ is needed to engineer the Majorana mass matrix
of the neutrinos; its coupling to leptons will be described in detail in the
next subsection. \newline Moreover, the trivial singlet $\Phi$ is needed to
reproduce the correct mass-squared difference\textrm{\ }$\Delta m_{31}^{2}%
\neq0$, while the nontrivial singlet $\Phi^{\prime}$\ has been added in order
to generate a nonzero mixing angle $\theta_{13}$. Notice also that the
discrete symmetry $A_{3}$ is required to satisfy the following:

\begin{description}
\item[$\left(  i\right)  $] Exclude unwanted terms that appear in $A_{4}%
$-invariant superpotentials for charged and chargeless leptons. Without the
extra $A_{3}$, generic $A_{4}$-invariant superpotentials $W\left(
\mathcal{\chi},\mathcal{\chi}^{\prime}\right)  $ would be invariant under the
exchange of the two flavon triplets, that is, by performing the permutation
\begin{equation}
\mathcal{\chi}_{i}\leftrightarrow \mathcal{\chi}_{i}^{\prime}. \label{ex}%
\end{equation}

\item[$\left(  ii\right)  $] Prevent $\mathcal{\chi \chi}^{\prime}$
interactions in the superpotential through other intermediate superfields, and
therefore between the charged and chargeless lepton subsectors of\textrm{\ }%
the\textrm{\ }supersymmetric $A_{4}\times A_{3}$ model. It happens that this
constraint coincides precisely with the so-called sequestering
problem\textrm{\ }\cite{30,31,32}. The $A_{3}$ subsymmetry is therefore a
requirement of the sequestering problem.
\end{description}

\subsection{Chargeless lepton sector}

Before implementing $A_{4}\times A_{3}$ invariance, it is interesting to
notice that without flavons, the part $W_{\mathrm{lep}^{0}}$ of the chiral
superpotential of the model\ that leads to the Majorana mass may be expressed
as%
\begin{align}
W_{\mathrm{lep}^{0}}  &  =\lambda_{\nu}^{ee}L_{e}\Delta_{d}L_{e}+\lambda_{\nu
}^{e\mu}L_{e}\Delta_{d}L_{\mu}+\lambda_{\nu}^{e\tau}L_{e}\Delta_{d}L_{\tau
}\nonumber \\
&  +\lambda_{\nu}^{\mu e}L_{\mu}\Delta_{d}L_{e}+\lambda_{\nu}^{\mu \mu}L_{\mu
}\Delta_{d}L_{\mu}+\lambda_{\nu}^{\mu \tau}L_{\mu}\Delta_{d}L_{\tau}%
\label{sup}\\
&  +\lambda_{\nu}^{\tau e}L_{\tau}\Delta_{d}L_{e}+\lambda_{\nu}^{\tau \mu
}L_{\tau}\Delta_{d}L_{\mu}+\lambda_{\nu}^{\tau \tau}L_{\tau}\Delta_{d}L_{\tau
},\nonumber
\end{align}
where $\lambda_{\nu}^{ij}=\lambda_{\nu}^{ji}$ are Yukawa coupling constants.
By using the $A_{4}$ quantum charges given in Tables \ref{z} and \ref{y}, it
follows that the three terms $L_{e}\Delta_{d}L_{e},$ $L_{\mu}\Delta_{d}%
L_{\tau},$ and $L_{\tau}\Delta_{d}L_{\mu}$ are invariant under $A_{4}$
transformations, but not the other terms of Eq. (\ref{sup}) due to the fusion
relation\textbf{\ }$\mathbf{1}_{\left(  1,\omega^{r}\right)  }\otimes
\mathbf{1}_{\left(  1,\omega^{s}\right)  }=\mathbf{1}_{\left(  1,\omega
^{r+s}\right)  }$ which in general is not a trivial singlet. For example, by
using Table \ref{z}, the superfield coupling $L_{\mu}\Delta_{d}L_{\mu}$
transforms under $A_{4}$ representation like
\begin{equation}
\mathbf{1}_{\left(  1,\omega^{2}\right)  }\otimes \mathbf{1}_{\left(
1,\omega^{2}\right)  }\otimes \mathbf{1}_{\left(  1,1\right)  },
\end{equation}
which behaves as a nontrivial singlet representation since it is given by
$\mathbf{1}_{\left(  1,\omega \right)  }$. To overcome this difficulty, we
introduce an extra flavon superfield that transforms as $\mathbf{1}_{\left(
1,\omega^{2}\right)  }$; by using the fusion algebra (\ref{re}), this
nontrivial singlet of $A_{4}$ can be thought of in terms of a composite of the
$\mathcal{\chi}^{\prime}$ triplet as
\begin{equation}
\left.  \left(  \mathcal{\chi}^{\prime}\mathcal{\chi}^{\prime}\right)
\right \vert _{\omega^{2}},
\end{equation}
where the notation (\ref{not}) has been used. The two other singlet composites
appearing in the reduction of the tensor product $\mathcal{\chi}^{\prime
}\otimes \mathcal{\chi}^{\prime},$ which are denoted as
\begin{equation}
\left.  \left(  \mathcal{\chi}^{\prime}\mathcal{\chi}^{\prime}\right)
\right \vert _{\omega}\qquad \text{and}\qquad \left.  \left(  \mathcal{\chi
}^{\prime}\mathcal{\chi}^{\prime}\right)  \right \vert _{\omega^{3}},
\end{equation}
are needed to recover $A_{4}$ invariance of the other couplings, as shown
below.\ Notice\ that if we use only the three $A_{4}$-invariant terms
described above, the neutrino mass matrix will not agree with the TBM matrix
and thus with the mixing angles $\theta_{12}$ and $\theta_{23}$; with the
three invariant terms $L_{e}\Delta_{d}L_{e},$\ $L_{\mu}\Delta_{d}L_{\tau}%
,$\ and $L_{\tau}\Delta_{d}L_{\mu}$\ the shape of neutrino mass matrix is
given by
\begin{equation}
\left(
\begin{array}
[c]{ccc}%
x & 0 & 0\\
0 & 0 & y\\
0 & y & 0
\end{array}
\right)  ,
\end{equation}
where the mixing matrix is
\begin{equation}
\left(
\begin{array}
[c]{ccc}%
1 & 0 & 0\\
0 & \frac{1}{\sqrt{2}} & -\frac{1}{\sqrt{2}}\\
0 & \frac{1}{\sqrt{2}} & \frac{1}{\sqrt{2}}%
\end{array}
\right)  ,
\end{equation}
which is clearly in conflict with the TBM matrix.\  \  \ 

\subsubsection{ Implementing the flavon triplet $\mathcal{\chi}_{i}^{\prime}$}

To restore $A_{4}$-invariance in the chargeless lepton subsector, we
add\footnote{The first triplet has been used in the charged lepton sector; see
Eq. (\ref{wlepton}).} the $A_{4}$ triplet\textrm{\ }$\mathcal{\chi}%
_{i}^{\prime}=(\mathcal{\chi}_{1}^{\prime},\mathcal{\chi}_{2}^{\prime
},\mathcal{\chi}_{3}^{\prime})$ and modify the superpotential $W_{\text{lep}%
^{0}}$ of Eq. (\ref{sup}) as%
\begin{equation}
\mathcal{W}_{\mathrm{lep}^{0}}=Tr_{A_{4}}\left[  W_{\mathrm{lep}^{0}}^{\prime
}\right]  \equiv \left.  W_{\mathrm{lep}^{0}}^{\prime}\right \vert _{1_{\left(
1,1\right)  }},
\end{equation}
with
\begin{align}
W_{\mathrm{lep}^{0}}^{\prime}  &  =\lambda_{\nu}^{ee}L_{e}\Delta_{d}%
L_{e}+\frac{\lambda_{\nu}^{\mu e}}{\Lambda^{2}}L_{e}\Delta_{d}L_{\mu}\left.
\left(  \mathcal{\chi}^{\prime}\mathcal{\chi}^{\prime}\right)  \right \vert
_{\omega}+\frac{\lambda_{\nu}^{e\tau}}{\Lambda^{2}}L_{e}\Delta_{d}L_{\tau
}\left.  \left(  \mathcal{\chi}^{\prime}\mathcal{\chi}^{\prime}\right)
\right \vert _{\omega^{2}}\nonumber \\
&  +\frac{\lambda_{\nu}^{e\mu}}{\Lambda^{2}}L_{\mu}\Delta_{d}L_{e}\left.
\left(  \mathcal{\chi}^{\prime}\mathcal{\chi}^{\prime}\right)  \right \vert
_{\omega}+\frac{\lambda_{\nu}^{\mu \mu}}{\Lambda^{2}}L_{\mu}\Delta_{d}L_{\mu
}\left.  \left(  \mathcal{\chi}^{\prime}\mathcal{\chi}^{\prime}\right)
\right \vert _{\omega^{2}}+\lambda_{\nu}^{\mu \tau}L_{\mu}\Delta_{d}L_{\tau
}\label{sp}\\
&  +\frac{\lambda_{\nu}^{\tau e}}{\Lambda^{2}}L_{\tau}\Delta_{d}L_{e}\left.
\left(  \mathcal{\chi}^{\prime}\mathcal{\chi}^{\prime}\right)  \right \vert
_{\omega^{2}}+\lambda_{\nu}^{\tau \mu}L_{\tau}\Delta_{d}L_{\mu}+\frac
{\lambda_{\nu}^{\tau \tau}}{\Lambda^{2}}L_{\tau}\Delta_{d}L_{\tau}\left.
\left(  \mathcal{\chi}^{\prime}\mathcal{\chi}^{\prime}\right)  \right \vert
_{\omega}.\nonumber
\end{align}
In this relation, the term $\left(  \mathcal{\chi}^{\prime}\mathcal{\chi
}^{\prime}\right)  $ stands for $\mathcal{\chi}^{\prime}\mathbf{\otimes
}\mathcal{\chi}^{\prime}$ transforming in the $\mathbf{3}_{\left(
-1,0\right)  }\otimes \mathbf{3}_{\left(  -1,0\right)  }$ representation of the
$A_{4}$ discrete symmetry whose reduction (\ref{re}) contains (amongst others)
three possible $A_{4}$ singlets. The notation $\left.  \left(  \mathcal{\chi
}^{\prime}\mathcal{\chi}^{\prime}\right)  \right \vert _{\xi}$ is as defined in
Eq. (\ref{not}), which for convenience we recall below:%
\begin{align}
&  \left.  \left(  \mathcal{\chi}^{\prime}\mathcal{\chi}^{\prime}\right)
\right \vert _{1_{\left(  1,1\right)  }}\text{ \ }\equiv \left.  \left(
\mathcal{\chi}^{\prime}\mathcal{\chi}^{\prime}\right)  \right \vert _{1}\text{
\ }=\mathcal{\chi}_{1}^{\prime2}+\mathcal{\chi}_{2}^{\prime2}+\mathcal{\chi
}_{3}^{\prime2},\nonumber \\
&  \left.  \left(  \mathcal{\chi}^{\prime}\mathcal{\chi}^{\prime}\right)
\right \vert _{1_{\left(  1,\omega \right)  }}\text{ }\equiv \left.  \left(
\mathcal{\chi}^{\prime}\mathcal{\chi}^{\prime}\right)  \right \vert _{\omega
}\text{ }=\mathcal{\chi}_{1}^{\prime2}+\omega \mathcal{\chi}_{2}^{\prime
2}+\omega^{2}\mathcal{\chi}_{3}^{\prime2},\\
&  \left.  \left(  \mathcal{\chi}^{\prime}\mathcal{\chi}^{\prime}\right)
\right \vert _{1_{\left(  1,\omega^{{\small 2}}\right)  }}\equiv \left.  \left(
\mathcal{\chi}^{\prime}\mathcal{\chi}^{\prime}\right)  \right \vert
_{\omega^{{\small 2}}}=\mathcal{\chi}_{1}^{\prime2}+\omega^{2}\mathcal{\chi
}_{2}^{\prime2}+\omega \mathcal{\chi}_{3}^{\prime2}.\nonumber
\end{align}

\subsubsection{Tribimaximal mixing matrix}

For the sake of the TBM matrix, the neutrino mass matrix must respect the
$\mu-\tau$ symmetry and the two following conditions \cite{A5,28}:%
\begin{equation}%
\begin{tabular}
[c]{lll}%
$\left(  M_{\upsilon}\right)  _{11}+\left(  M_{\upsilon}\right)  _{12}$ & $=$
& $\left(  M_{\upsilon}\right)  _{22}+\left(  M_{\upsilon}\right)  _{23},$\\
$\left(  M_{\upsilon}\right)  _{12}$ & $=$ & $\left(  M_{\upsilon}\right)
_{13}.$%
\end{tabular}
\  \  \  \label{co}%
\end{equation}
The implementation of the form of the (TBM) matrix for generating neutrino
masses requires\ vacuum alignment of the $A_{4}$ triplet $\mathcal{\chi
}^{\prime}$\ and for $\Delta_{d}$ as follows\footnote{To avoid heavy
notations, we denote the leading scalar components with the same letter as the
superfields; see also the comment after Eq.(\ref{22}).}:%
\begin{equation}
\left \langle \mathcal{\chi}^{\prime}\right \rangle =\upsilon_{\chi^{\prime}%
}(1,0,0),\qquad \left \langle \Delta_{d}\right \rangle =\upsilon_{\Delta_{d}}.
\label{vacua}%
\end{equation}
Hence the neutrino mass matrix is%
\begin{equation}
M_{\upsilon}=\upsilon_{\Delta_{d}}\left(
\begin{array}
[c]{ccc}%
\lambda_{\nu}^{ee} & \lambda_{\nu}^{e\mu}b & \lambda_{\nu}^{e\tau}b\\
\lambda_{\nu}^{e\mu}b & \lambda_{\nu}^{\mu \mu}b & \lambda_{\nu}^{\mu \tau}\\
\lambda_{\nu}^{e\tau}b & \lambda_{\nu}^{\mu \tau} & \lambda_{\nu}^{\tau \tau}b
\end{array}
\right)  , \label{nmatrix}%
\end{equation}
where we have set%
\begin{equation}
\frac{\upsilon_{\chi^{\prime}}^{2}}{\Lambda^{2}}\equiv \beta^{2}=b. \label{n}%
\end{equation}
Since the higher-dimensional operators involving $\left(  \mathcal{\chi
}^{\prime}\mathcal{\chi}^{\prime}\right)  $\ contribute to the tiny mass of
the neutrinos, the VEV of the flavon $\chi^{\prime}$\ should be small and
close to the cutoff scale $\upsilon_{\chi^{\prime}}\lesssim \Lambda$ which
means that $b\lesssim1$. Assuming for simplicity that the Yukawa couplings
$\lambda_{\nu}^{ij}$ are of the\textrm{\ }order of unity\footnote{We can get
the TBM matrix without assuming the Yukawa coupling of $\mathcal{O(}%
1\mathcal{)}$, but to do so we have to impose some conditions on them in order
to satisfy the relations (\ref{co}); hence, for the matrix (\ref{nmatrix}) we
impose the following: $\lambda_{\nu}^{e\mu}=\lambda_{\nu}^{e\tau}$,
$\lambda_{\nu}^{\mu \mu}=\lambda_{\nu}^{\tau \tau}$ and $\lambda_{\nu}%
^{ee}+\lambda_{\nu}^{e\mu}b=\lambda_{\nu}^{\mu \mu}b+\lambda_{\nu}^{\mu \tau}$%
.}, and using the usual tribimaximal mixing matrix $U$, it results that the
above mass matrix $M_{\upsilon}$ is diagonalized as $\mathcal{M}_{\upsilon
}=U^{T}M_{\upsilon}U$ with%
\begin{equation}
\mathcal{M}_{\upsilon}=\upsilon_{\Delta_{d}}\left(
\begin{array}
[c]{ccc}%
1-b & 0 & 0\\
0 & 1+2b & 0\\
0 & 0 & -1+b
\end{array}
\right)  . \label{diff mass}%
\end{equation}
Recall that the TBM mixing matrix has the form
\begin{equation}
U=\left(
\begin{array}
[c]{ccc}%
-\sqrt{\frac{2}{3}} & \frac{1}{\sqrt{3}} & 0\\
\frac{1}{\sqrt{6}} & \frac{1}{\sqrt{3}} & -\frac{1}{\sqrt{2}}\\
\frac{1}{\sqrt{6}} & \frac{1}{\sqrt{3}} & \frac{1}{\sqrt{2}}%
\end{array}
\right)  .
\end{equation}
It predicts the mixing angles as follows:%
\begin{equation}
\sin^{2}\theta_{12}=\frac{1}{3},\qquad \sin^{2}\theta_{23}=\frac{1}{2}%
,\qquad \sin^{2}\theta_{13}=0.
\end{equation}
However, a careful inspection of the eigenvalues of $\mathcal{M}_{\upsilon}$
reveals that we have $\Delta m_{31}^{2}=0,$ which is in conflict with the data
in Table \ref{t}. For this reason, we need to correct the mass matrix
(\ref{nmatrix}), a correction that we realize by further enlarging the flavon
spectrum of the model as described below.\  \  \  \ 

\subsubsection{An extra flavon singlet\emph{\ }$\Phi$}

To generate appropriate masses for the neutrinos, we deform the superpotential
(\ref{sp}) by adding $\delta W_{\text{lep}^{0}}$ contributions inducing
off-diagonal elements in the matrix $\mathcal{M}_{\upsilon}$ as a perturbation
so that we can preserve the form of the matrix (\ref{nmatrix}), which respects
the $\mu-\tau$ symmetry and the conditions in Eq. (\ref{co}) where the $A_{4}$
trivial singlet $\Phi$ is sufficient to solve the problem. Since the
superpotential (\ref{sp}) is $A_{4}$ invariant, if we add one nontrivial
singlet (such as $\Phi^{\prime}\sim1_{(1,\omega)}$ or $\Phi^{\prime \prime}%
\sim1_{(1,\omega^{2})}$) we do not obtain invariant terms; this is why in the
case of one singlet, the trivial $1_{(1,1)}\sim \Phi=\zeta+\theta \psi_{\zeta
}+\theta^{2}F_{\zeta}$ is the only representation that reproduces the TBM
matrix. Hence, the desired deformed chiral superpotential reads as%
\begin{equation}
\mathcal{W}_{\mathrm{lep}^{0}}^{\prime \prime}=\mathcal{W}_{\mathrm{lep}^{0}%
}^{\prime}+\delta \mathcal{W}_{\mathrm{lep}^{0}}, \label{ww}%
\end{equation}
with an additional $\delta \mathcal{W}_{\mathrm{lep}^{0}}=Tr_{A_{4}}\left[
\delta W_{\mathrm{lep}^{0}}\right]  $ term given by
\begin{equation}%
\begin{tabular}
[c]{lll}%
$\delta W_{\mathrm{lep}^{0}}$ & $=$ & $\frac{\lambda_{\nu}^{e\mu}}{\Lambda
^{3}}\left[  L_{e}\Delta_{d}L_{\mu}+L_{\mu}\Delta_{d}L_{e}\right]  \left(
\Phi \left.  \left(  \mathcal{\chi}^{\prime}\mathcal{\chi}^{\prime}\right)
\right \vert _{\omega}\right)  $\\
&  & $+\frac{\lambda_{\nu}^{e\tau}}{\Lambda^{3}}\left[  L_{e}\Delta_{d}%
L_{\tau}+L_{\tau}\Delta_{d}L_{e}\right]  \left(  \Phi \left.  \left(
\mathcal{\chi}^{\prime}\mathcal{\chi}^{\prime}\right)  \right \vert
_{\omega^{2}}\right)  $\\
&  & $+\frac{\lambda_{\nu}^{\tau \mu}}{\Lambda^{3}}\left[  L_{\mu}\Delta
_{d}L_{\tau}+L_{\tau}\Delta_{d}L_{\mu}\right]  \left(  \Phi \left.  \left(
\mathcal{\chi}^{\prime}\mathcal{\chi}^{\prime}\right)  \right \vert
_{\omega^{3}}\right)  ,$%
\end{tabular}
\  \label{dl}%
\end{equation}
where the scale $\Lambda$ is the cutoff introduced before. Since the flavon
$\Phi$ is introduced only to resolve the problem of the zero squared-mass
difference $\Delta m_{31}^{2}=0$ its presence does not change the mixing
angles, and also because it transforms trivially under $A_{4}$ its VEV does
not break $A_{4}$. Accordingly we have two possible routes: $\left(  i\right)
$ either we assume that $\left \langle \Phi \right \rangle =\upsilon_{\Phi}$ is
much smaller than the cutoff scale $\upsilon_{\Phi}\ll \Lambda$ where invariant
terms like the series $\sum \nolimits_{n}L_{e}\Delta_{d}L_{e}\left(  \frac
{\Phi}{\Lambda}\right)  ^{n}$ may be suppressed by the factor of
$\frac{\upsilon_{\Phi}}{\Lambda}<<1$, or $\left(  ii\right)  $ the VEV
$\upsilon_{\Phi}$\ is of the order of the cutoff scale $\left(  \upsilon
_{\Phi}\sim \Lambda \right)  $ where the terms $\lambda_{\nu}^{ee}L_{e}%
\Delta_{d}L_{e}\left(  \frac{\Phi}{\Lambda}\right)  ^{n}$\ are comparable to
$\lambda_{\nu}^{ee}L_{e}\Delta_{d}L_{e}$. In this way, we assume that
\textrm{the additional factor}\ coming from the combination of these operators
is absorbed into the coupling constants $\lambda_{\nu}^{ee}$. The previous
neutrino mass matrix $M_{\upsilon}$ [Eq.(\ref{nmatrix})] gets corrected like
$M_{\upsilon}^{\prime}=M_{\upsilon}+\delta M_{\upsilon},$ whose expression can
be put into the form%
\begin{equation}
M_{\upsilon}^{\prime}=\upsilon_{\Delta_{d}}\left(
\begin{array}
[c]{ccc}%
1 & b+c & b+c\\
b+c & b & 1+c\\
b+c & 1+c & b
\end{array}
\right)  , \label{NMM}%
\end{equation}
where $b$ is as in Eq. (\ref{n}) and where we have set%
\begin{equation}
c=\frac{\upsilon_{\chi^{\prime}}^{2}}{\Lambda^{2}}\frac{\upsilon_{\Phi}%
}{\Lambda}=b\frac{\upsilon_{\Phi}}{\Lambda}.
\end{equation}
\textrm{Therefore, the convergence of the geometric series }$L_{e}\Delta
_{d}L_{e}\sum \nolimits_{n}\left(  \frac{\Phi}{\Lambda}\right)  ^{n}%
$\textrm{\ turns into the condition }$\left \vert c\right \vert <\left \vert
b\right \vert .$ The new mass matrix $M_{\upsilon}^{\prime}$ is diagonalized by
the TBM mixing matrix $U$ as $\mathcal{M}_{\upsilon}^{\prime}=\mathrm{diag}%
\left(  m_{1},m_{2},m_{2}\right)  ,$\ with neutrino mass eigenvalues (in units
of $\upsilon_{\Delta_{d}})$ given as%
\begin{equation}%
\begin{tabular}
[c]{lll}%
$m_{1}$ & $=$ & $1-c-b,$\\
$m_{2}$ & $=$ & $2b+2c+1,$\\
$m_{3}$ & $=$ & $b-c-1.$%
\end{tabular}
\end{equation}
From these new eigenvalues we learn that $\Delta m_{31}^{2}=-4c\left(
b-1\right)  $ is no longer vanishing provided that we have $b\neq1$ and
$c\neq0$. Notice that the same constraint on the parameter $b$\  \ ($b\lesssim
1$) holds for the parameter $c$\ for the same reasons we mentioned in the
previous subsection; thus, $c\lesssim1,$\ which means that $\upsilon
_{\chi^{\prime}}^{2}\upsilon_{\mathbf{\zeta}}\lesssim \Lambda^{3}$.

\subsection{$A_{4}\times A_{3}$-invariant scalar potential}

Here we study the $A_{4}\times A_{3}$-invariant scalar potential; the $A_{3}$
symmetry is needed for the reasons mentioned in Sec. III A.

\subsubsection{Higgs and flavon sector}

By using the notation of Ref. \textrm{\cite{32}} for monomials of flavons (in
particular, the quadratic $\mathcal{\chi}^{\prime2}\equiv \mathcal{\chi
}^{\prime}\otimes \mathcal{\chi}^{\prime}$ and the cubic $\mathcal{\chi
}^{\prime3}\equiv \mathcal{\chi}^{\prime}\otimes \mathcal{\chi}^{\prime2}$), the
$A_{4}\times A_{3}$-invariant superpotential restricted to the Higgs
isodoublet $H_{u,d}$, isotriplet $\Delta_{u,d}$, and flavon superfields
$\mathcal{\chi},$ $\mathcal{\chi}^{\prime},\Phi$ is given by%
\begin{equation}%
\begin{tabular}
[c]{lll}%
$W_{H\text{-}F}$ & $=$ & $\mu H_{u}H_{d}+\mu_{\Delta}Tr(\Delta_{u}\Delta
_{d})+\lambda_{u}H_{u}\Delta_{u}H_{u}+\lambda_{d}H_{d}\Delta_{d}H_{d}$\\
&  & $+\mu_{\chi}\mathcal{\chi}^{\prime2}+\lambda_{\zeta \chi}\Phi
\mathcal{\chi}^{\prime2}+\mu_{\zeta}\Phi^{2}+\lambda \mathcal{\chi}^{3}%
+\lambda^{\prime}\mathcal{\chi}^{\prime3}+\lambda_{\zeta}\Phi^{3}+k_{\zeta
}\Phi$\\
&  & $+h_{\zeta}H_{u}\Phi H_{d}+\delta_{\zeta}\Phi Tr(\Delta_{u}\Delta_{d}),$%
\end{tabular}
\end{equation}
where $\mu,$ $\mu_{\Delta},$ $\mu_{\zeta},$ $\mu_{\chi}$ are mass parameters
and $\lambda_{x},$ $h_{\zeta}$, $\delta_{\zeta}$ are coupling constants. To
justify the choice of the $A_{3}$ symmetry instead of just $Z_{2}$ to
discriminate the two flavon triplets, we need to analyze the scalar potential.

\subsubsection{Scalar potential}

Gathering all the contributions from $F$, $D,$ and soft terms, the scalar
potential $\mathcal{V}_{\text{tot}}$ of the model is given by
\begin{equation}
\mathcal{V}_{\mathrm{tot}}=V_{\mathrm{SUSY}}+V_{\mathrm{soft}}, \label{Vt}%
\end{equation}
with%
\begin{equation}%
\begin{tabular}
[c]{lll}%
$V_{\mathrm{SUSY}}$ & $=$ & $\left \vert F_{u}\right \vert ^{2}+\left \vert
F_{d}\right \vert ^{2}+\left \vert F_{\Delta_{d}}\right \vert ^{2}+\left \vert
F_{\Delta_{u}}\right \vert ^{2}$\\
&  & $+\left \vert F_{\chi}\right \vert ^{2}+\left \vert F_{\chi^{\prime}%
}\right \vert ^{2}+\left \vert F_{\Phi}\right \vert ^{2}$\\
&  & $+\vec{D}^{2}+D^{2},$%
\end{tabular}
\end{equation}
where the explicit forms of $V_{\mathrm{SUSY}}$ and $V_{\mathrm{soft}}$ are
given in Appendix B. So the $A_{4}\times A_{3}$-invariant scalar potential is
as follows%
\begin{equation}%
\begin{tabular}
[c]{lll}%
$\mathcal{V}$ & $=$ & $9\lambda^{2}\left \vert \mathcal{\chi}\right \vert
^{4}+4\left \vert \mu_{\chi}\right \vert ^{2}\left \vert \chi^{\prime}\right \vert
^{2}+4\lambda_{\zeta \chi}^{2}\left \vert \chi^{\prime}\right \vert
^{2}\left \vert \Phi \right \vert ^{2}+9\lambda^{\prime2}\left \vert \chi^{\prime
}\right \vert ^{4}+8\mu_{\chi}\lambda_{\zeta \chi}\left \vert \chi^{\prime
}\right \vert ^{2}\Phi$\\
&  & $+12\mu_{\chi}\lambda^{\prime}\left \vert \chi^{\prime}\right \vert
^{3}+12\lambda_{\zeta \chi}\lambda^{\prime}\left \vert \chi^{\prime}\right \vert
^{3}\Phi+\lambda_{\zeta \chi}^{2}\left \vert \chi^{\prime}\right \vert
^{4}+2k_{\zeta}\lambda_{\zeta \chi}\left \vert \chi^{\prime}\right \vert ^{2}$\\
&  & $+6\lambda_{\zeta \chi}\lambda_{\zeta}\left \vert \chi^{\prime}\right \vert
^{2}\left \vert \Phi \right \vert ^{2}+2h_{\zeta}\lambda_{\zeta \chi}H_{u}%
H_{d}\left \vert \chi^{\prime}\right \vert ^{2}+2\delta_{\zeta}\lambda
_{\zeta \chi}Tr\left(  \Delta_{u}\Delta_{d}\right)  \left \vert \chi^{\prime
}\right \vert ^{2}$\\
&  & $+4\mu_{\zeta}\lambda_{\zeta \chi}\Phi \left \vert \chi^{\prime}\right \vert
^{3}+m_{\chi}^{2}\left \vert \mathcal{\chi}\right \vert ^{2}+m_{\chi^{\prime}%
}^{2}\left \vert \chi^{\prime}\right \vert ^{2}+2b_{\chi^{\prime}}\left \vert
\chi^{\prime}\right \vert ^{2}$\\
&  & $+2A_{\zeta \chi^{\prime}}\Phi \left \vert \chi^{\prime}\right \vert
^{2}+2A_{\chi}\left \vert \mathcal{\chi}\right \vert ^{3}+2A_{\chi^{\prime}%
}\left \vert \chi^{\prime}\right \vert ^{3}+\mathcal{V}_{\mathrm{ind}},$%
\end{tabular}
\  \  \label{fp}%
\end{equation}
where $\mathcal{V}_{\mathrm{ind}}$\ consists of terms that are irrelevant with
two $A_{4}$ triplets. The tensor products for all possible $A_{4}$-invariant
terms are reported in Appendix C.\newline As stated before, in order to avoid
the communication between the charged and chargeless sectors (and thus the
interaction between the two $A_{4}$\ triplets $\chi_{i}$ and $\chi_{i}%
^{\prime}$), we impose invariance under the additional $A_{3}$\ symmetry given
in Table \ref{y}. It is easy to check that without the charges of this
symmetry, we can add to $W_{H\text{-}F}$ other $A_{4}$-invariant terms like
\begin{equation}
\lambda_{_{\zeta \chi}}\text{ }\Phi \mathcal{\chi}^{2}. \label{sc}%
\end{equation}
But because of Eq. (\ref{ex}), the $W_{H\text{-}F}$ will also have
$\lambda_{\zeta \chi}\Phi \mathcal{\chi}^{\prime}{}^{2},$ and thus an induced
interaction between $\mathcal{\chi}$ and $\mathcal{\chi}^{\prime}$ through
$\Phi$. This feature can be checked by first computing the $F_{\Phi}$ term of
the singlet superfield $\Phi$ singlet and then $\left \vert F_{\Phi}\right \vert
^{2}$. The resulting term
\begin{equation}
\lambda_{\chi \chi^{\prime}}\left \vert \chi \right \vert ^{2}\left \vert
\chi^{\prime}\right \vert ^{2} \label{ct}%
\end{equation}
spoils the vacuum alignment of the triplets (\ref{va}) and (\ref{vacua}). To
prevent the existence of the term (\ref{ct}) in the scalar potential, one of
the triplet-singlet interactions should be excluded; this has been achieved by
the $A_{3}$ charges given in Table \ref{y} [excluding thus the term
(\ref{sc})]. It is possible to choose $\mathcal{\chi^{\prime}}$ to carry a
nonzero charge under $A_{3}$ instead of $\mathcal{\chi}$; this eliminates the
term $\lambda_{\zeta \chi}\Phi \mathcal{\chi}^{\prime}{}^{2}$ from
$W_{H\text{-}F}$ instead of $\lambda_{_{\zeta \chi}}\Phi \mathcal{\chi}^{2}%
$,\textrm{\ }but this choice would take apart the invariance of the
superpotential (\ref{dl}) needed to obtain the TBM matrix consistent with the
data. Therefore, the absence of the term (\ref{sc}) in $W_{H\text{-}F}$
implies the absence of the term (\ref{ct}) in $\mathcal{V}$, thus allowing us
to get the desired vacuum alignment in Eqs. (\ref{va}) and (\ref{vacua}) after
breaking the $A_{4}$ symmetry; see Appendix B for the details. \newline In
addition, if we consider the interchange between $\chi_{i}$\ and $\chi
_{i}^{\prime}$\ for instance in Eq. (\ref{wlepton}), one generates the new
gauge-invariant term
\begin{equation}
\tilde{W}_{lep^{+}}^{\prime}=\frac{y^{ijk}}{\Lambda}\mathcal{\chi}_{i}%
^{\prime}R_{j}^{c}L_{k}H_{d}, \label{u}%
\end{equation}
which is also invariant under\textrm{\ }$A_{4}$. This extra term could be
excluded with a\textrm{\ }$Z_{2}$\textrm{\ }symmetry acting differently on the
two\textrm{\ }$A_{4}$ triplets like
\begin{equation}%
\begin{array}
[c]{c}%
\mathcal{\chi}_{i}\rightarrow+\mathcal{\chi}_{i},\\
\mathcal{\chi}_{i}^{\prime}\rightarrow-\mathcal{\chi}_{i}^{\prime},
\end{array}
\qquad \text{or\qquad}%
\begin{array}
[c]{c}%
\mathcal{\chi}_{i}\rightarrow-\mathcal{\chi}_{i},\\
\mathcal{\chi}_{i}^{\prime}\rightarrow+\mathcal{\chi}_{i}^{\prime}.
\end{array}
\label{gi}%
\end{equation}
One may also assign\textrm{\ }$Z_{2}$\textrm{\ }charges $\left(  +1,-1\right)
$ for the rest of the superfields so that the superpotentials (\ref{wlepton})
and (\ref{ww}) are invariant under $Z_{2}$\ symmetry while preventing Eq.
(\ref{u}). However, within this picture the term\textrm{\ }$\lambda
_{\zeta \Upsilon}\Phi \mathcal{\chi}^{2}$\textrm{\ }cannot be banned with the
two possible assignments in Eq. (\ref{gi}), thus allowing for the existence of
Eq. (\ref{ct}) in the scalar potential which would spoil the vacuum alignment
of the\textrm{\ }$A_{4}$\textrm{ }triplets as mentioned before. This is why we
choose the\textrm{\ }$A_{3}$\textrm{\ }symmetry to exclude the unwanted terms
(\ref{sc}) and (\ref{u}) while keeping the required ones (\ref{wlepton}),
(\ref{ww}), and \ref{fp}) with respect to $A_{3}$ charges assigned to the
various superfields listed in Tables \ref{z} and \ref{y}.\newline As stated in
Sec. III B 2, another chiral superfield is needed to study the deviation from
TBM, so one may ask how this new flavon $\Phi^{\prime}$\ will affect the
scalar potential (\ref{fp}). Since our aim is to study the vacuum alignment of
the $A_{4}$ triplets (\ref{va}) and (\ref{vacua}) and (as we presented above)
only one triplet is allowed to interact with the singlet $\Phi$ in order to
avoid the sequestering problem thanks to the $A_{3}$\ symmetry we have
imposed, as the $A_{3}$ charge assignment for $\Phi^{\prime}$ is the same as
$\Phi$ only one triplet is able to interact with $\Phi^{\prime},$ allowing for
the vacuum alignment to be satisfied also with the presence of this extra flavon.

\section{Deviation from TBM matrix}

In this section we study the angle deviation from TBM in order to reconcile
the reactor angle $\theta_{13}$\ with the recent data collected in Table
\ref{t}. First, we present the perturbation of the neutrino mass matrix
(\ref{NMM}); this perturbation is captured by the VEV\ of the extra chiral
superfield singlet $\Phi^{\prime}$ of the spectrum in Table \ref{y}
transforming as $\mathbf{1}_{(1,\omega)}$ under $A_{4}$.\ Then we study the
effect of this deviation on the mixing angles $\theta_{13}$ and $\theta_{23}$.

\subsection{Deviation by $A_{4}$ singlet $1_{1,\omega}$}

Using the chiral superfield $\Phi^{\prime}$ of Table \ref{y} and the cutoff
$\Lambda$, we see that we can perform a symmetric perturbation of the
superpotential (\ref{sup}) that induces a deviation of the mass matrix
$M_{\upsilon}^{\prime}$ of Eq. (\ref{NMM}). At leading order, the linear
deviation in $\Phi^{\prime}$ that respects the symmetries of the model is as
follows%
\begin{equation}
\delta W_{\nu}^{\prime}=\frac{\Phi^{\prime}}{\Lambda}\left(  L_{e}\Delta
_{d}L_{\mu}+L_{\mu}\Delta_{d}L_{e}+L_{\tau}\Delta_{d}L_{\tau}\right)  ,
\label{h}%
\end{equation}
where the deviation parameter where $\varepsilon=\frac{\left \langle
\Phi^{\prime}\right \rangle }{\Lambda}<<1$. While local gauge and discrete
$A_{3}$ symmetries are manifest, invariance may be explicitly exhibited by
using the $A_{4}$ representation language,%
\begin{equation}%
\begin{tabular}
[c]{lll}%
$L_{e}\Delta_{d}L_{\mu}\frac{\Phi^{\prime}}{\Lambda}$ & $\sim$ &
$\mathbf{1}_{(1,1)}\otimes \mathbf{1}_{(1,1)}\otimes \mathbf{1}_{(1,\omega^{2}%
)}\otimes \mathbf{1}_{(1,\omega)},$\\
$L_{\tau}\Delta_{d}L_{\tau}\frac{\Phi^{\prime}}{\Lambda}$ & $\sim$ &
$\mathbf{1}_{(1,\omega)}\otimes \mathbf{1}_{(1,1)}\otimes \mathbf{1}%
_{(1,\omega)}\otimes \mathbf{1}_{(1,\omega)}.$%
\end{tabular}
\end{equation}
With this correction, the previous neutrino mass matrix $M_{\upsilon}^{\prime
}$ gets deformed as
\begin{equation}
M_{\upsilon}^{\prime \prime}=\upsilon_{\Delta_{d}}\left(
\begin{array}
[c]{ccc}%
1 & b+c+\mathrm{\varepsilon} & b+c\\
b+c+\mathrm{\varepsilon} & b & 1+c\\
b+c & 1+c & b+\mathrm{\varepsilon}%
\end{array}
\right)  . \label{nd}%
\end{equation}
This is a symmetric matrix that can be diagonalized by a similarity
transformation like $M_{\mathrm{diag}}=\tilde{U}^{T}M_{\upsilon}^{\prime
\prime}\tilde{U}$. The system of eigenvalues m$_{i}$ and eigenvectors
$\vec{\upsilon}_{i}$ can be computed perturbatively; we find up to $o\left(
\varepsilon^{2}\right)  $, the eigenvalues (in units of\textrm{\ }%
$\upsilon_{\Delta_{d}}$)
\begin{equation}%
\begin{tabular}
[c]{lll}%
$m_{1}$ & $=$ & $1-c-b-\frac{\varepsilon}{2}+o\left(  \varepsilon^{2}\right)
,$\\
$m_{2}$ & $=$ & $2b+2c+1+\varepsilon,$\\
$m_{3}$ & $=$ & $b-c-1+\frac{\varepsilon}{2}+o\left(  \varepsilon^{2}\right)
,$%
\end{tabular}
\  \label{EVa}%
\end{equation}
and eigenvectors%
\[
\upsilon_{1}=\left(
\begin{array}
[c]{c}%
-\sqrt{\frac{2}{3}}\\
\frac{1}{\sqrt{6}}+\frac{\sqrt{3}\varepsilon}{4\sqrt{2}(b-1)}\\
\frac{1}{\sqrt{6}}-\frac{\sqrt{3}\varepsilon}{4\sqrt{2}(b-1)}%
\end{array}
\right)  ,\quad \upsilon_{2}=\frac{1}{\sqrt{3}}\left(
\begin{array}
[c]{c}%
1\\
1\\
1
\end{array}
\right)  ,\quad \upsilon_{3}=\left(
\begin{array}
[c]{c}%
-\frac{\varepsilon}{2\sqrt{2}(b-1)}\\
-\frac{1}{\sqrt{2}}+\frac{\varepsilon}{4\sqrt{2}(b-1)}\\
\frac{1}{\sqrt{2}}+\frac{\varepsilon}{4\sqrt{2}(b-1)}%
\end{array}
\right)  ,
\]
with the condition $b\neq1$ imposed previously. From these eigenvectors, we
get the unitary matrix $\tilde{U}$ diagonalizing $M_{\upsilon}^{\prime \prime}%
$; it reads, up to order $O(\varepsilon^{2})$,%
\begin{equation}
\tilde{U}=\left(
\begin{array}
[c]{ccc}%
-\sqrt{\frac{2}{3}} & \frac{1}{\sqrt{3}} & -\frac{\varepsilon}{2\sqrt{2}%
(b-1)}\\
\frac{1}{\sqrt{6}}+\frac{\sqrt{3}\varepsilon}{4\sqrt{2}(b-1)} & \frac{1}%
{\sqrt{3}} & -\frac{1}{\sqrt{2}}+\frac{\varepsilon}{4\sqrt{2}(b-1)}\\
\frac{1}{\sqrt{6}}-\frac{\sqrt{3}\varepsilon}{4\sqrt{2}(b-1)} & \frac{1}%
{\sqrt{3}} & \frac{1}{\sqrt{2}}+\frac{\varepsilon}{4\sqrt{2}(b-1)}%
\end{array}
\right)  +O(\varepsilon^{2}) \label{dev}%
\end{equation}
and coincides with TBM in the limit $\varepsilon \rightarrow0$. The unitary
property of the above matrix holds up to second order in the deformation
parameter, i.e., $\tilde{U}^{\dagger}\tilde{U}\simeq I+O\left(  \varepsilon
^{2}\right)  $. Notice, by the way, that\textrm{\ }Eq. (\ref{dev}) depends on
two free parameters $\varepsilon,$ $b$, in particular on $\frac{\varepsilon
}{b-1}$ (which will be used later on). Notice also from Eq. (\ref{dev}) that
the parameter of deviation $\varepsilon$\ does not affect the mixing angle
$\theta_{12},$\ where we have the same value as in the case of TBM,\textrm{\ }%
$\sin \theta_{12}=\frac{1}{\sqrt{3}}$. Moreover, by using the usual
relationships $\sin \theta_{13}=\left \vert U_{e3}\right \vert $ and $\cos
\theta_{13}\sin \theta_{23}=\left \vert U_{\mu3}\right \vert $, we get the link
between the $\theta_{13}$ reactor and the $\theta_{23}$ atmospheric angles and
$b,$ $\varepsilon$ as given below (see also Figs. 1-3):%
\begin{equation}%
\begin{tabular}
[c]{lll}%
$\sin \theta_{13}$ & $=$ & $\left \vert \frac{\varepsilon}{2\sqrt{2}%
(b-1)}\right \vert ,$\\
$\sin \theta_{23}$ & $=$ & $\left \vert \frac{\varepsilon}{4\sqrt{2}(b-1)}%
-\frac{1}{\sqrt{2}}\right \vert .$%
\end{tabular}
\  \  \label{ss}%
\end{equation}
The deviation of the atmospheric angle $\theta_{23}$\ from its TBM value can
be seen as
\begin{equation}
\sin^{2}\theta_{23}=\frac{1}{2}-\frac{\varepsilon}{4(b-1)}+O(\varepsilon^{2}),
\end{equation}
where, by looking at the Table \ref{t}, we understand that%
\begin{align}
-0.143  &  \leq \frac{\varepsilon}{4(b-1)}\leq0.108\qquad \qquad \text{for
NH,}\nonumber \\
-0.14  &  \leq \frac{\varepsilon}{4(b-1)}\leq0.097\qquad \qquad \text{for IH.}%
\end{align}
Using Eq. (\ref{EVa}), the parameter $c$ may be related to the neutrino
mass-squared differences,%
\begin{equation}%
\begin{tabular}
[c]{lll}%
$\Delta m_{31}^{2}$ & $=$ & $4v_{\Delta_{d}}^{2}\left(  1-b-\frac{\varepsilon
}{2}\right)  c,$\\
$\Delta m_{21}^{2}$ & $=$ & $3v_{\Delta_{d}}^{2}\left[  (b+c)\left(
b+c+2+\varepsilon \right)  +\varepsilon \right]  .$%
\end{tabular}
\  \  \label{dd}%
\end{equation}
In the next subsection, we use the experimental values of $\sin \theta_{ij}$
and $\Delta m_{ij}^{2}$ to make predictions concerning numerical estimations
of the parameters $\varepsilon,$ $b,$ and $c$ capturing data on the VEVs of flavons.

\subsection{Normal hierarchy}

\begin{figure}[ptbh]
\begin{minipage}[c]{.45\linewidth}
\begin{center}
\includegraphics[scale=0.62]{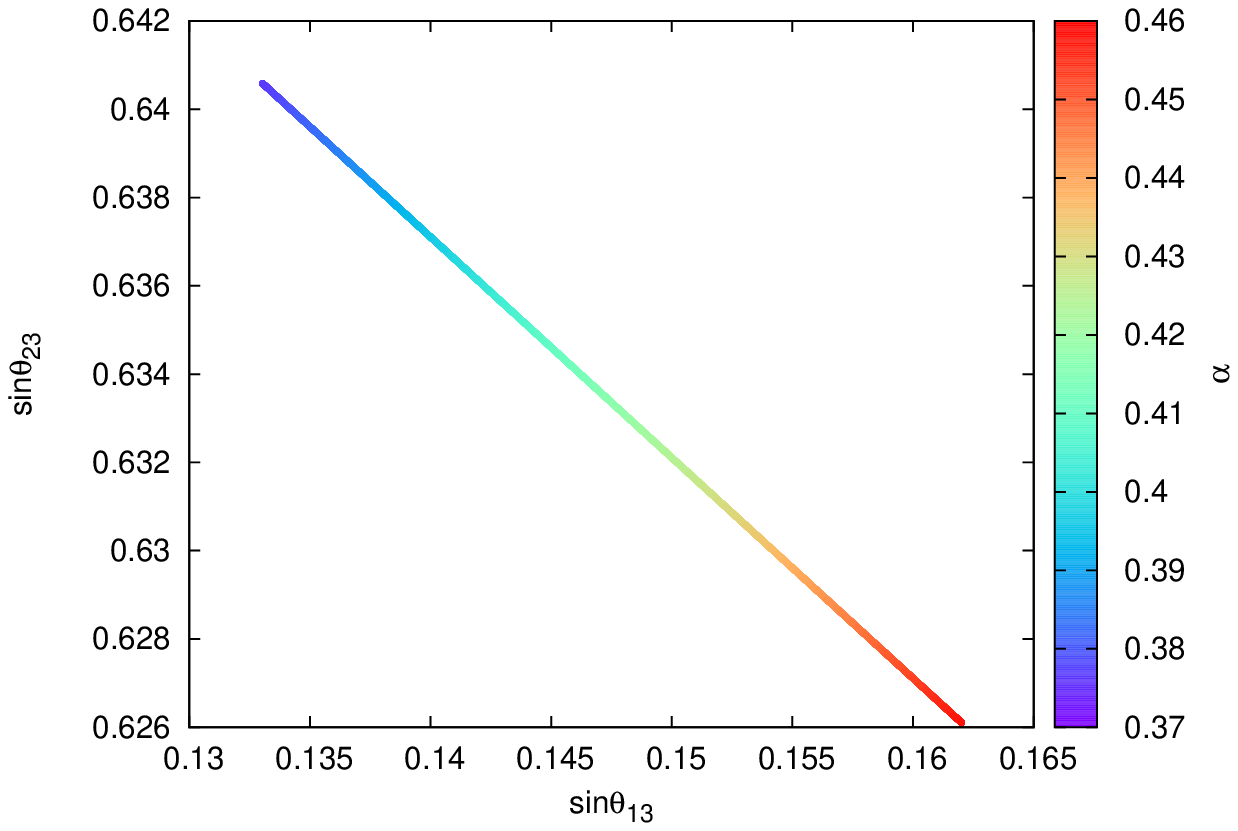}
\end{center}
\end{minipage}
\hfill \begin{minipage}[c]{.45\linewidth}
\begin{center}
\includegraphics[scale=0.62]{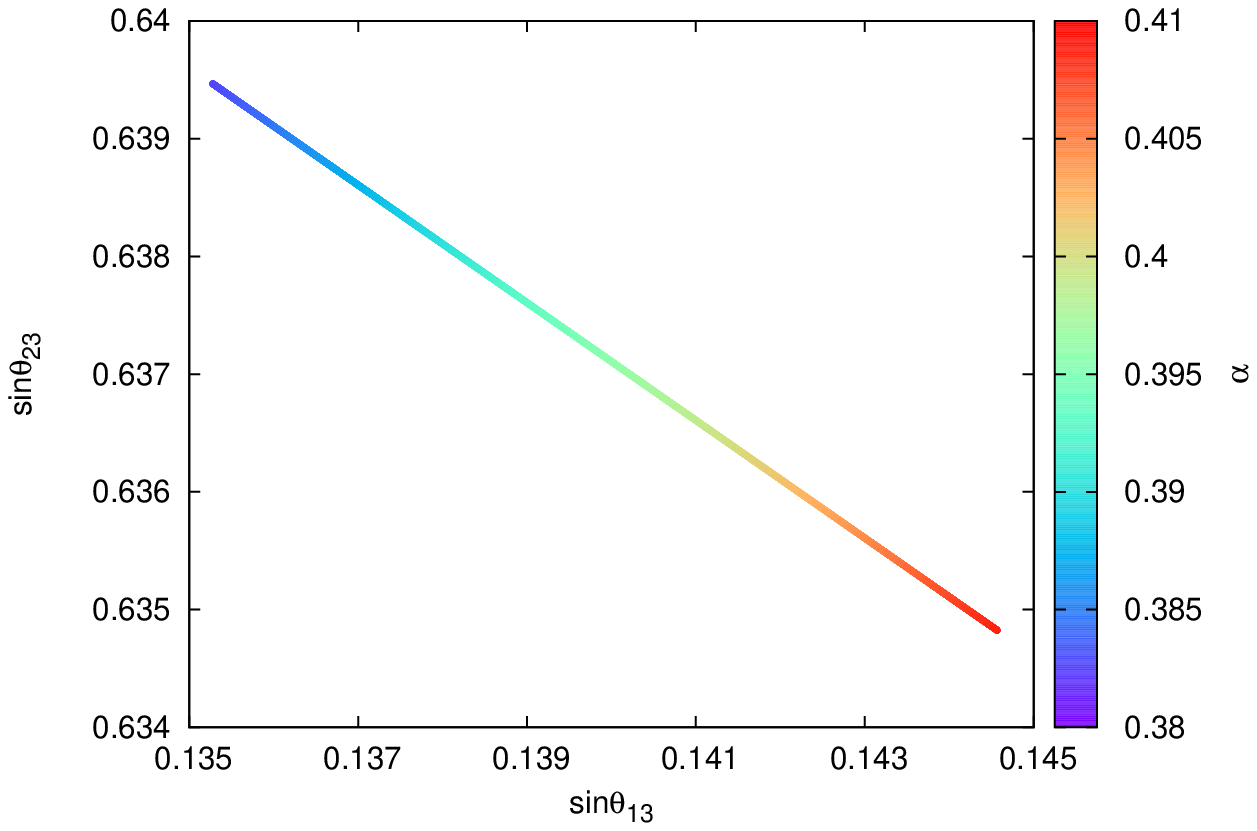}
\end{center}
\end{minipage}
\vspace{0cm}\caption{Left: $\sin \theta_{23}$ as a function of $\sin \theta
_{13}$ with the relative parameter $\alpha=\frac{\varepsilon}{b-1}$ shown in
the palette. Right: The same variation as in the left panel but for inverted
hierarchy.}%
\label{01}%
\end{figure}

Focusing on relations in Eq. (\ref{ss}), we plot in Fig. \ref{01} (left panel)
$\sin \theta_{23}$\ as a function of $\sin \theta_{13}$\ in terms of the ratio
\begin{equation}
\frac{\varepsilon}{b-1}=\alpha
\end{equation}
induced by the VEV of the singlet $\Phi^{\prime}$ (provided the condition
$b\neq1$ holds) and from Eq. (\ref{n}) the relations%
\begin{equation}
\frac{\upsilon_{\chi^{\prime}}^{2}}{\Lambda^{2}}\neq1,\qquad \qquad
\frac{\upsilon_{\chi^{\prime}}}{\Lambda}\neq \pm1.
\end{equation}
Notice that although the matrix (\ref{dev}) involves two free parameters, the
true dependence is only through their ratio $\alpha$ which generates the
deviation of TBM we are interested in. Notice also that to draw this
variation, we have assumed that $\varepsilon$ and $b$ are real parameters, and
by using Eq. (\ref{ss}) we find the linear deviations
\begin{equation}
\sin \theta_{13}=\pm \frac{1}{2\sqrt{2}}\alpha.
\end{equation}
The values of the parameter $\alpha$\ that are compatible with both
$\sin \theta_{13}$\ and $\sin \theta_{23}$\ are shown in the left panel of Fig.
\ref{01} within their $3\sigma$\ allowed range for the normal hierarchy (
$\Delta m_{31}^{2}>0$) case \textrm{\cite{A2}}; see Table \ref{t}. We observe
that the best fit for $\theta_{13},$%
\begin{equation}
\sin \theta_{13}=0.1529,
\end{equation}
corresponds to
\begin{equation}
\alpha \simeq0.43,
\end{equation}
while for $\theta_{23},$\ we have%
\begin{equation}
0.626\leq \sin \theta_{23}\lesssim0.641,
\end{equation}
\textrm{\ }which is in the $\left[  -2\sigma,-3\sigma \right]  $ range (as can
be read from Table \ref{t}), and the interval of $\sin \theta_{23}$ corresponds
to%
\begin{equation}
0.37\leq \alpha \lesssim0.452.
\end{equation}

\subsubsection{Allowed interval for b}

\begin{figure}[ptbh]
\begin{minipage}[c]{.45\linewidth}
\begin{center}
\includegraphics[scale=0.62]{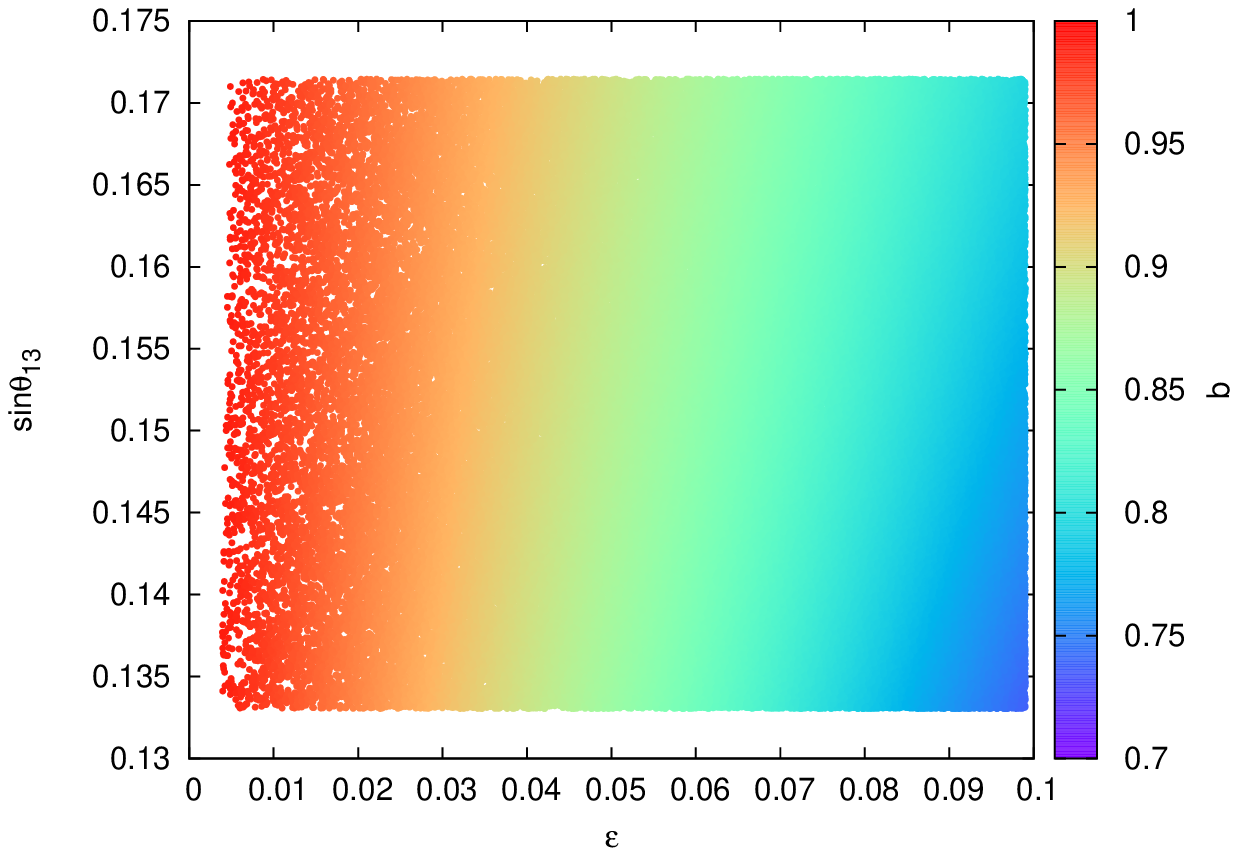}
\end{center}
\end{minipage}
\hfill \begin{minipage}[c]{.45\linewidth}
\begin{center}
\includegraphics[scale=0.52]{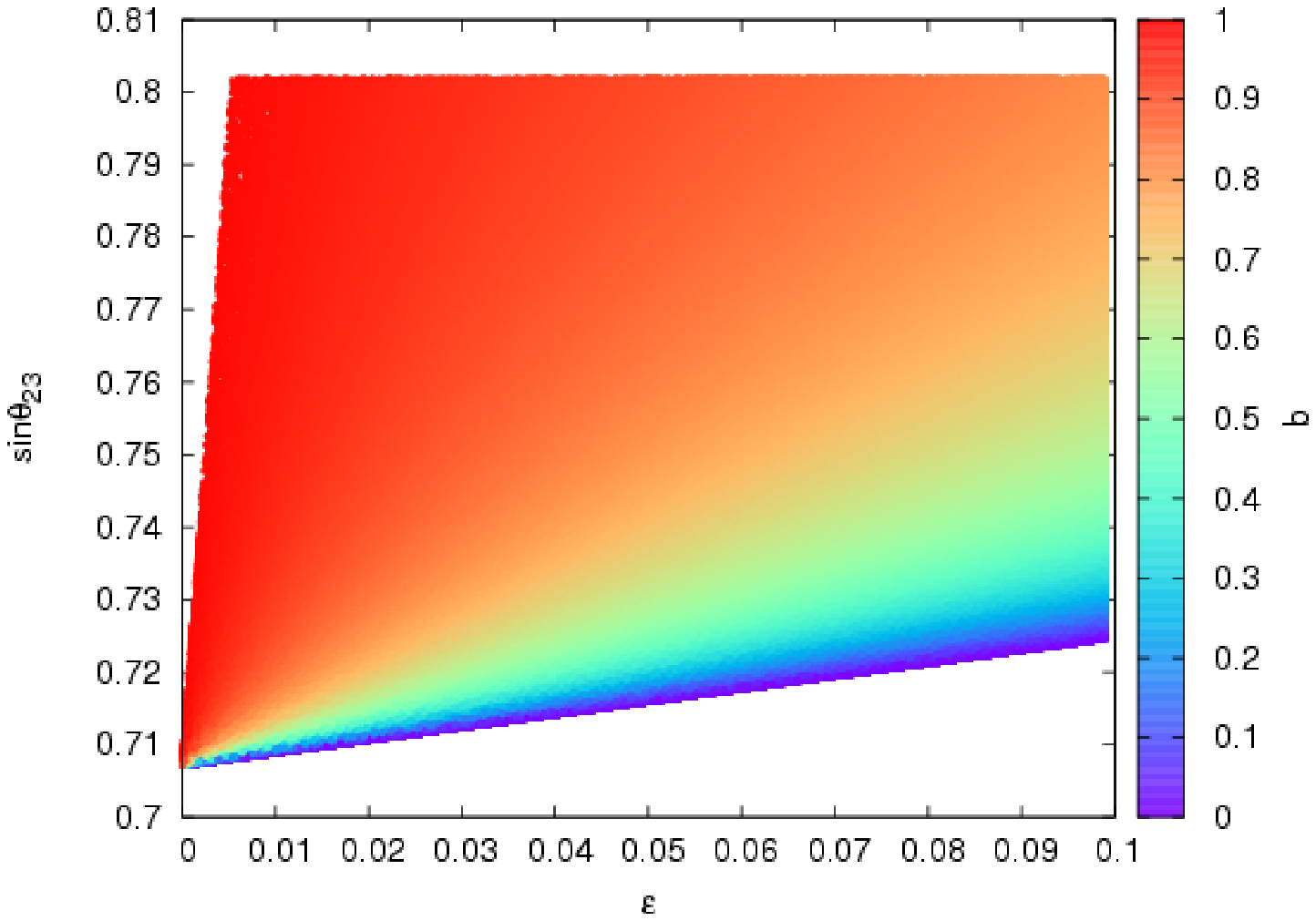}
\end{center}
\end{minipage}
\vspace{0cm}\caption{Left: $\sin \theta_{13}$ as a function of $\varepsilon$
with $b$ shown in the palette on the right. Right: $\sin \theta_{23}$ as a
function of $\varepsilon$ with $b$ shown in the palette on the right.}%
\label{02}%
\end{figure}

Since the parameter of deviation $\varepsilon$\ should be small we fix its
value in the range of $O\left(  \frac{1}{10}\right)  ,$\ and from the
equations in Eq. (\ref{ss}) we plot in the left panel in Fig. \ref{02}
$\sin \theta_{13}$\ as a function of $\varepsilon$ with the parameter
$b$\ presented in the palette on the right. We plot the same variation in the
right panel but for $\sin \theta_{23}$\ instead of $\sin \theta_{13}$. We
observe with the color palettes on the right of both panels in Fig. \ref{02}
that $b$\ is large for different values of $\varepsilon$. Moreover, as we
discussed previously in Sec. III B 2, in order to have a tiny masses for
neutrinos the parameter $b$\ should be less than approximately $1$
$(b\lesssim1)$. Hence, with the order $O\left(  \frac{1}{10}\right)  $\ used
for the range of $\varepsilon$, we read from Fig. \ref{02} that $b$\ is
positive and closely framed as
\begin{equation}
0.005\lesssim b=\frac{\upsilon_{\chi^{\prime}}^{2}}{\Lambda^{2}}<1, \label{8}%
\end{equation}
and by using Eq. (\ref{n}) we conclude that the value of the cutoff $\Lambda$
is around the value $\upsilon_{\chi^{\prime}}$, the VEV of the flavon triplet
$\mathcal{\chi}^{\prime}$. 

\subsubsection{Allowed intervals for c}

\begin{figure}[ptbh]
\begin{minipage}[c]{.45\linewidth}
\begin{center}
\includegraphics[scale=0.62]{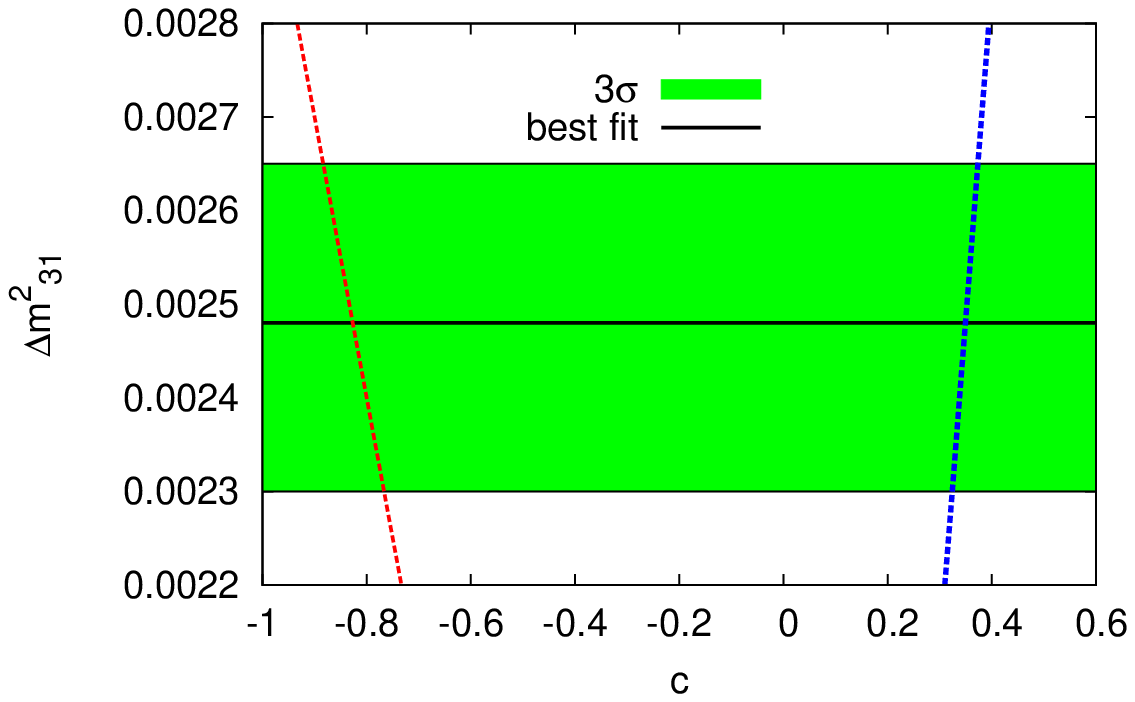}
\end{center}
\end{minipage}
\hfill \begin{minipage}[c]{.45\linewidth}
\begin{center}
\includegraphics[scale=0.62]{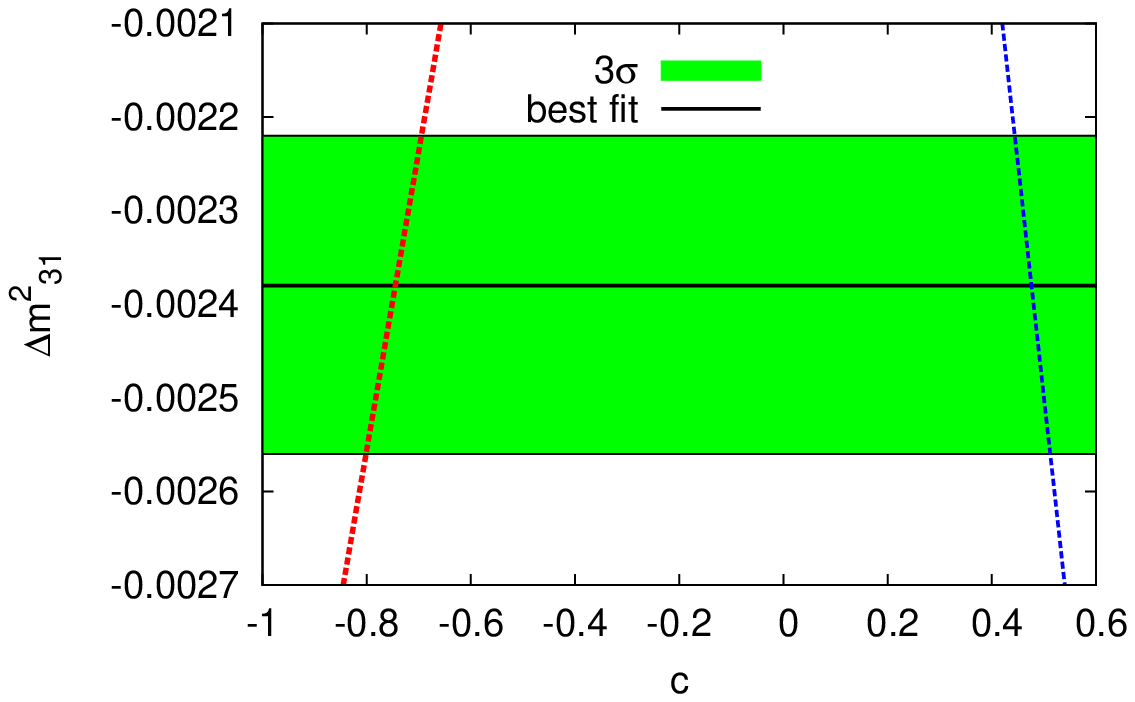}
\end{center}
\end{minipage}
\vspace{-1cm}\caption{Left (Right): Variation of $\Delta m_{31}^{2}$ as a
function of the parameter $c$ for different inputs $(v_{\Delta_{d}}%
^{2},b,\varepsilon)$ for NH (IH).}%
\label{03}%
\end{figure}

To get the allowed interval of the parameter $c$, we shall think of $\left(
\upsilon_{\Delta_{d}}^{2},b,\varepsilon \right)  $ as spectral parameters and
consider the first equation in Eq. (\ref{dd}) with the $3\sigma$ to express
$\Delta m_{31}^{2}$ as a function of $c$. For $\varepsilon \sim \mathcal{O}%
\left(  \frac{1}{10}\right)  $ the parameter $b$ is as in Eq. (\ref{8}), while
in models with an extra Higgs triplet $\Delta_{d}$ the $\upsilon_{\Delta_{d}}$
is fixed by using the relation $\upsilon_{\Delta_{d}}\sim \frac{m_{\nu}%
}{\lambda_{\nu}^{ij}}$ ($\lambda_{\nu}^{ij}$ are the Yukawa couplings). By
using this relation, and the recent cosmological upper bound on the sum of the
neutrino masses (which is constrained to $\sum m_{\nu}<0.23$eV \cite{34}), the
forthcoming inputs for $\upsilon_{\Delta_{d}}^{2}$\ are reasonable. \newline
In the left panel of Fig. \ref{03} we plot the variation of $\Delta m_{31}%
^{2}$ as a function of $c$ in the case of normal hierarchy ($\Delta m_{31}%
^{2}>0$) for two inputs:
\begin{equation}
\upsilon_{\Delta_{d}}^{2}\simeq0.01\mathrm{eV}^{2},\qquad \qquad b\simeq
0.8,\qquad \qquad \varepsilon \simeq0.09,
\end{equation}
for the blue dashed line, and
\begin{equation}
\upsilon_{\Delta_{d}}^{2}\simeq0.3\mathrm{eV}^{2},\qquad \qquad b\simeq
0.98,\qquad \qquad \varepsilon \simeq0.045
\end{equation}
for the red dashed line. It is clear from the equation for $\Delta m_{31}^{2}$ in Eq.
(\ref{dd}) that the sign of $c$ depends only on the value of $b$, which we
found to be positive from Fig. \ref{02}, because $\Delta m_{31}^{2}$ and
$v_{\Delta_{d}}^{2}$ are positive-definite parameters. We observe in the left
panel that $c$ varies in the range
\begin{equation}
0.32\text{ \ }\lesssim \text{ \ }c\text{ \ }\lesssim \text{ \ }0.38
\end{equation}
for the blue dashed line, and\textrm{\ }%
\begin{equation}
-0.83\text{\ }\lesssim \text{ \ }c\text{ \ }\lesssim \text{ \ }-0.78
\end{equation}
for the red dashed line. Notice that the NH depends strongly on the parameter
$b$; for example, for values $0.96\leq b<1$\ we remark that the factor
$\left(  1-b-\frac{\varepsilon}{2}\right)  $ in the first equation of Eq.
(\ref{dd}) is negative, so $c$\ has to be negative as well in order to respect
$\Delta m_{31}^{2}>0$ (red line in left panel of Fig. \ref{03}). On the other
hand, for $0.005\lesssim b\leq0.95,$\ the factor $\left(  1-b-\frac
{\varepsilon}{2}\right)  $\ is positive for any allowed value of $\varepsilon
$; this requires $c$\ to be positive in order to respect $\Delta m_{31}^{2}>0$
(blue line in left panel of Fig. \ref{03}).

\subsection{Inverted hierarchy}

We represent in the right panel of Fig. \ref{01} the same parameters
$\sin \theta_{13}$, $\sin \theta_{23},$ and $\frac{\varepsilon}{b-1}=\alpha$ as
in the left panel of the same figure, but this time for the inverted hierarchy
with ($\Delta m_{31}^{2}<0$). The allowed region for $\alpha$ is constrained
by the values of the mixing angles $\sin \theta_{13}$ and $\sin \theta_{23}$ at
$3\sigma$; we observe that for the$\ $mixing angles $\theta_{23}%
\ $and$\  \theta_{13}$ \textrm{we have}%
\begin{equation}
0.6348\lesssim \sin \theta_{23}\lesssim0.6394,
\end{equation}
\textrm{\ }which is in the range $\left[  -2\sigma,-3\sigma \right]  $ (as can
be read from Table \ref{t}) and%
\begin{equation}
0.1348\lesssim \sin \theta_{13}\lesssim0.1354
\end{equation}
where this intervals corresponds to%
\begin{equation}
0.385\leq \alpha \lesssim0.408.
\end{equation}
We show in the right panel of Fig. \ref{03} the variation of $\Delta
m_{31}^{2}$ as a function of the parameter $c,$ where the latter is
constrained by the $3\sigma$ allowed region of $\Delta m_{31}^{2}$. The input
parameters $b$, $\varepsilon,$\ and $v_{\Delta_{d}}^{2}$\ are as follows:%
\begin{equation}
\upsilon_{\Delta_{d}}^{2}\simeq0.5\mathrm{eV}^{2},\qquad \qquad b\simeq
0.98,\qquad \qquad \varepsilon \simeq0.045,
\end{equation}
for the blue dashed line, and%
\begin{equation}
\upsilon_{\Delta_{d}}^{2}\simeq0.0045\mathrm{eV}^{2},\qquad \qquad
b\simeq0.8,\qquad \qquad \varepsilon \simeq0.08
\end{equation}
for the red dashed line. Thus, we observe that $c$\ varies in the range%
\begin{equation}
0.42\text{ \ }\lesssim \text{ \ }c\text{ \ }\lesssim \text{ \ }0.5
\end{equation}
for the blue dashed line and%
\begin{equation}
-0.8\text{\ }\lesssim \text{ \ }c\text{ \ }\lesssim \text{ \ }-0.7
\end{equation}
for the red dashed line.

\section{LFV to constrain masses}

In this section, we study (LFV) in the charged lepton sector in order to
provide estimations on the mass of the flavon $\chi_{i}$\ and the cutoff scale
$\Lambda$ used in Eqs. (\ref{wlepton}) and (\ref{sp}). First, we break the
$A_{4}$\ symmetry down to $Z_{3}$\ in order to induce LFV in the charged
lepton sector; then, we calculate the analytic flavon masses. Next, we use the
branching ratio of the allowed lepton-flavor-violating decays to give
numerical lower bound estimations on the flavon masses and an upper bound on
the cutoff scale $\Lambda$.

\subsection{Breaking $A_{4}$ to $Z_{3}$}

The discovery of neutrino oscillations provides clear evidence of lepton
flavor violation in the chargeless lepton sector; however, in the charged
sector LFV have not been yet observed. In this subsection, we study the
breaking of the $A_{4}$\ group to its subgroup $Z_{3}$\ in order to get the
allowed lepton-flavor-violating decays mediated by the flavon $\chi_{i}$\ in
the charged lepton sector. \newline To start recall that in Sec. II B 2 the
VEV of the flavon triplet\textrm{\ }was taken as\textrm{\ }$\left \langle
\chi \right \rangle =\upsilon_{\chi}\left(  1,1,1\right)  $\textrm{\ }%
[Eq\textrm{. }(\ref{va})], and because we are working in a basis of $A_{4}%
$\ where the matrix generator $S_{ij}$\ is diagonal\footnote{The alternating
group $A_{4}$ has two noncommuting generators $S$ and $T$ with the property
$S^{2}=T^{3}=I$; because of the noncommutativity $ST\neq TS,$ only one of them
can be chosen diagonal. In Eqs. (\ref{m0}) and (\ref{m00}), the diagonal $S$
and nondiagonal $T$ are, respectively given by the matrices $a_{2}$ and
$b_{1}$.}\textrm{ }this structure of the triplet VEV breaks $A_{4}$\ down to
its subgroup $Z_{3},$\ with the matrix $T_{ij}$ as a generator,
\begin{equation}
T_{ij}\left \langle \chi_{j}\right \rangle =0\qquad,\qquad S_{ij}\left \langle
\chi_{j}\right \rangle \neq0.
\end{equation}
By looking at the characters of the $S$\ and $T$\ generators of $A_{4}$\ for
the lepton superfields (\ref{tt}), it is not difficult to check that leptons
$l_{i}$ transform in different manners under the three possible
representations\textrm{\ }$\mathbf{1}_{\omega^{r}}$\textrm{\ }of the residual
symmetry\textrm{\ }$Z_{3}$ characterized by the phases $\omega^{r}%
=e^{\frac{2i\pi r}{3}},$ with $r=0,1,2$ and sum $1+\omega+\omega^{2}=0$.
Indeed, because $A_{4}$ singlets are also singlets of its subgroup $Z_{3}$,
the left-handed charged leptons $L_{x}$ live in the representations%
\begin{equation}
L_{e}\sim \mathbf{1}_{1},\qquad \qquad L_{\mu}\sim \mathbf{1}_{\omega^{2}}%
,\qquad \qquad L_{\tau}\sim \mathbf{1}_{\omega}, \label{zz}%
\end{equation}
and because of the decomposition of the $A_{4}$ triplet $\mathbf{3}$ in terms
of irreducible $Z_{3}$ representations (namely, $\mathbf{3}_{0}=\mathbf{1}%
_{1}\oplus \mathbf{1}_{\omega}\oplus \mathbf{1}_{\omega^{2}}$), the right-handed
$A_{4}$ triplet $\left(  e_{i}^{c}\right)  \sim \mathbf{3}$ is now combined
into three $Z_{3}$ singlets with different characters as follows%
\begin{equation}%
\begin{array}
[c]{cccc}%
e^{c}= & \frac{1}{\sqrt{3}}\left(  e_{1}^{c}+e_{2}^{c}+e_{3}^{c}\right)
\text{ \  \  \  \ } & \sim & \mathbf{1}_{1},\\
\mu^{c}= & \frac{1}{\sqrt{3}}\left(  e_{1}^{c}+\omega e_{2}^{c}+\omega
^{2}e_{3}^{c}\right)  & \sim & \mathbf{1}_{\omega},\\
\tau^{c}= & \frac{1}{\sqrt{3}}\left(  e_{1}^{c}+\omega^{2}e_{2}^{c}+\omega
e_{3}^{c}\right)  & \sim & \mathbf{1}_{\omega^{2}}.
\end{array}
\label{gg}%
\end{equation}
Consequently, the radiative decays $l_{i}\rightarrow l_{j}\gamma$\ ($i\neq j$)
are all excluded in our model by the residual symmetry $Z_{3}$; this is
because $l_{i}$ and $l_{j}$ live in different representations $\mathbf{1}%
_{\omega^{i}}$ and $\mathbf{1}_{\omega^{j}},$ and the photon $\gamma$ is a
singlet of $Z_{3}$. On the other hand, by using Eqs. (\ref{zz})and (\ref{gg}),
the LFV three-body decays\textrm{\ }%
\begin{equation}
\tau^{+}\rightarrow e^{+}e^{+}\mu^{-},\qquad \qquad \tau^{+}\rightarrow \mu
^{+}\mu^{+}e^{-} \label{vv}%
\end{equation}
and their charged conjugates are allowed due to the representation character
property $\mathbf{1}_{\omega^{n}}\otimes \mathbf{1}_{\omega^{m}}=\mathbf{1}%
_{\omega^{n+m}}$.\textrm{ }As these decay modes are mediated by the flavon
triplet $\chi_{i},$\ we start by calculating its mass.

\subsection{Mass matrix of flavons}

In order to calculate the mass matrix of field modes ${\small \xi}_{i}$
describing the $\chi_{i}$ fluctuations near the\ vacuum expectation value
$\left(  \upsilon_{\chi},\upsilon_{\chi},\upsilon_{\chi}\right)  $\ of the
flavon triplet $\chi_{i}$, we proceed as follows. First, we consider the pure
$\chi$ contribution $\mathcal{V}_{\chi}$ to the full scalar potential
(\ref{fp}) of the model; it is given by $\mathcal{V}_{\chi}=\mathrm{Tr}%
_{A_{4}}V_{\chi}$ with%
\begin{equation}
V_{\chi}=\left(  \left \vert 3\lambda \chi^{2}\right \vert ^{2}+m_{\chi}%
^{2}\left \vert \chi \right \vert ^{2}+2A_{\chi}\chi^{3}\right)
\end{equation}
[where $\chi^{2}$ stands for $\chi \otimes \chi \equiv \left(  \chi_{i}\chi
_{j}\right)  ],$ and a similar relation for the other $\chi^{3}$ and $\chi
^{4}$ terms. \newline Second, we use $A_{4}$ representation properties to
decompose these tensor products into sums over irreducible representations of
$A_{4}$ and take the trace afterwards; the explicit expression of
$\mathrm{Tr}_{A_{4}}V_{\chi}$ can be read by substituting Eqs. (\ref{m1}) and
(\ref{m3}) from Appendix C. Then, we expand the flavon field triplet $\left(
\chi_{1},\chi_{2},\chi_{3}\right)  $ around the vacuum expectation value as
follows:%
\begin{equation}%
\begin{tabular}
[c]{lll}%
$\chi_{1}$ & $=$ & $\upsilon_{\chi}+{\small \xi}_{1},$\\
$\chi_{2}$ & $=$ & $\upsilon_{\chi}+{\small \xi}_{2},$\\
$\chi_{3}$ & $=$ & $\upsilon_{\chi}+{\small \xi}_{3},$%
\end{tabular}
\  \  \label{ee}%
\end{equation}
where the ${\small \xi}_{i}$'s are field fluctuations; they will be thought of
as real fields. This step, which breaks $A_{4}$ to its subgroup $Z_{3}$, leads
to a quartic scalar potential $\mathcal{V}_{\chi}=\mathcal{V}\left(
{\small \xi}_{1},{\small \xi}_{2},{\small \xi}_{3}\right)  $ from which we can
determine the mass matrix%
\begin{equation}
\left(  m_{\xi}^{2}\right)  _{ij}=\frac{1}{2}\left.  \frac{\partial
^{2}\mathcal{V}_{\chi}}{\partial{\small \xi}_{i}\partial{\small \xi}_{j}%
}\right \vert _{\xi=0}.
\end{equation}
It reads explicitly as follows:%
\begin{equation}
\left(  m_{\xi}^{2}\right)  _{ij}=\frac{1}{2}\left(
\begin{array}
[c]{ccc}%
m_{\chi}^{2}+234\lambda^{2}\upsilon_{\chi}^{2} & 144\lambda^{2}\upsilon_{\chi
}^{2}+12A_{\chi}\upsilon_{\chi} & 144\lambda^{2}\upsilon_{\chi}^{2}+12A_{\chi
}\upsilon_{\chi}\\
144\lambda^{2}\upsilon_{\chi}^{2}+12A_{\chi}\upsilon_{\chi} & m_{\chi}%
^{2}+234\lambda^{2}\upsilon_{\chi}^{2} & 144\lambda^{2}\upsilon_{\chi}%
^{2}+12A_{\chi}\upsilon_{\chi}\\
144\lambda^{2}\upsilon_{\chi}^{2}+12A_{\chi}\upsilon_{\chi} & 144\lambda
^{2}\upsilon_{\chi}^{2}+12A_{\chi}\upsilon_{\chi} & m_{\chi}^{2}%
+234\lambda^{2}\upsilon_{\chi}^{2}%
\end{array}
\right)  . \label{mm}%
\end{equation}
The next step is to diagonalize the above mass matrix; we find%
\begin{equation}%
\begin{tabular}
[c]{lll}%
$m_{{\small \xi}_{1}}^{2}$ & $=$ & $\frac{1}{2}m_{\chi}^{2}+45\lambda
^{2}\upsilon_{\chi}^{2}-6A_{\chi}\upsilon_{\chi},$\\
$m_{{\small \xi}_{2}}^{2}$ & $=$ & $m_{{\small \xi}_{1}}^{2},$\\
$m_{{\small \xi}_{3}}^{2}$ & $=$ & $\frac{1}{2}m_{\chi}^{2}+261\lambda
^{2}\upsilon_{\chi}^{2}+12A_{\chi}\upsilon_{\chi},$%
\end{tabular}
\  \  \  \label{mf}%
\end{equation}
with two degenerate values.

\subsection{Mass scale $\Lambda$}

\begin{figure}[h]
\begin{center}
\includegraphics[scale=0.6]{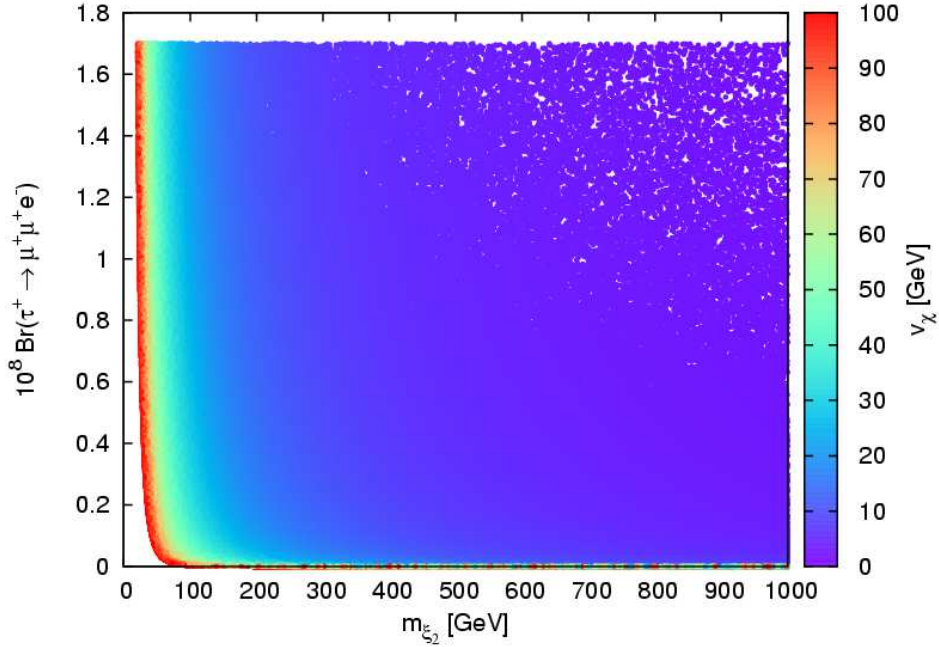}
\end{center}
\par
\vspace{-1.5cm} \hfill \caption{{}$\mathrm{Br}\left(  \tau^{+}\rightarrow
\mu^{+}\mu^{+}e^{-}\right)  $ as a function of $m_{{\protect \small \xi}_{2}}$
with $\upsilon_{\chi}$ shown in the palette on the right.}%
\label{04}%
\end{figure}

To get the order of magnitude of the cutoff scale, we need extra information
in addition to the above flavon masses (\ref{mf}), in particular the structure
of the flavon Yukawa couplings $\left.  L_{\mathrm{Yuk}}\right \vert _{\xi}$ in
the charged lepton sector. To be able to use the experimental results on
branching ratios (\ref{vv}), the explicit expression of $\left.
L_{\mathrm{Yuk}}\right \vert _{\xi}$ is also needed to extract information
about which of the fields $\xi_{i}$ is exchanged in lepton-flavor-violating
decays. \textrm{The} \textrm{fields }$\xi_{i}$\textrm{ transform under }%
$Z_{3}$ \textrm{symmetry like}%
\begin{equation}
{\small \xi}_{1}\sim \mathbf{1}_{1,}\qquad \qquad{\small \xi}_{2}\sim
\mathbf{1}_{\omega},\qquad \qquad{\small \xi}_{3}\sim \mathbf{1}_{\omega^{2}}.
\label{ii}%
\end{equation}
\textrm{Hence,} we obtain the desired expression for $\left.  L_{\mathrm{Yuk}%
}\right \vert _{\xi}$ which, by using Eqs. (\ref{zz}), (\ref{gg}), and
(\ref{ii}) reads as follows:%
\begin{equation}%
\begin{array}
[c]{ccc}%
\left.  L_{\mathrm{Yuk}}\right \vert _{\xi} & = & \frac{y_{e}\upsilon_{d}%
}{\Lambda}(e^{c}{\small \xi}_{1}+\mu^{c}{\small \xi}_{3}+\tau^{c}{\small \xi
}_{2})L_{e}\\
& + & \frac{y_{\mu}\upsilon_{d}}{\Lambda}(e^{c}{\small \xi}_{2}+\mu
^{c}{\small \xi}_{1}+\tau^{c}{\small \xi}_{3})L_{\mu}\\
& + & \frac{y_{\tau}\upsilon_{d}}{\Lambda}(e^{c}{\small \xi}_{3}+\mu
^{c}{\small \xi}_{2}+\tau^{c}{\small \xi}_{1})L_{\tau}.
\end{array}
\end{equation}
Moreover, by substituting\textrm{\ }the expression for the lepton masses we
obtained in Sec. II B 2 [Eq. (\ref{ml})],\textrm{\ }the flavon Yukawa
interactions of the charged leptons in terms of the flavons $\xi_{i}$\ are
given by%
\begin{equation}%
\begin{array}
[c]{ccc}%
\left.  L_{\mathrm{Yuk}}\right \vert _{\xi} & = & \left(  \frac{m_{e}}{\sqrt
{3}\upsilon_{\chi}}e^{c}L_{e}+\frac{m_{\mu}}{\sqrt{3}\upsilon_{\chi}}\mu
^{c}L_{\mu}+\frac{m_{\tau}}{\sqrt{3}\upsilon_{\chi}}\tau^{c}L_{\tau}\right)
{\small \xi}_{1}\\
& + & \left(  \frac{m_{e}}{\sqrt{3}\upsilon_{\chi}}\tau^{c}L_{e}+\frac{m_{\mu
}}{\sqrt{3}\upsilon_{\chi}}e^{c}L_{\mu}+\frac{m_{\tau}}{\sqrt{3}\upsilon
_{\chi}}\mu^{c}L_{\tau}\right)  {\small \xi}_{2}\\
& + & \left(  \frac{m_{e}}{\sqrt{3}\upsilon_{\chi}}\mu^{c}L_{e}+\frac{m_{\mu}%
}{\sqrt{3}\upsilon_{\chi}}\tau^{c}L_{\mu}+\frac{m_{\tau}}{\sqrt{3}%
\upsilon_{\chi}}e^{c}L_{\tau}\right)  {\small \xi}_{3}%
\end{array}
\end{equation}
Accordingly, \textrm{we find that the flavon exchange }$\xi_{1}$\textrm{\ does
not lead to flavor violation while the flavons} $\xi_{2}$ and $\xi_{3}$
contribute to the lepton flavor violation processes (\ref{vv}).\textrm{
}Following Ref. \cite{FL} and assuming that the contribution of supersymmetric
particles in the decay modes (\ref{vv}) is negligible, the branching ratios of
the these decays are as follows:%
\begin{equation}%
\begin{array}
[c]{ccc}%
\mathrm{Br}\left(  \tau^{+}\rightarrow e^{+}e^{+}\mu^{-}\right)  & = &
t_{\tau}\frac{m_{\tau}^{5}}{3072\pi^{3}}\left(  \left \vert \frac{m_{\tau}%
m_{e}}{3\upsilon_{\chi}^{2}m_{{\small \xi}_{3}}^{2}}\right \vert ^{2}%
+\left \vert \frac{m_{e}m_{\mu}}{3\upsilon_{\chi}^{2}m_{{\small \xi}_{2}}^{2}%
}\right \vert ^{2}\right)  ,\\
\mathrm{Br}\left(  \tau^{+}\rightarrow \mu^{+}\mu^{+}e^{-}\right)  & = &
t_{\tau}\frac{m_{\tau}^{5}}{3072\pi^{3}}\left(  \left \vert \frac{m_{\tau
}m_{\mu}}{3\upsilon_{\chi}^{2}m_{{\small \xi}_{2}}^{2}}\right \vert
^{2}+\left \vert \frac{m_{\mu}m_{e}}{3\upsilon_{\chi}^{2}m_{{\small \xi}_{3}%
}^{2}}\right \vert ^{2}\right)  ,
\end{array}
\label{BR}%
\end{equation}
where $t_{\tau}$\ is the mean life of the tau lepton. To get an estimate on
$m_{{\small \xi}_{2}}^{2}$, we consider the second equation in Eq. (\ref{BR})
and we assume that all terms proportional to $m_{e}^{2}m_{\mu}^{2}$\ and
$m_{\tau}^{2}m_{e}^{2}$\ are negligible because $m_{e}<<m_{\mu}<<m_{\tau}$; we
obtain the branching ratio
\begin{equation}
\mathrm{Br}\left(  \tau^{+}\rightarrow \mu^{+}\mu^{+}e^{-}\right)  \simeq
t_{\tau}\frac{m_{\tau}^{7}m_{\mu}^{2}}{27648\pi^{3}\upsilon_{\chi}^{4}}%
\frac{1}{m_{{\small \xi}_{2}}^{4}}%
\end{equation}
which, after substituting $t_{\tau}$\ as well as the numerical values of the
leptons masses from the Particle Data Group (PDG) \cite{DG}, we obtain%
\begin{equation}
\mathrm{Br}\left(  \tau^{+}\rightarrow \mu^{+}\mu^{+}e^{-}\right)  \simeq
\frac{3.21}{\upsilon_{\chi}^{4}m_{{\small \xi}_{2}}^{4}}\times10^{5}%
\mathrm{GeV}^{8} \label{br}%
\end{equation}
Using the current upper bound of the branching ratio (\ref{br}), which is
$\mathrm{Br}\left(  \tau^{+}\rightarrow \mu^{+}\mu^{+}e^{-}\right)
<1.7\times10^{-8}$\ at $90\%$ C.L. \cite{DG}, we get the following lower bound
on the mass:%
\begin{equation}
m_{{\small \xi}_{2}}^{2}\gtrsim \frac{10^{2}}{\upsilon_{\chi}^{2}}\sqrt
{t_{\tau}\frac{m_{\tau}^{7}m_{\mu}^{2}}{4.7\pi^{3}}}. \label{lb}%
\end{equation}
If we assume that the mass of the flavon $\xi_{2}$ is of same order of
magnitude as $\upsilon_{\chi}$--say, $m_{{\small \xi}_{2}}\simeq \upsilon
_{\chi}$--we get a lower bound on its mass $m_{{\small \xi}_{2}}%
\gtrsim45.6\mathrm{GeV}$, which is surprisingly very light. With this limit,
such kind of flavons could be generated through several decays; for instance,
if the flavon mass $m_{{\small \xi}_{2}}$\ could be lighter than the $Z^{0}%
$-\ boson, the decay $Z^{0}\rightarrow f\bar{f}\xi_{2}$\ could occur at tree
level.\textrm{\ }Moreover, using Eq. (\ref{cu}), by giving a lower bound on
the ratio of the flavon VEV with respect to the cutoff scale (namely
$\frac{\upsilon_{\chi}}{\Lambda}>0.004$) and taking $m_{{\small \xi}_{2}%
}\simeq \upsilon_{\chi}$, we find an upper bound for the cutoff scale given by%
\begin{equation}
\Lambda \lesssim1.14\times10^{4}\mathrm{GeV.}%
\end{equation}
Notice that in Eq. (\ref{mf}) if the flavon trilinear coupling $A_{\chi}\geq
0$, the mass of the flavon $\xi_{3}$\ could be heavier than $m_{{\small \xi
}_{2}}=m_{{\small \xi}_{1}}$. However, the lower bound of the flavon mass in
Eq. (\ref{lb}) depends on $\upsilon_{\chi}$\ and is specific for our model; in
general, such as constraint is\ model dependent\textrm{. }To illustrate the
relationship between the mass $m_{{\small \xi}_{2}}$\ and the VEV
$\upsilon_{\chi}$, we plot in Fig. \ref{04} the branching ratio $\mathrm{Br}%
\left(  \tau^{+}\rightarrow \mu^{+}\mu^{+}e^{-}\right)  $ as a function of
$m_{{\small \xi}_{2}}$\ for $\upsilon_{\chi}<10^{2}$GeV represented by the
color palette on the right of the figure. We observe that for $\upsilon_{\chi
}\in \lbrack40-100]$ GeV the mass $m_{{\small \xi}_{2}}$ is less than $100$ GeV
including the value we find above for $m_{{\small \xi}_{2}}\simeq
\upsilon_{\chi}$; on the other hand, when the value of $\upsilon_{\chi}$\ goes
down to $40$ GeV, $m_{{\small \xi}_{2}}$\ rises up until $1$ TeV which
corresponds to $\upsilon_{\chi}\simeq10$ GeV and\ to an upper bound of the
cutoff scale of the order $\Lambda \lesssim2.5\times10^{3}$ $\mathrm{GeV}$.
Hence, as $m_{{\small \xi}_{2}}$ increases both $\Lambda$ and $\upsilon_{\chi
}$\ decrease.

As a general comment, since the four flavon superfields we added
in our model are all gauge singlets, they do not contribute to the mass of
$W^{\pm}$ and $Z^{0}$ bosons. However, in the scalar potential (\ref{fp}) we
notice that the flavon $\chi^{\prime}$\ mixes with the Higgs doublets $H_{u}$
and $H_{d}$; thus, they might contribute to the so-called\ $S$\ and\ $T$%
\ oblique parameters\  \cite{PE}. Moreover, because some of the flavons could
be lighter than the Higgs or the $Z^{0}$ boson, they will open new decay
channels for these particles; as these two final points requires examining the
collider phenomenology of the flavons, we leave the detailed investigations to
future work.

\section{Conclusion and discussion}

In this paper, we have constructed a supersymmetric neutrino model based on
$A_{4}\times A_{3}$ discrete symmetry. In this model, neutrinos acquire a
Majorana mass via the type II seesaw mechanism, and TBM acquires an
appropriate deviation with $\theta_{13}\neq0$.\newline First, we showed that
it is possible to obtain the TBM pattern with only one {$A$}$_{4}$ triplet;
however, we found that the physical observable $\Delta m_{31}^{2}=0$ which is
in conflict with the present data. We then allowed for the presence of an
extra {$A$}$_{4}$ scalar singlet $\Phi \sim1_{1,1}$ which successfully
reproduced the TBM matrix with $\Delta m_{31}^{2}\neq0$, see Eq(\ref{NMM}). We
have studied the scalar potential of the supersymmetric model where we allowed
the addition of an extra $A_{3}$ discrete symmetry, which is necessary to
forbid the terms coming from the interchange between the TBM {$A$}$_{4}$
triplet and the one involved in the charged lepton sector, and also to avoid
the sequestering problem.\newline We next studied the perturbation of the
neutrino mass matrix that induces a deviation from the TBM matrix leading
therefore to a nonzero $\theta_{13}$ as proved by many experiments recently.
This deviation is made with the help of a nontrivial $A_{4}$ singlet
$\Phi^{\prime}$ which transforms under it as $1_{1,\omega}$. In the beginning,
we gave the resulting neutrino mass matrix (\ref{nd}) which received a new
contribution from the VEV singlet $\Phi^{\prime}$. Then, we gave the deformed
TBM matrix where the reactor angle $\theta_{13}\neq0$ [Eq. (\ref{EVa})]. Next,
we showed numerically by means of scatter plots the allowed regions of the
parameters of the model which we have constrained by using the $3\sigma$
ranges of the neutrino oscillation parameters $\sin \theta_{31},$\ $\sin
\theta_{23},$\ and $\Delta m_{31}^{2}$. Moreover, we gave the allowed regions
of the parameter $c$ where we found that the normal and inverted hierarchies
are both permitted in our model. Finally, after discussing how the VEV
alignment of the flavon triplet in the charged lepton sector breaks $A_{4}$ to
$Z_{3}$, we studied the LFV in this sector and we found that only the
three-body decays $\tau \rightarrow ee\mu$ and $\tau \rightarrow \mu \mu e$ are
possible under the residual symmetry $Z_{3}$. We also found that these decays
are mediated by the flavons\ $\xi_{2}$\ and $\xi_{3}$; therefore, we
calculated the lower bound of the flavon mass $m_{\xi_{2}}$\ by using the
experimental branching ratio of the decay $\tau \rightarrow \mu \mu e$ where we
found that $m_{{\small \xi}_{2}}$\ is very light $\left(  m_{{\small \xi}_{2}%
}\gtrsim45.6\mathrm{GeV}\right)  $ if we assume $m_{{\small \xi}_{2}}%
\simeq \upsilon_{\chi}$.\ We then used the relation between the cutoff scale
$\Lambda$ and $\upsilon_{\chi}$ (namely $\frac{\upsilon_{\chi}}{\Lambda
}>0.004$) to get an estimation on the upper bound of the cutoff scale which we
found to be of the order of $1.14\times10^{4}$GeV. Nevertheless, we showed in
Fig. \ref{04} that the bound of $m_{{\small \xi}_{2}}$ increases when
$\upsilon_{\chi}$\ decreases, and therefore, the cutoff scale also decreases,
giving its relation with\textrm{\ }$\upsilon_{\chi}$\textrm{.}\newline We end
this conclusion by making a comment on the TBM deviation using the other
non-$A_{4}$ singlet $\mathbf{1}_{(1,\omega^{2})}\sim \Phi^{\prime \prime}$
instead of $\mathbf{1}_{(1,\omega)}\sim \Phi^{\prime}$.\ The new contributions
added to the superpotential (\ref{sup}) are given by%
\begin{equation}
\delta W_{\nu}=\frac{\Phi^{\prime \prime}}{\Lambda}\left(  L_{e}\Delta
_{d}L_{\tau}+L_{\mu}\Delta_{d}L_{\mu}+L_{\tau}\Delta_{d}L_{e}\right)  ,
\end{equation}
where the cutoff $\Lambda$ is the same as before. The invariance of the above
$\delta W_{\nu}$ under $A_{4}$ may be exhibited explicitly by using%
\begin{equation}%
\begin{tabular}
[c]{lll}%
$L_{e}\Delta_{d}L_{\tau}\frac{\Phi^{^{\prime \prime}}}{\Lambda}$ & $\sim$ &
$1_{(1,1)}\otimes1_{(1,1)}\otimes1_{(1,\omega)}\otimes1_{(1,\omega^{2})},$\\
$L_{\mu}\Delta_{d}L_{\mu}\frac{\Phi^{^{\prime \prime}}}{\Lambda}$ & $\sim$ &
$1_{(1,\omega^{2})}\otimes1_{(1,1)}\otimes1_{(1,\omega^{2})}\otimes
1_{(1,\omega^{2})}.$%
\end{tabular}
\end{equation}
With this $\Phi^{\prime}$- correction, the previous neutrino mass matrix
$M_{\upsilon}^{\prime}$ gets deformed as
\begin{equation}
\hat{M}_{\upsilon}=\upsilon_{\Delta_{d}}\left(
\begin{array}
[c]{ccc}%
1 & b+c & b+c+\mathrm{\varepsilon}\\
b+c & b+\mathrm{\varepsilon} & 1+c\\
b+c+\mathrm{\varepsilon} & 1+c & b
\end{array}
\right)  .
\end{equation}
We repeat the same study as in the case of the singlet $\Phi^{\prime}.$\ We
find that the eigenvectors at first order of $\varepsilon$\ are as follows:%
\begin{equation}
\tilde{U}^{\prime}=\left(
\begin{array}
[c]{ccc}%
-\sqrt{\frac{2}{3}} & \frac{1}{\sqrt{3}} & \frac{\varepsilon}{2\sqrt{2}%
(b-1)}\\
\frac{1}{\sqrt{6}}-\frac{\sqrt{3}\varepsilon}{4\sqrt{2}(b-1)} & \frac{1}%
{\sqrt{3}} & -\frac{1}{\sqrt{2}}-\frac{\varepsilon}{4\sqrt{2}(b-1)}\\
\frac{1}{\sqrt{6}}+\frac{\sqrt{3}\varepsilon}{4\sqrt{2}(b-1)} & \frac{1}%
{\sqrt{3}} & \frac{1}{\sqrt{2}}-\frac{\varepsilon}{4\sqrt{2}(b-1)}%
\end{array}
\right)  +O(\varepsilon^{2}),
\end{equation}
where after diagonalizing $\hat{M}_{\upsilon}$ by the transformation
$M_{\mathrm{diag}}$=$\tilde{U}^{\prime T}\hat{M}_{\upsilon}\tilde{U}^{\prime}%
$, we obtain the same mass eigenvalues as in the case of the singlet
$\Phi^{\prime}$ [Eq. (\ref{EVa})] and therefore the same neutrino mass-squared
differences $\Delta m_{ij}^{2}$\ as in Eq.(\ref{dd}).\ The mixing angles in
the case of $\Phi^{\prime \prime}$\ are given by%
\begin{equation}%
\begin{tabular}
[c]{lll}%
$\sin \theta_{13}$ & $=$ & $\left \vert \frac{\varepsilon}{2\sqrt{2}%
(b-1)}\right \vert ,$\\
$\sin \theta_{23}$ & $=$ & $\left \vert -\frac{1}{\sqrt{2}}-\frac{\varepsilon
}{4\sqrt{2}(b-1)}\right \vert .$%
\end{tabular}
\end{equation}
The deviation of the atmospheric angle $\theta_{23}$\ from its TBM value can
be seen as
\begin{equation}
\sin^{2}\theta_{23}=\frac{1}{2}+\frac{\varepsilon}{4(b-1)}+O(\varepsilon^{2}),
\end{equation}
where the sign in front of $\frac{\varepsilon}{4(b-1)}$ is changed compared to
the case of the singlet $\Phi^{\prime}$. Therefore, the signs of its intervals
are reversed as follows\textrm{:}%
\begin{align}
-0.108  &  \leq \frac{\varepsilon}{4(b-1)}\leq0.143\qquad \text{for
NH.}\nonumber \\
-0.097  &  \leq \frac{\varepsilon}{4(b-1)}\leq0.14\qquad \text{for IH.}%
\end{align}

\section{Appendices}

We here provide three appendices. Appendix A contains useful aspects of the
alternating $A_{4}.$ Appendix B concerns the explicit derivation of the vacuum
alignment property. Appendix C concerns properties of the tensor algebra of
flavon superfield triplets used in the computation of the scalar potential.

\subsection{Appendix A: Discrete alternating $A_{4}$}

The alternating $A_{4}$ group has 12 elements that can be generated by two
noncommuting basic ones that we denote by $S$ and $T,$ satisfying the
periodicity relations $S^{2}=I_{id}\equiv e$ and $T^{3}=I_{id}$. In terms of
these generators, we have \cite{A10}%
\begin{equation}%
\begin{tabular}
[c]{lllll}%
$a_{1}=e,$ &  & $a_{2}=S,$ &  & $a_{3}=TST^{2},$\\
$a_{4}=T^{2}ST,$ &  & $b_{1}=T,$ &  & $b_{2}=ST,$\\
$b_{3}=TS,$ &  & $b_{4}=STS,$ &  & $c_{1}=T^{2},$\\
$c_{2}=ST^{2},$ &  & $c_{3}=TST,$ &  & $c_{4}=T^{2}S.$%
\end{tabular}
\end{equation}
This discrete group has four irreducible representations; three of them have
one dimension, while the nontrivial fourth one has three dimensions. A
realization of these elements in terms of 3$\times$3 matrices is given by%
\begin{align}
&
\begin{array}
[c]{ccccc}%
{\small a}_{1}{\small =}\left(
\begin{array}
[c]{ccc}%
{\small 1} & {\small 0} & {\small 0}\\
{\small 0} & {\small 1} & {\small 0}\\
{\small 0} & {\small 0} & {\small 1}%
\end{array}
\right)  {\small ,} &  & {\small a}_{2}{\small =}\left(
\begin{array}
[c]{ccc}%
{\small 1} & {\small 0} & {\small 0}\\
{\small 0} & {\small -1} & {\small 0}\\
{\small 0} & {\small 0} & {\small -1}%
\end{array}
\right)  {\small ,} &  & {\small a}_{3}{\small =}\left(
\begin{array}
[c]{ccc}%
{\small -1} & {\small 0} & {\small 0}\\
{\small 0} & {\small 1} & {\small 0}\\
{\small 0} & {\small 0} & {\small -1}%
\end{array}
\right)  ,
\end{array}
\nonumber \\
&
\begin{array}
[c]{ccccc}%
{\small a}_{4}{\small =}\left(
\begin{array}
[c]{ccc}%
{\small -1} & {\small 0} & {\small 0}\\
{\small 0} & {\small -1} & {\small 0}\\
{\small 0} & {\small 0} & {\small 1}%
\end{array}
\right)  {\small ,} &  & {\small b}_{1}{\small =}\left(
\begin{array}
[c]{ccc}%
{\small 0} & {\small 0} & {\small 1}\\
{\small 1} & {\small 0} & {\small 0}\\
{\small 0} & {\small 1} & {\small 0}%
\end{array}
\right)  {\small ,} &  & {\small b}_{2}{\small =}\left(
\begin{array}
[c]{ccc}%
{\small 0} & {\small 0} & {\small 1}\\
{\small -1} & {\small 0} & {\small 0}\\
{\small 0} & {\small -1} & {\small 0}%
\end{array}
\right)  ,
\end{array}
\label{m0}%
\end{align}
and%
\begin{align}
&
\begin{array}
[c]{ccccc}%
{\small b}_{3}{\small =}\left(
\begin{array}
[c]{ccc}%
{\small 0} & {\small 0} & {\small -1}\\
{\small 1} & {\small 0} & {\small 0}\\
{\small 0} & {\small -1} & {\small 0}%
\end{array}
\right)  {\small ,} &  & {\small b}_{4}{\small =}\left(
\begin{array}
[c]{ccc}%
{\small 0} & {\small 0} & {\small -1}\\
{\small -1} & {\small 0} & {\small 0}\\
{\small 0} & {\small 1} & {\small 0}%
\end{array}
\right)  {\small ,} &  & {\small c}_{1}{\small =}\left(
\begin{array}
[c]{ccc}%
{\small 0} & {\small 1} & {\small 0}\\
{\small 0} & {\small 0} & {\small 1}\\
{\small 1} & {\small 0} & {\small 0}%
\end{array}
\right)  ,
\end{array}
\nonumber \\
&
\begin{array}
[c]{ccccc}%
{\small c}_{2}{\small =}\left(
\begin{array}
[c]{ccc}%
{\small 0} & {\small 1} & {\small 0}\\
{\small 0} & {\small 0} & {\small -1}\\
{\small -1} & {\small 0} & {\small 0}%
\end{array}
\right)  {\small ,} &  & {\small c}_{3}{\small =}\left(
\begin{array}
[c]{ccc}%
{\small 0} & {\small -1} & {\small 0}\\
{\small 0} & {\small 0} & {\small 1}\\
{\small -1} & {\small 0} & {\small 0}%
\end{array}
\right)  {\small ,} &  & {\small c}_{4}{\small =}\left(
\begin{array}
[c]{ccc}%
{\small 0} & {\small -1} & {\small 0}\\
{\small 0} & {\small 0} & {\small -1}\\
{\small 1} & {\small 0} & {\small 0}%
\end{array}
\right)  .
\end{array}
\label{m00}%
\end{align}%
\[
\]
Recall that $A_{4}$ is a subgroup of the symmetric $S_{4}$ consisting of only
even permutations; a canonical representation of $A_{4}$ elements is naturally
obtained by considering $4\times4$ matrices acting on four elements $x_{i}$
and we choose the generators as $S=\left(  12\right)  (34)$, $T=(123)\left(
4\right)  ,$ with matrix representations as follows:%
\begin{equation}
\left(
\begin{array}
[c]{cccc}%
0 & 1 & 0 & 0\\
1 & 0 & 0 & 0\\
0 & 0 & 0 & 1\\
0 & 0 & 1 & 0
\end{array}
\right)  \left(
\begin{array}
[c]{c}%
x_{1}\\
x_{2}\\
x_{3}\\
x_{4}%
\end{array}
\right)  =\left(
\begin{array}
[c]{c}%
x_{2}\\
x_{1}\\
x_{4}\\
x_{3}%
\end{array}
\right)  ,\qquad \left(
\begin{array}
[c]{cccc}%
0 & 1 & 0 & 0\\
0 & 0 & 1 & 0\\
1 & 0 & 0 & 0\\
0 & 0 & 0 & 1
\end{array}
\right)  \left(
\begin{array}
[c]{c}%
x_{1}\\
x_{2}\\
x_{3}\\
x_{4}%
\end{array}
\right)  =\left(
\begin{array}
[c]{c}%
x_{2}\\
x_{3}\\
x_{1}\\
x_{4}%
\end{array}
\right)  .
\end{equation}
Recall also that the discrete group $A_{4}$ has four irreducible
representations $\boldsymbol{R}_{i}$ with properties encoded in the
orthogonality character relations; in particular, in the formula
$12=1^{2}+1^{2}+1^{2}+3^{2}$. It also has four conjugacy classes
$\mathcal{C}_{i}$ given by%
\begin{equation}%
\begin{tabular}
[c]{lll}%
$\mathcal{C}_{1}$ & $=$ & $\left \{  e\right \}  ,$\\
$\mathcal{C}_{3}$ & $=$ & $\left \{  S,TST^{2},T^{2}ST\right \}  ,$\\
$\mathcal{C}_{4}$ & $=$ & $\left \{  T,TS,ST,STS\right \}  ,$\\
$\mathcal{C}_{4^{\prime}}$ & $=$ & $\left \{  T^{2},ST^{2},T^{2}S,TST\right \}
,$%
\end{tabular}
\  \label{c4}%
\end{equation}
and it is used in building the character table $\chi_{ij}$ which reads as
follows:%
\begin{equation}%
\begin{tabular}
[c]{|l|l|l|l|l|}\hline
$\chi_{ij}\left(  A_{4}\right)  $ & $\boldsymbol{R}_{1}$ & $\boldsymbol{R}%
_{1^{\prime}}$ & $\boldsymbol{R}_{1^{\prime \prime}}$ & $\boldsymbol{R}_{3}%
$\\ \hline
$\mathcal{C}_{1}$ & $1$ & $1$ & $1$ & $3$\\ \hline
$\mathcal{C}_{2}$ & $1$ & $1$ & $1$ & $-1$\\ \hline
$\mathcal{C}_{3}$ & $1$ & $\omega$ & $\omega^{2}$ & $0$\\ \hline
$\mathcal{C}_{4}$ & $1$ & $\omega^{2}$ & $\omega$ & $0$\\ \hline
\end{tabular}
\  \label{c5}%
\end{equation}

\subsection{Appendix B: Vacuum alignment}

The scalar potential (\ref{fp}) is derived from the usual $F$, $D$ and
$\mathrm{soft}$ terms of the supersymmetric minimal standard model and its
extensions. The F terms are given by%
\begin{equation}%
\begin{tabular}
[c]{lll}%
$\left \vert F_{u}\right \vert ^{2}$ & $=$ & $\left \vert \mu H_{d}+\lambda
_{u}\Delta_{u}H_{u}+h_{\zeta}\Phi H_{d}\right \vert ^{2},$\\
$\left \vert F_{d}\right \vert ^{2}$ & $=$ & $\left \vert \mu H_{u}+\lambda
_{d}\Delta_{d}H_{d}+h_{\zeta}H_{u}\Phi \right \vert ^{2},$\\
$\left \vert F_{\Delta_{u}}\right \vert ^{2}$ & $=$ & $\left \vert \mu_{\Delta
}\Delta_{d}+\lambda_{u}H_{u}H_{u}+\delta_{\zeta}\Phi \Delta_{d}\right \vert
^{2},$\\
$\left \vert F_{\Delta_{d}}\right \vert ^{2}$ & $=$ & $\left \vert \mu_{\Delta
}\Delta_{u}+\lambda_{d}H_{d}H_{d}+\delta_{\zeta}\Delta_{u}\Phi \right \vert
^{2},$\\
$\left \vert F_{\chi}\right \vert ^{2}$ & $=$ & $\left \vert 3\lambda \chi
^{2}\right \vert ^{2},$\\
$\left \vert F_{\chi^{\prime}}\right \vert ^{2}$ & $=$ & $\left \vert 2\mu_{\chi
}\chi^{\prime}+2\lambda_{\zeta \chi}\chi^{\prime}\Phi+3\lambda^{\prime}%
\chi^{\prime2}\right \vert ^{2},$\\
$\left \vert F_{\Phi}\right \vert ^{2}$ & $=$ & $\left \vert h_{\zeta}H_{u}%
H_{d}+\delta_{\zeta}Tr(\Delta_{u}\Delta_{d})+2\mu_{\zeta}\Phi+k_{\zeta
}+\lambda_{\zeta \chi}\chi^{\prime^{2}}+3\lambda_{\zeta}\Phi^{2}\right \vert
^{2}.$%
\end{tabular}
\  \label{B1}%
\end{equation}
The D terms are
\begin{align*}
D^{2}  &  =\frac{g_{1}^{2}}{2}\left[  \frac{1}{2}\left(  H_{u}^{\dag}%
H_{u}-H_{d}^{\dag}H_{d}\right)  +Tr\left(  \Delta_{d}^{\dag}\Delta_{d}\right)
-Tr\left(  \Delta_{u}^{\dag}\Delta_{u}\right)  \right]  ^{2},\\
\vec{D}^{2}  &  =\frac{g_{2}^{2}}{2}\sum_{a=1}^{3}\left[  \frac{1}{2}\left(
H_{u}^{\dag}\sigma^{a}H_{u}+H_{d}^{\dag}\sigma^{a}H_{d}\right)  +\frac{1}%
{2}Tr\left(  \Delta_{d}^{\dag}[\sigma^{a},\Delta_{d}]\right)  +\frac{1}%
{2}Tr\left(  \Delta_{u}^{\dag}[\sigma^{a},\Delta_{u}]\right)  \right]  ^{2},
\end{align*}
and for the soft terms we have%
\begin{equation}%
\begin{tabular}
[c]{lll}%
$V_{\mathrm{soft}}$ & $=$ & $m_{H_{d}}^{2}\left \vert H_{d}\right \vert
^{2}+m_{H_{u}}^{2}\left \vert H_{u}\right \vert ^{2}+m_{\Delta_{d}}%
^{2}\left \vert \Delta_{d}\right \vert ^{2}+m_{\Delta_{u}}^{2}\left \vert
\Delta_{u}\right \vert ^{2}+$\\
&  & $m_{\chi}^{2}\left \vert \chi \right \vert ^{2}+m_{\chi^{\prime}}%
^{2}\left \vert \mathcal{\chi}\mathbf{^{\prime}}\right \vert ^{2}+m_{\zeta}%
^{2}\left \vert \Phi \right \vert ^{2}+\left(  b_{H}H_{u}H_{d}+\mathrm{H.c.}%
\right)  +$\\
&  & $\left(  b_{\Delta}Tr(\Delta_{u}\Delta_{d})+\mathrm{H.c.}\right)
+\left(  b_{\chi^{\prime}}\mathcal{\chi}\mathbf{^{\prime}}^{2}+h.c\right)
+\left(  b_{\zeta}\Phi^{2}+\mathrm{H.c.}\right)  +$\\
&  & $\left[  (A_{u}H_{u}\Delta_{u}H_{u}+A_{d}H_{d}\Delta_{d}H_{d}+A_{H\zeta
}H_{u}\Phi H_{d})+\mathrm{H.c.}\right]  +$\\
&  & $\left(  A_{\Delta \zeta}\Phi Tr(\Delta_{u}\Delta_{d})+\mathrm{H.c.}%
\right)  +$\\
&  & $\left(  A_{\zeta \chi^{\prime}}\mathcal{\chi}\mathbf{^{\prime}}^{2}%
\Phi+A_{\chi}\chi^{3}+A_{\chi^{\prime}}\mathcal{\chi}\mathbf{^{\prime}}%
^{3}+A_{\zeta}\Phi^{3}+\mathrm{H.c.}\right)  .$%
\end{tabular}
\  \label{B3}%
\end{equation}
To break the flavor and electroweak symmetries, we give nonzero VEVs to the
neutral fields of the Higgs doublets, the triplets, and the flavons. Focusing
on the $A_{4}$- triplets $\mathcal{\chi}$ and $\mathcal{\chi}^{\prime}$ , and
denoting by
\[
\left \langle \mathcal{\chi}\right \rangle =(\upsilon_{\chi_{1}},\upsilon
_{\chi_{2}},\upsilon_{\chi_{3}})\qquad \left \langle \mathcal{\chi
}\mathbf{^{\prime}}\right \rangle =(\upsilon_{\chi_{1}^{\prime}},\upsilon
_{\chi_{2}^{\prime}},\upsilon_{\chi_{3}^{\prime}})
\]
the VEVs solve the minimum conditions
\begin{equation}
\frac{\partial \mathcal{V}}{\partial \mathcal{\chi}_{i}}=0,\qquad \qquad
\frac{\partial \mathcal{V}}{\partial \mathcal{\chi}_{i}^{\prime}}=0 \label{mc}%
\end{equation}
with $\mathcal{V}$ as in Eq. (\ref{fp}) and the VEVs of the triplets are as in
Eqs. (\ref{va}) and (\ref{vacua}). To get these VEVs, we should take into
account all possible $A_{4}$-invariant contributions coming from the tensor
products of three and four triplets of $A_{4}$ as they appear in the
computation of $\left \vert \mathcal{\chi}\right \vert ^{4}$ and $\left \vert
\mathcal{\chi}\right \vert ^{3}$; see also Appendix C for more details. By
using the fusion operator algebra of $A_{4}$, we have for the tensor product
$\left(  3_{-1,0}\right)  ^{\otimes4}$ the following expression%
\[%
\begin{tabular}
[c]{lll}%
$\left(  3_{-1,0}\otimes3_{-1,0}\right)  ^{\otimes2}$ & $\rightarrow$ &
$\left(  1_{1,1}\otimes1_{1,1}\right)  \oplus(1_{1,\omega}\otimes
1_{1,\omega^{2}})$\\
&  & $\oplus(1_{1,\omega^{2}}\otimes1_{1,\omega})\oplus \left(  3_{-1,0}%
^{s}\otimes3_{-1,0}^{s}\right)  $\\
&  & $\oplus \left(  3_{-1,0}^{s}\otimes3_{-1,0}^{a}\right)  \oplus \left(
3_{-1,0}^{a}\otimes3_{-1,0}^{s}\right)  $\\
&  & $\oplus \left(  3_{-1,0}^{a}\otimes3_{-1,0}^{a}\right)  ,$%
\end{tabular}
\
\]
which can be reduced further. Using the method of Ref. \cite{32}, we can
approach the solution of the minimum conditions $\mathcal{V}$ for the $A_{4}$
triplet $\chi$ through the relations%
\begin{equation}%
\begin{tabular}
[c]{lll}%
$\upsilon_{\chi_{2}}\frac{\partial \mathcal{V}}{\partial \upsilon_{\chi_{1}}%
}-\upsilon_{\chi_{1}}\frac{\partial \mathcal{V}}{\partial \upsilon_{\chi_{2}}}$
& $=$ & $0,$\\
$\upsilon_{\chi_{3}}\frac{\partial \mathcal{V}}{\partial \upsilon_{\chi_{1}}%
}-\upsilon_{\chi_{1}}\frac{\partial \mathcal{V}}{\partial \upsilon_{\chi_{3}}}$
& $=$ & $0,$\\
$\upsilon_{\chi_{3}}\frac{\partial \mathcal{V}}{\partial \upsilon_{\chi_{2}}%
}-\upsilon_{\chi_{2}}\frac{\partial \mathcal{V}}{\partial \upsilon_{\chi_{3}}}$
& $=$ & $0,$%
\end{tabular}
\end{equation}
they read explicitly as%
\begin{equation}%
\begin{tabular}
[c]{lll}%
$0$ & $=$ & $36\lambda^{2}\upsilon_{\chi_{1}}\upsilon_{\chi_{2}}\left(
\upsilon_{\chi_{1}}^{2}-\upsilon_{\chi_{2}}^{2}\right)  +12A_{\chi}%
\upsilon_{\chi_{3}}\left(  \upsilon_{\chi_{2}}^{2}-\upsilon_{\chi_{1}}%
^{2}\right)  ,$\\
$0$ & $=$ & $36\lambda^{2}\upsilon_{\chi_{1}}\upsilon_{\chi_{3}}\left(
\upsilon_{\chi_{1}}^{2}-\upsilon_{\chi_{3}}^{2}\right)  +12A_{\chi}%
\upsilon_{\chi_{2}}\left(  \upsilon_{\chi_{3}}^{2}-\upsilon_{\chi_{1}}%
^{2}\right)  ,$\\
$0$ & $=$ & $36\lambda^{2}\upsilon_{\chi_{2}}\upsilon_{\chi_{3}}\left(
\upsilon_{\chi_{2}}^{2}-\upsilon_{\chi_{3}}^{2}\right)  +12A_{\chi}%
\upsilon_{\chi_{1}}\left(  \upsilon_{\chi_{3}}^{2}-\upsilon_{\chi_{2}}%
^{2}\right)  .$%
\end{tabular}
\end{equation}
Clearly, the solution for the last three equations is given by
\begin{equation}
\upsilon_{\chi_{1}}=\upsilon_{\chi_{2}}=\upsilon_{\chi_{3}}=\upsilon_{\chi}%
\end{equation}
It is precisely the VEV structure we choose in Eq. (\ref{vacua}) to produce
the TBM matrix pattern. The same method applies for the minimum conditions
coming from the triplet $\mathcal{\chi}$\textbf{$^{\prime}$}; we have
\[%
\begin{tabular}
[c]{lll}%
$\upsilon_{\chi_{2}^{\prime}}\frac{\partial \mathcal{V}}{\partial \upsilon
_{\chi_{1}^{\prime}}}-\upsilon_{\chi_{1}^{\prime}}\frac{\partial \mathcal{V}%
}{\partial \upsilon_{\chi_{2}^{\prime}}}$ & $=$ & $\mathrm{0,}$\\
$\upsilon_{\chi_{3}^{\prime}}\frac{\partial \mathcal{V}}{\partial
\upsilon_{\mathbf{\chi}_{1}^{\prime}}}-\upsilon_{\chi_{1}^{\prime}}%
\frac{\partial \mathcal{V}}{\partial \upsilon_{\chi_{3}^{\prime}}}$ & $=$ &
$\mathrm{0,}$\\
$\upsilon_{\chi_{3}^{\prime}}\frac{\partial \mathcal{V}}{\partial \upsilon
_{\chi_{2}^{\prime}}}-\upsilon_{\chi_{2}^{\prime}}\frac{\partial \mathcal{V}%
}{\partial \upsilon_{\chi_{3}^{\prime}}}$ & $=$ & $\mathrm{0.}$%
\end{tabular}
\
\]
Explicitly,%
\begin{equation}%
\begin{tabular}
[c]{lll}%
$0$ & $=$ & $36\lambda^{\prime2}\upsilon_{\chi_{1}^{\prime}}\upsilon_{\chi
_{2}^{\prime}}\left(  \upsilon_{\chi_{1}^{\prime}}^{2}-\upsilon_{\chi
_{2}^{\prime}}^{2}\right)  +72\lambda^{\prime}\upsilon_{\chi_{3}^{\prime}%
}\left(  \upsilon_{\chi_{2}^{\prime}}^{2}-\upsilon_{\chi_{1}^{\prime}}%
^{2}\right)  \left(  \mu_{\chi}+\lambda_{\zeta \chi}\upsilon_{\Phi}\right)  $\\
&  & $+4\lambda_{\zeta \chi}^{2}\upsilon_{\chi_{1}^{\prime}}\upsilon_{\chi
_{2}^{\prime}}\left(  \upsilon_{\chi_{1}^{\prime}}^{2}-\upsilon_{\chi
_{2}^{\prime}}^{2}\right)  +12A_{\chi^{\prime}}\upsilon_{\chi_{3}^{\prime}%
}\left(  \upsilon_{\chi_{2}^{\prime}}^{2}-\upsilon_{\chi_{1}^{\prime}}%
^{2}\right)  $%
\end{tabular}
\end{equation}
and%
\[%
\begin{array}
[c]{ccc}%
0 & = & 36\lambda^{\prime2}\upsilon_{\mathbf{\chi}_{1}^{\prime}}%
\upsilon_{\mathbf{\chi}_{3}^{\prime}}\left(  \upsilon_{\mathbf{\chi}%
_{1}^{\prime}}^{2}-\upsilon_{\mathbf{\chi}_{3}^{\prime}}^{2}\right)
+72\lambda^{\prime}\upsilon_{\mathbf{\chi}_{2}^{\prime}}\left(  \upsilon
_{\mathbf{\chi}_{3}^{\prime}}^{2}-\upsilon_{\mathbf{\chi}_{1}^{\prime}}%
^{2}\right)  \left(  \mu_{\chi}+\lambda_{\zeta \chi}\upsilon_{\Phi}\right) \\
&  & +4\lambda_{\zeta \chi}^{2}\upsilon_{\mathbf{\chi}_{1}^{\prime}}%
\upsilon_{\mathbf{\chi}_{3}^{\prime}}\left(  \upsilon_{\mathbf{\chi}%
_{1}^{\prime}}^{2}-\upsilon_{\mathbf{\chi}_{3}^{\prime}}^{2}\right)
+12A_{\chi^{\prime}}\upsilon_{\mathbf{\chi}_{2}^{\prime}}\left(
\upsilon_{\mathbf{\chi}_{3}^{\prime}}^{2}-\upsilon_{\mathbf{\chi}_{1}^{\prime
}}^{2}\right)  ,\qquad \qquad \quad
\end{array}
\]
as well as%
\[%
\begin{array}
[c]{ccc}%
0 & = & 36\lambda^{\prime2}\upsilon_{\mathbf{\chi}_{2}^{\prime}}%
\upsilon_{\mathbf{\chi}_{3}^{\prime}}\left(  \upsilon_{\mathbf{\chi}%
_{2}^{\prime}}^{2}-\upsilon_{\mathbf{\chi}_{3}^{\prime}}^{2}\right)
+72\lambda^{\prime}\upsilon_{\mathbf{\chi}_{1}^{\prime}}\left(  \upsilon
_{\mathbf{\chi}_{3}^{\prime}}^{2}-\upsilon_{\mathbf{\chi}_{2}^{\prime}}%
^{2}\right)  \left(  \mu_{\chi}+\lambda_{\zeta \chi}\upsilon_{\Phi}\right) \\
&  & +4\lambda_{\zeta \chi}^{2}\upsilon_{\mathbf{\chi}_{2}^{\prime}}%
\upsilon_{\mathbf{\chi}_{3}^{\prime}}\left(  \upsilon_{\mathbf{\chi}%
_{2}^{\prime}}^{2}-\upsilon_{\mathbf{\chi}_{3}^{\prime}}^{2}\right)
+12A_{\chi^{\prime}}\upsilon_{\mathbf{\chi}_{1}^{\prime}}\left(
\upsilon_{\mathbf{\chi}_{3}^{\prime}}^{2}-\upsilon_{\mathbf{\chi}_{2}^{\prime
}}^{2}\right)  .\qquad \qquad \quad
\end{array}
\]
These equations have three solutions: we choose one to produce the neutrino
mass matrix $\left \langle \mathcal{\chi}\mathbf{^{\prime}}\right \rangle
=(\upsilon_{\chi_{1}^{\prime}},0,0),$ and the other two possibilities are
$\left \langle \mathcal{\chi}\mathbf{^{\prime}}\right \rangle =(0,\upsilon
_{\chi_{2}^{\prime}},0)$ and $\left \langle \mathcal{\chi}\mathbf{^{\prime}%
}\right \rangle =(0,0,\upsilon_{\chi_{3}^{\prime}})$.\newline

\subsection{Appendix C: Tensor product of $A_{4}$ triplets}

Here we give useful tools for the computation of the tensor product of $A_{4}$
triplets. For the case of two $A_{4}$ triplets taken as $\mathbf{a}%
=(a_{1},a_{2},a_{3})$ and $\mathbf{b}=(b_{1},b_{2},b_{3})$, their tensor
product is reducible with irreducible components given by the following
decomposition relation:%
\begin{equation}
3\otimes3=1\oplus1^{\prime}\oplus1^{\prime \prime}\oplus3_{S}\oplus3_{A}.
\end{equation}
Expressing this product as%
\begin{equation}
\mathbf{a\otimes b}=\oplus_{i}\left(  \left.  \left(  \mathbf{a\otimes
b}\right)  \right \vert _{R_{i}}\right)  ,
\end{equation}
the irreducible components are given by%
\begin{align}
\left.  \left(  \mathbf{a\otimes b}\right)  \right \vert _{1}  &  =a_{1}%
b_{1}+a_{2}b_{2}+a_{3}b_{3},\nonumber \\
\left.  \left(  \mathbf{a\otimes b}\right)  \right \vert _{1^{\prime}}  &
=a_{1}b_{1}+\omega a_{2}b_{2}+\omega^{2}a_{3}b_{3},\nonumber \\
\left.  \left(  \mathbf{a\otimes b}\right)  \right \vert _{1^{\prime \prime}}
&  =a_{1}b_{1}+\omega^{2}a_{2}b_{2}+\omega a_{3}b_{3},\label{tp}\\
\left.  \left(  \mathbf{a\otimes b}\right)  \right \vert _{3_{S}}  &
=(a_{2}b_{3}+a_{3}b_{2},a_{3}b_{1}+a_{1}b_{3},a_{1}b_{2}+a_{2}b_{1}%
),\nonumber \\
\left.  \left(  \mathbf{a\otimes b}\right)  \right \vert _{3_{A}}  &
=(a_{2}b_{3}-a_{3}b_{2},a_{3}b_{1}-a_{1}b_{3},a_{1}b_{2}-a_{2}b_{1}).\nonumber
\end{align}
As an application, we present all possible $A_{4}$-invariant terms for the
monomials $\mathcal{\chi}^{2}$, $\mathcal{\chi}^{3},$ and $\mathcal{\chi}^{4}$
which we encounter in the scalar potential (\ref{fp}) by using Eq. (\ref{tp}).
For the case $\mathcal{\chi}^{2}$, the previous $\mathbf{a}$ and $\mathbf{b}$
are identical, so we have
\begin{equation}%
\begin{array}
[c]{ccc}%
\left.  \left(  \mathcal{\chi \otimes \chi}\right)  \right \vert _{1} & = &
\mathcal{\chi}_{1}^{2}+\mathcal{\chi}_{2}^{2}+\mathcal{\chi}_{3}^{2}.
\end{array}
\label{m1}%
\end{equation}
The other $\left.  \left(  \mathcal{\chi \otimes \chi}\right)  \right \vert
_{R_{i}}$ are directly obtained from Eq. (\ref{tp}). For \textbf{$\chi$}$^{3}%
$, we have for the example of $\left.  \left(  \mathcal{\chi \otimes \chi
\otimes \chi}\right)  \right \vert _{1}$ the following expression:%
\begin{equation}%
\begin{array}
[c]{ccc}%
\left.  \left(  \mathcal{\chi \otimes \chi \otimes \chi}\right)  \right \vert _{1}
& = & \left.  \left[  \left(
\begin{array}
[c]{c}%
\mathcal{\chi}_{1}\\
\mathcal{\chi}_{2}\\
\mathcal{\chi}_{3}%
\end{array}
\right)  \otimes \left(
\begin{array}
[c]{c}%
\mathcal{\chi}_{1}\\
\mathcal{\chi}_{2}\\
\mathcal{\chi}_{3}%
\end{array}
\right)  \right]  _{3}\otimes \left(
\begin{array}
[c]{c}%
\mathcal{\chi}_{1}\\
\mathcal{\chi}_{2}\\
\mathcal{\chi}_{3}%
\end{array}
\right)  _{3}\right \vert _{1}\\
&  & \\
& = & \left.  \left[  \left(
\begin{array}
[c]{c}%
2\mathcal{\chi}_{2}\mathcal{\chi}_{3}\\
2\mathcal{\chi}_{1}\mathcal{\chi}_{3}\\
2\mathcal{\chi}_{1}\mathcal{\chi}_{2}%
\end{array}
\right)  _{S}+\left(
\begin{array}
[c]{c}%
0\\
0\\
0
\end{array}
\right)  _{A}\right]  \otimes \left(
\begin{array}
[c]{c}%
\mathcal{\chi}_{1}\\
\mathcal{\chi}_{2}\\
\mathcal{\chi}_{3}%
\end{array}
\right)  ,\right \vert _{1}%
\end{array}
\end{equation}
leading to%
\begin{equation}
\left.  \left(  \mathcal{\chi \otimes \chi \otimes \chi}\right)  \right \vert
_{1}=6\mathcal{\chi}_{1}\mathcal{\chi}_{2}\mathcal{\chi}_{3}. \label{m2}%
\end{equation}
Similar expressions can be written down for the other $\left.  \left(
\mathcal{\chi \otimes \chi \otimes \chi}\right)  \right \vert _{R_{i}}$; they are
not relevant for our study. To determine $\left.  \left(  \mathcal{\chi
\otimes \chi \otimes \chi \otimes \chi}\right)  \right \vert _{1}$, we start from
\begin{equation}%
\begin{array}
[c]{ccc}%
\left.  \left(  \mathcal{\chi \otimes \chi \otimes \chi \otimes \chi}\right)
\right \vert _{1} & = & \left.  \left[  \left(
\begin{array}
[c]{c}%
\mathcal{\chi}_{1}\\
\mathcal{\chi}_{2}\\
\mathcal{\chi}_{3}%
\end{array}
\right)  \otimes \left(
\begin{array}
[c]{c}%
\mathcal{\chi}_{1}\\
\mathcal{\chi}_{2}\\
\mathcal{\chi}_{3}%
\end{array}
\right)  \right]  \otimes \left[  \left(
\begin{array}
[c]{c}%
\mathcal{\chi}_{1}\\
\mathcal{\chi}_{2}\\
\mathcal{\chi}_{3}%
\end{array}
\right)  \otimes \left(
\begin{array}
[c]{c}%
\mathcal{\chi}_{1}\\
\mathcal{\chi}_{2}\\
\mathcal{\chi}_{3}%
\end{array}
\right)  .\right]  \right \vert _{1}%
\end{array}
\end{equation}
Then, using%
\begin{equation}
\left.  \left(  3\otimes3\otimes3\otimes3\right)  \right \vert _{1}=\left.
\left[  1\oplus1^{\prime}\oplus1^{\prime \prime}\oplus3_{S}\oplus3_{A}\right]
\otimes \left[  1\oplus1^{\prime}\oplus1^{\prime \prime}\oplus3_{S}\oplus
3_{A}\right]  \right \vert _{1}%
\end{equation}
and by setting
\begin{align}
1\times1  &  =X,\nonumber \\
1^{\prime}\times1^{\prime \prime}  &  =Y,\\
1^{\prime \prime}\times1^{\prime}  &  =Z,\nonumber
\end{align}
we have%
\begin{align}
X  &  =\left[  \left(  \mathcal{\chi}_{1}\right)  ^{2}+\left(  \mathcal{\chi
}_{2}\right)  ^{2}+\left(  \mathcal{\chi}_{3}\right)  ^{2}\right]  _{1}%
\times \left[  \left(  \mathcal{\chi}_{1}\right)  ^{2}+\left(  \mathcal{\chi
}_{2}\right)  ^{2}+\left(  \mathcal{\chi}_{3}\right)  ^{2}\right]
_{1},\nonumber \\
Y  &  =\left[  \left(  \mathcal{\chi}_{1}\right)  ^{2}+\omega \left(
\mathcal{\chi}_{2}\right)  ^{2}+\omega^{2}\left(  \mathcal{\chi}_{3}\right)
^{2}\right]  _{1^{\prime}}\times \left[  \left(  \mathcal{\chi}_{1}\right)
^{2}+\omega^{2}\left(  \mathcal{\chi}_{2}\right)  ^{2}+\omega \left(
\mathcal{\chi}_{3}\right)  ^{2}\right]  _{1^{\prime \prime}},\\
Z  &  =\left[  \left(  \mathcal{\chi}_{1}\right)  ^{2}+\omega^{2}\left(
\mathcal{\chi}_{2}\right)  ^{2}+\omega \left(  \mathcal{\chi}_{3}\right)
^{2}\right]  _{1^{\prime \prime}}\times \left[  \left(  \mathcal{\chi}%
_{1}\right)  ^{2}+\omega \left(  \mathcal{\chi}_{2}\right)  ^{2}+\omega
^{2}\left(  \mathcal{\chi}_{3}\right)  ^{2}\right]  _{1^{\prime}}.\nonumber
\end{align}
We also have%
\begin{equation}%
\begin{array}
[c]{cccc}%
3_{S}\times3_{S}= & \left(
\begin{array}
[c]{c}%
2\mathcal{\chi}_{2}\mathcal{\chi}_{3}\\
2\mathcal{\chi}_{1}\mathcal{\chi}_{3}\\
2\mathcal{\chi}_{1}\mathcal{\chi}_{2}%
\end{array}
\right)  _{S} & \times & \left(
\begin{array}
[c]{c}%
2\mathcal{\chi}_{2}\mathcal{\chi}_{3}\\
2\mathcal{\chi}_{1}\mathcal{\chi}_{3}\\
2\mathcal{\chi}_{1}\mathcal{\chi}_{2}%
\end{array}
\right)  _{S},\\
&  &  & \\
3_{S}\times3_{A}= & \left(
\begin{array}
[c]{c}%
2\mathcal{\chi}_{2}\mathcal{\chi}_{3}\\
2\mathcal{\chi}_{1}\mathcal{\chi}_{3}\\
2\mathcal{\chi}_{1}\mathcal{\chi}_{2}%
\end{array}
\right)  _{S} & \times & \left(
\begin{array}
[c]{c}%
0\\
0\\
0
\end{array}
\right)  _{A},\\
&  &  & \\
3_{A}\times3_{A}= & \left(
\begin{array}
[c]{c}%
0\\
0\\
0
\end{array}
\right)  _{A} & \times & \left(
\begin{array}
[c]{c}%
0\\
0\\
0
\end{array}
\right)  _{A},\\
&  &  & \\
3_{A}\times3_{S}= & \left(
\begin{array}
[c]{c}%
0\\
0\\
0
\end{array}
\right)  _{A} & \times & \left(
\begin{array}
[c]{c}%
2\mathcal{\chi}_{2}\mathcal{\chi}_{3}\\
2\mathcal{\chi}_{1}\mathcal{\chi}_{3}\\
2\mathcal{\chi}_{1}\mathcal{\chi}_{2}%
\end{array}
\right)  _{S}.
\end{array}
\end{equation}
We end with%
\begin{equation}
\left.  \mathcal{\chi}^{4}\right \vert _{1}=3\left[  \left(  \mathcal{\chi}%
_{1}\right)  ^{4}+\left(  \mathcal{\chi}_{2}\right)  ^{4}+\left(
\mathcal{\chi}_{3}\right)  ^{4}\right]  +4\left[  \left(  \mathcal{\chi}%
_{1}\right)  ^{2}\left(  \mathcal{\chi}_{2}\right)  ^{2}+\left(
\mathcal{\chi}_{1}\right)  ^{2}\left(  \mathcal{\chi}_{3}\right)  ^{2}+\left(
\mathcal{\chi}_{2}\right)  ^{2}\left(  \mathcal{\chi}_{3}\right)  ^{2}\right]
. \label{m3}%
\end{equation}
Analogously, the exact calculations for the triplet $\mathcal{\chi}%
$\textbf{$^{\prime}$} lead to%
\begin{align}
\left.  \mathcal{\chi}\mathbf{^{\prime}}^{2}\right \vert _{1}  &
=\mathcal{\chi}_{1}^{\prime2}+\mathcal{\chi}_{2}^{\prime2}+\mathcal{\chi}%
_{3}^{\prime2},\nonumber \\
\left.  \mathcal{\chi}\mathbf{^{\prime}}^{3}\right \vert _{1}  &
=6\mathcal{\chi}_{1}^{\prime}\mathcal{\chi}_{2}^{\prime}\mathcal{\chi}%
_{3}^{\prime},\\
\left.  \mathcal{\chi}\mathbf{^{\prime}}^{4}\right \vert _{1}  &  =3\left[
\left(  \mathcal{\chi}_{1}^{\prime}\right)  ^{4}+\left(  \mathcal{\chi}%
_{2}^{\prime}\right)  ^{4}+\left(  \mathcal{\chi}_{3}^{\prime}\right)
^{4}\right]  +4\left[  \left(  \mathcal{\chi}_{1}^{\prime}\right)  ^{2}\left(
\mathcal{\chi}_{2}^{\prime}\right)  ^{2}+\left(  \mathcal{\chi}_{1}^{\prime
}\right)  ^{2}\left(  \mathcal{\chi}_{3}^{\prime}\right)  ^{2}+\left(
\mathcal{\chi}_{2}^{\prime}\right)  ^{2}\left(  \mathcal{\chi}_{3}^{\prime
}\right)  ^{2}\right]  .\nonumber
\end{align}
After substituting the above results into the scalar potential (\ref{fp}), the
minimum conditions (\ref{mc}) are as follows:%
\begin{equation}%
\begin{array}
[c]{ccc}%
\left.  \frac{\partial \mathcal{V}}{\partial \mathcal{\chi}_{1}}\right \vert
_{\left \langle \mathcal{\chi}_{i}\right \rangle =\upsilon_{\chi_{i}}} & = &
0,\\
\left.  \frac{\partial \mathcal{V}}{\partial \mathcal{\chi}_{2}}\right \vert
_{\left \langle \mathcal{\chi}_{i}\right \rangle =\upsilon_{\chi_{i}}} & = &
0,\\
\left.  \frac{\partial \mathcal{V}}{\partial \mathcal{\chi}_{3}}\right \vert
_{\left \langle \mathcal{\chi}_{i}\right \rangle =\upsilon_{\chi_{i}}} & = & 0,
\end{array}
\end{equation}
leading to%
\begin{equation}%
\begin{array}
[c]{ccc}%
108\lambda^{2}\upsilon_{\mathbf{\chi}_{1}}^{3}+72\lambda^{2}\upsilon
_{\mathbf{\chi}_{1}}\upsilon_{\mathbf{\chi}_{2}}^{2}+72\lambda^{2}%
\upsilon_{\mathbf{\chi}_{1}}\upsilon_{\mathbf{\chi}_{3}}^{2}+2m_{\chi}%
^{2}\upsilon_{\mathbf{\chi}_{1}}+12A_{\chi}\upsilon_{\mathbf{\chi}_{2}%
}\upsilon_{\mathbf{\chi}_{3}} & = & 0,\\
108\lambda^{2}\upsilon_{\mathbf{\chi}_{2}}^{3}+72\lambda^{2}\upsilon
_{\mathbf{\chi}_{2}}\upsilon_{\mathbf{\chi}_{1}}^{2}+72\lambda^{2}%
\upsilon_{\mathbf{\chi}_{2}}\upsilon_{\mathbf{\chi}_{3}}^{2}+2m_{\chi}%
^{2}\upsilon_{\mathbf{\chi}_{2}}+12A_{\chi}\upsilon_{\mathbf{\chi}_{1}%
}\upsilon_{\mathbf{\chi}_{3}} & = & 0,\\
108\lambda^{2}\upsilon_{\mathbf{\chi}_{3}}^{3}+72\lambda^{2}\upsilon
_{\mathbf{\chi}_{3}}\upsilon_{\mathbf{\chi}_{1}}^{2}+72\lambda^{2}%
\upsilon_{\mathbf{\chi}_{3}}\upsilon_{\mathbf{\chi}_{2}}^{2}+2m_{\chi}%
^{2}\upsilon_{\mathbf{\chi}_{3}}+12A_{\chi}\upsilon_{\mathbf{\chi}_{1}%
}\upsilon_{\mathbf{\chi}_{2}} & = & 0.
\end{array}
\end{equation}
We also have%
\begin{equation}%
\begin{array}
[c]{ccc}%
\left.  \frac{\partial \mathcal{V}}{\partial \mathcal{\chi}_{1}^{\prime}%
}\right \vert _{\left \langle \mathcal{\chi}_{i}^{\prime}\right \rangle
=\upsilon_{\chi_{i}^{\prime}}} & = & 0,\\
\left.  \frac{\partial \mathcal{V}}{\partial \mathcal{\chi}_{2}^{\prime}%
}\right \vert _{\left \langle \mathcal{\chi}_{i}^{\prime}\right \rangle
=\upsilon_{\chi_{i}^{\prime}}} & = & 0,\\
\left.  \frac{\partial \mathcal{V}}{\partial \mathcal{\chi}_{3}^{\prime}%
}\right \vert _{\left \langle \mathcal{\chi}_{i}^{\prime}\right \rangle
=\upsilon_{\chi_{i}^{\prime}}} & = & 0,
\end{array}
\end{equation}
giving%
\begin{align}
0  &  =8\left \vert \mu_{\chi}\right \vert ^{2}\upsilon_{\mathbf{\chi}%
_{1}^{\prime}}+8\lambda^{\prime2}\upsilon_{\mathbf{\chi}_{1}^{\prime}}%
\upsilon_{\Phi}^{2}+108\lambda^{\prime2}\upsilon_{\mathbf{\chi}_{1}^{\prime}%
}^{3}+72\lambda^{\prime2}\upsilon_{\mathbf{\chi}_{1}^{\prime}}\upsilon
_{\mathbf{\chi}_{2}^{\prime}}^{2}+72\lambda^{\prime2}\upsilon_{\mathbf{\chi
}_{1}^{\prime}}\upsilon_{\mathbf{\chi}_{3}^{\prime}}^{2}+16\mu_{\chi}%
\lambda_{\zeta \chi}\upsilon_{\mathbf{\chi}_{1}^{\prime}}\nonumber \\
&  +72\mu_{\chi}\lambda^{\prime}\upsilon_{\mathbf{\chi}_{2}^{\prime}}%
\upsilon_{\mathbf{\chi}_{3}^{\prime}}+72\lambda_{\zeta \chi}\lambda^{\prime
}\upsilon_{\Phi}\upsilon_{\mathbf{\chi}_{2}^{\prime}}\upsilon_{\mathbf{\chi
}_{3}^{\prime}}+12\lambda_{\zeta \chi}^{2}\upsilon_{\mathbf{\chi}_{1}^{\prime}%
}^{3}+8\lambda_{\zeta \chi}^{2}\upsilon_{\mathbf{\chi}_{1}^{\prime}}%
\upsilon_{\mathbf{\chi}_{2}^{\prime}}^{2}+8\lambda_{\zeta \chi}^{2}%
\upsilon_{\mathbf{\chi}_{1}^{\prime}}\upsilon_{\mathbf{\chi}_{3}^{\prime}}%
^{2}\nonumber \\
&  +4k_{\zeta}\lambda_{\zeta \chi}\upsilon_{\mathbf{\chi}_{1}^{\prime}%
}+12\lambda_{\zeta \chi}\lambda_{\zeta}\upsilon_{\mathbf{\chi}_{1}^{\prime}%
}\upsilon_{\Phi}^{2}+4h_{\zeta}\lambda_{\zeta \chi}\upsilon_{u}\upsilon
_{d}\upsilon_{\mathbf{\chi}_{1}^{\prime}}+4\delta_{\zeta}\lambda_{\zeta \chi
}\upsilon_{\Delta_{u}}\upsilon_{\Delta_{d}}\upsilon_{\mathbf{\chi}_{1}%
^{\prime}}\\
&  +8\mu_{\chi}\lambda_{\zeta \chi}\upsilon_{\Phi}\upsilon_{\mathbf{\chi}%
_{1}^{\prime}}+2m_{\chi^{\prime}}^{2}\upsilon_{\mathbf{\chi}_{1}^{\prime}%
}+4b_{\chi^{\prime}}\upsilon_{\mathbf{\chi}_{1}^{\prime}}+4A_{\zeta
\chi^{\prime}}\upsilon_{\Phi}\upsilon_{\mathbf{\chi}_{1}^{\prime}}%
+12A_{\chi^{\prime}}\upsilon_{\mathbf{\chi}_{3}^{\prime}}\upsilon
_{\mathbf{\chi}_{2}^{\prime}}\nonumber
\end{align}
and%
\begin{align}
0  &  =8\left \vert \mu_{\chi}\right \vert ^{2}\upsilon_{\mathbf{\chi}%
_{2}^{\prime}}+8\lambda^{\prime2}\upsilon_{\mathbf{\chi}_{2}^{\prime}}%
\upsilon_{\Phi}^{2}+108\lambda^{\prime2}\upsilon_{\mathbf{\chi}_{2}^{\prime}%
}^{3}+72\lambda^{\prime2}\upsilon_{\mathbf{\chi}_{1}^{\prime}}^{2}%
\upsilon_{\mathbf{\chi}_{2}^{\prime}}+72\lambda^{\prime2}\upsilon
_{\mathbf{\chi}_{2}^{\prime}}\upsilon_{\mathbf{\chi}_{3}^{\prime}}^{2}%
+16\mu_{\chi}\lambda_{\zeta \chi}\upsilon_{\mathbf{\chi}_{2}^{\prime}%
}\nonumber \\
&  +72\mu_{\chi}\lambda^{\prime}\upsilon_{\mathbf{\chi}_{1}^{\prime}}%
\upsilon_{\mathbf{\chi}_{3}^{\prime}}+72\lambda_{\zeta \chi}\lambda^{\prime
}\upsilon_{\Phi}\upsilon_{\mathbf{\chi}_{1}^{\prime}}\upsilon_{\mathbf{\chi
}_{3}^{\prime}}+12\lambda_{\zeta \chi}^{2}\upsilon_{\mathbf{\chi}_{2}^{\prime}%
}^{3}+8\lambda_{\zeta \chi}^{2}\upsilon_{\mathbf{\chi}_{1}^{\prime}}%
^{2}\upsilon_{\mathbf{\chi}_{2}^{\prime}}+8\lambda_{\zeta \chi}^{2}%
\upsilon_{\mathbf{\chi}_{2}^{\prime}}\upsilon_{\mathbf{\chi}_{3}^{\prime}}%
^{2}\nonumber \\
&  +4k_{\zeta}\lambda_{\zeta \chi}\upsilon_{\mathbf{\chi}_{2}^{\prime}%
}+12\lambda_{\zeta \chi}\lambda_{\zeta}\upsilon_{\mathbf{\chi}_{2}^{\prime}%
}\upsilon_{\Phi}^{2}+4h_{\zeta}\lambda_{\zeta \chi}\upsilon_{u}\upsilon
_{d}\upsilon_{\mathbf{\chi}_{2}^{\prime}}+4\delta_{\zeta}\lambda_{\zeta \chi
}\upsilon_{\Delta_{u}}\upsilon_{\Delta_{d}}\upsilon_{\mathbf{\chi}_{2}%
^{\prime}}\\
&  +8\mu_{\chi}\lambda_{\zeta \chi}\upsilon_{\Phi}\upsilon_{\mathbf{\chi}%
_{2}^{\prime}}+2m_{\chi^{\prime}}^{2}\upsilon_{\mathbf{\chi}_{2}^{\prime}%
}+4b_{\chi^{\prime}}\upsilon_{\mathbf{\chi}_{2}^{\prime}}+4A_{\zeta
\chi^{\prime}}\upsilon_{\Phi}\upsilon_{\mathbf{\chi}_{2}^{\prime}}%
+12A_{\chi^{\prime}}\upsilon_{\mathbf{\chi}_{3}^{\prime}}\upsilon
_{\mathbf{\chi}_{1}^{\prime}},\nonumber
\end{align}
as well as
\begin{align}
0  &  =8\left \vert \mu_{\chi}\right \vert ^{2}\upsilon_{\mathbf{\chi}%
_{3}^{\prime}}+8\lambda^{\prime2}\upsilon_{\mathbf{\chi}_{3}^{\prime}}%
\upsilon_{\Phi}^{2}+108\lambda^{\prime2}\upsilon_{\mathbf{\chi}_{3}^{\prime}%
}^{3}+72\lambda^{\prime2}\upsilon_{\mathbf{\chi}_{3}^{\prime}}\upsilon
_{\mathbf{\chi}_{1}^{\prime}}^{2}+72\lambda^{\prime2}\upsilon_{\mathbf{\chi
}_{3}^{\prime}}\upsilon_{\mathbf{\chi}_{2}^{\prime}}^{2}+16\mu_{\chi}%
\lambda_{\zeta \chi}\upsilon_{\mathbf{\chi}_{3}^{\prime}}\nonumber \\
&  +72\mu_{\chi}\lambda^{\prime}\upsilon_{\mathbf{\chi}_{1}^{\prime}}%
\upsilon_{\mathbf{\chi}_{2}^{\prime}}+72\lambda_{\zeta \chi}\lambda^{\prime
}\upsilon_{\Phi}\upsilon_{\mathbf{\chi}_{1}^{\prime}}\upsilon_{\mathbf{\chi
}_{2}^{\prime}}+12\lambda_{\zeta \chi}^{2}\upsilon_{\mathbf{\chi}_{3}^{\prime}%
}^{3}+8\lambda_{\zeta \chi}^{2}\upsilon_{\mathbf{\chi}_{3}^{\prime}}%
\upsilon_{\mathbf{\chi}_{1}^{\prime}}^{2}+8\lambda_{\zeta \chi}^{2}%
\upsilon_{\mathbf{\chi}_{3}^{\prime}}\upsilon_{\mathbf{\chi}_{2}^{\prime}}%
^{2}\nonumber \\
&  +4k_{\zeta}\lambda_{\zeta \chi}\upsilon_{\mathbf{\chi}_{3}^{\prime}%
}+12\lambda_{\zeta \chi}\lambda_{\zeta}\upsilon_{\mathbf{\chi}_{3}^{\prime}%
}\upsilon_{\Phi}^{2}+4h_{\zeta}\lambda_{\zeta \chi}\upsilon_{u}\upsilon
_{d}\upsilon_{\mathbf{\chi}_{3}^{\prime}}+4\delta_{\zeta}\lambda_{\zeta \chi
}\upsilon_{\Delta_{u}}\upsilon_{\Delta_{d}}\upsilon_{\mathbf{\chi}_{3}%
^{\prime}}\\
&  +8\mu_{\chi}\lambda_{\zeta \chi}\upsilon_{\Phi}\upsilon_{\mathbf{\chi}%
_{3}^{\prime}}+2m_{\chi^{\prime}}^{2}\upsilon_{\mathbf{\chi}_{3}^{\prime}%
}+4b_{\chi^{\prime}}\upsilon_{\mathbf{\chi}_{3}^{\prime}}+4A_{\zeta
\chi^{\prime}}\upsilon_{\Phi}\upsilon_{\mathbf{\chi}_{3}^{\prime}}%
+12A_{\chi^{\prime}}\upsilon_{\mathbf{\chi}_{1}^{\prime}}\upsilon
_{\mathbf{\chi}_{2}^{\prime}}.\nonumber
\end{align}%
\[
\]

\end{document}